\documentclass[reprint,superscriptaddress,nofootinbib,amsmath,amssymb,
 aps,twocolumn,prd]{revtex4-2}

\usepackage{natbib}
\usepackage{aas_macros}
\usepackage{graphicx}
\usepackage{dcolumn}
\usepackage{bm}
\usepackage{txfonts}
\usepackage{threeparttable}
\usepackage{latexsym}
\usepackage{amsbsy}
\usepackage{float}
\usepackage{mathrsfs}
\usepackage{color}
\usepackage{array}

\newcommand{\BibitemShut}[1]{}

\begin{document}


\title{Detection of Cosmic Magnification via Galaxy Shear - Galaxy Number Density Correlation \\from HSC Survey Data}

\author{Xiangkun Liu}
\email[]{liuxk@ynu.edu.cn}
\affiliation{South-Western Institute for Astronomy Research, Yunnan University, Kunming 650500, P.R.China}
\author{Dezi Liu}
\affiliation{South-Western Institute for Astronomy Research, Yunnan University, Kunming 650500, P.R.China}
\author{Zucheng Gao}
\affiliation{School of Physics, Peking University, Beijing 100871, P.R.China}
\author{Chengliang Wei}
\affiliation{Purple Mountain Observatory, Chinese Academy of Sciences, Nanjing, 210023, P.R.China}
\author{Guoliang Li}
\affiliation{Purple Mountain Observatory, Chinese Academy of Sciences, Nanjing, 210023, P.R.China}
\affiliation{School of Astronomy and Space Science, University of Science and Technology of China, Hefei, 230026, P.R.China}
\author{Liping Fu}
\affiliation{The Shanghai Key Lab for Astrophysics, Shanghai Normal University, 100 Guilin Road, Shanghai 200234, P.R.China}
\author{Toshifumi Futamase}
\affiliation{Department of Astrophysics and Meteorology, Kyoto Sangyo University, Kita-ku, Kyoto 603-8555, Japan}
\author{Zuhui Fan}
\email[]{zuhuifan@ynu.edu.cn}
\affiliation{South-Western Institute for Astronomy Research, Yunnan University, Kunming 650500, P.R.China}

\date{\today}

\begin{abstract}
We propose a novel method to detect cosmic magnification signals by cross-correlating foreground convergence fields constructed from galaxy shear measurements with background galaxy positional distributions, namely shear-number density correlation. 
We apply it to the Hyper Suprime-Cam Subaru Strategic Program (HSC-SSP) survey data. With 27 non-independent data points and their full covariance, $\chi_0^2\approx 34.1$ 
and $\chi_T^2\approx 24.0$ with respect to the null and the cosmological model with the parameters from HSC shear correlation analyses in \citet{Hamana2020}, respectively. 
The Bayes factor of the two is $\log_{10}B_{T0}\approx 2.2$ assuming equal model probabilities of null and HSC cosmology, showing a clear detection of the magnification signals. 
Theoretically, the ratio of the shear-number density and shear-shear correlations can provide a constraint on the effective multiplicative shear bias $\bar m$ using internal data themselves. 
We demonstrate the idea with the signals from our HSC-SSP mock simulations and rescaling the statistical uncertainties to a survey of $15000\deg^2$. 
For two-bin analyses with background galaxies brighter than $m_{lim}=23$, the combined analyses lead to a forecasted constraint of $\sigma(\bar m) \sim 0.032$, 
$2.3$ times tighter than that of using the shear-shear correlation alone. 
Correspondingly, $\sigma(S_8)$ with $S_8=\sigma_8(\Omega_\mathrm{m}/0.3)^{0.5}$ is tightened by $\sim 2.1$ times.
Importantly, the joint constraint on $\bar m$ is nearly independent of cosmological parameters.
Our studies therefore point to the importance of including the shear-number density correlation in weak lensing analyses, which can provide
valuable consistency tests of observational data, and thus to solidify the derived cosmological constraints.
\end{abstract}

\maketitle

\section{Introduction}
Weak gravitational lensing effects (WL), mainly consisting of cosmic shear and magnification, probe directly the matter distribution 
of the Universe, and are uniquely important in cosmological studies \citep{BS2001, HJ2008, van2010a, FuFan2014, Kil2015}. 

Cosmic shear induces tiny image distortions on distant galaxies. By accurately measuring their shapes, shear signals can be 
extracted statistically. This has been the main stream of current weak lensing studies. 
With large surveys, such as the on-going Dark Energy Survey \citep[DES,][]{DES2016}, the Kilo-Degree Survey \citep[KiDS,][]{KiDS2013}, 
and the Hyper Suprime-Cam Subaru Strategic Program survey\citep[HSC-SSP,][]{HSC2018}, the derived cosmological constraints are improving significantly
\citep[e.g.,][]{Kil2013, Hilde2017, Abbott2018, Hamana2020}.

Cosmic magnification, on the other hand, leads to slight changes of the galaxy spatial distribution due both to the lensing magnification of galaxy flux and the lensing increase
of solid angle \citep{BN1992, BN1995}. In principle, cosmic magnification can be extracted from photometric surveys without the need of shape measurements.  
There have been extensive theoretical investigations on the complementarity of cosmic magnification and shear and how to accurately reconstruct cosmic magnification power spectrum
purely from the spatial distribution of galaxies \citep[e.g.,][]{Menard2002, Menard2003, Menard2010,van2010b, Yangxj2017,Zhangpj2018}, where the galaxy intrinsic clustering and the bias are the main concerns. 
Observational detections of cosmic magnification using galaxy distributions have been reported \citep[e.g.,][]{MS2012, FH2014,GS2018}. 

Currently, most of the studies concentrate either on cosmic shear or on cosmic magnification. Combined analyses employing both sides of the signals are done for 
clusters of galaxies \citep{Umetsu2014}, but are still lacking on cosmic scales.
In this paper, we put forward a method to cross-correlate foreground shear and background position of galaxies. 
This not only can lead to clean detections of cosmic magnification signals without being affected by galaxy bias, but also can, by combining with shear-shear correlation,
provide important consistency tests of observational data themselves and further constrain the shear measurement bias. Validated by mock simulations, we apply the analyses 
to HSC-SSP data, and show cross-correlation signals with high significance. We also forecast the potential constraint on the multiplicative bias parameter $m$ 
of shear measurement from future large surveys.

The paper is organized as follows. In Sec. \ref{method}, we present our cross-correlation methodology theoretically. Sec. \ref{Obsdata} describes the HSC-SSP data used in our analyses. Sec. \ref{CCanalyses} contains details of the cross-correlation analyses. In Sec. \ref{results}, we show the results measured from our HSC mock simulations and from observations. The cosmological potential of combining the shear-shear and shear-number density correlations is shown in Sec. \ref{forecast}. Summary and discussions are given in Sec. \ref{final}.   

\section{Method}
\label{method}
In our analyses, we cross-correlate the foreground convergence field constructed from the shear measurements and the spatial distribution of background galaxies (hereafter, shear-number density correlation). Specifically, 
we have
\begin{equation}
\omega_{ij}(\theta)=\langle{\kappa_i(\hat{\mathbf{n}})}\delta_j(\hat{\mathbf{n}}',m_{lim})\rangle_\theta,
\label{2ptwij}
\end{equation}
where $\theta$ is the angular separation of the two directional vectors $\hat{\mathbf{n}}$, $\hat{\mathbf{n}}'$, $\kappa_i$ is the foreground lensing convergence field from the
galaxy shear measurements, and $\delta_j$ is the fluctuation field of background galaxy number count with the limiting magnitude $m_{lim}$.  The latter is given by    
\begin{equation}
\delta_j(\hat{\mathbf{n}}',m_{lim})=\delta_{gj}(\hat{\mathbf{n}}',m_{lim})+\delta_{{\mu}j}(\hat{\mathbf{n}}',m_{lim}),
\label{deltaO}
\end{equation}
where $\delta_{gj}$ describes the intrinsic clustering of the galaxy sample, which is related the density fluctuation field with a bias factor involved.
The second term $\delta_{{\mu}j}$ is the fluctuation arising from the lensing magnification effect. 

Assuming the foreground and the background are well separated, the intrinsic clustering term does not contribute to the correlation of Eqn.(\ref{2ptwij}), and therefore 
eliminating the impact of the galaxy bias. For $\delta_{{\mu}j}$, it is \citep[e.g.,][]{BS2001, GS2018}   
\begin{equation}
\delta_{{\mu}j}(\hat{\mathbf{n}}',m_{lim})=\mu^{\alpha(m_{lim})-1}-1\approx 2\kappa_j(\hat{\mathbf{n}}')[\alpha(m_{lim})-1],
\label{magcontrast3}
\end{equation}
where $\alpha(m_{lim})=2.5\mathrm{d}[\mathrm{log}_{10}N(m_{lim})]/{\mathrm{d}m_{lim}}$ is the local slope of the galaxy cumulative number count $N(m_{lim})$ in log scale, 
and $\mu$ is the lensing magnification. The second expression in Eqn.(\ref{magcontrast3}) is the approximation in the weak lensing limit with $\kappa\ll 1$ and $\mu\approx 1+2\kappa$.
We then have
\begin{equation}
\begin{aligned}
\omega_{ij}(\theta)
&=\langle 2[\alpha(m_{lim})-1]\kappa_i(\hat{\mathbf{n}})\kappa_j(\hat{\mathbf{n}}')\rangle_\theta\\
&=2[\alpha(m_{lim})-1]\frac{1}{2\pi}\int_0^\infty l\mathrm{d}l P_\kappa^{ij}(l)J_0(l\theta),
\end{aligned}
\label{2ptwij_mid}
\end{equation}
where $J_0$ is the zero-th order Bessel function and $P_\kappa^{ij}$ is the convergence cross power spectrum between $i$ and $j$ given by 
\begin{equation}
P_\kappa^{ij}(l)=\int{\mathrm{d}w\frac{q_i(w)q_j(w)}{f_K^2(w)}}P_\delta\left(\frac{l}{f_K(w)},w\right)
\label{Pij},
\end{equation}
with 
\begin{equation}
q_{i(j)}(w)=\frac{3H_0^2\Omega_\mathrm{m}}{2a(w)c^2}\int_w^{w_H}\mathrm{d}w'p_{i(j)}(w')\frac{f_K(w'-w)}{f_K(w')}.
\label{qij}
\end{equation}
Here $w$, $f_K(w)$, $P_\delta$ and $p_{i(j)}(w)$ are the comoving radial distance, comoving angular diameter distance, 3-dimensional matter power spectrum, 
and the radial distribution calculated from the redshift distribution for $i$-th ($j$-th) redshift bin, respectively. 
$\Omega_\mathrm{m}$, $H_0$, $a$ and $c$ are the present matter density parameter, Hubble constant, cosmic scale factor, and the speed of light, respectively.
We note that the shear-number density correlation here is, in some sense, the inverse of the galaxy-galaxy lensing analyses which correlate foreground galaxy positions
with background shear signals.

For the shear-shear correlation between $i$ and $j$, we have \citep{Kil2013} 
\begin{equation}
\xi_{+ij}(\theta)=\langle\gamma_{ti}(\hat{\mathbf{n}})\gamma_{tj}(\hat{\mathbf{n}}')+\gamma_{\times i}(\hat{\mathbf{n}})\gamma_{\times j}(\hat{\mathbf{n}}')\rangle=\frac{1}{2\pi}\int_0^\infty l\mathrm{d}l P_\kappa^{ij}(l)J_0(l\theta).
\label{2ptfinal}
\end{equation}

It is seen from Eqns.(\ref{2ptwij_mid}) and (\ref{2ptfinal}) that theoretically, $\omega_{ij}$ and $\xi_{+ij}$ have exactly the same cosmology dependence with the ratio of the two 
being a constant $2[\alpha(m_{lim})-1]$, independent of $\theta$. Observationally, the shear measurements often involve a multiplicative bias and an additive bias. 
While the additive bias can be estimated from the data, the determination of the multiplicative bias $m$ needs extra calibrations. Assuming that we have 
no knowledge of $m$, then the two correlations calculated from the observational data should be, to a very good approximation, $\hat \omega_{ij}=(1+\bar m_{i})\omega_{ij}$ and 
$\hat \xi_{+ij}=(1+\bar m_i)(1+\bar m_j)\xi_{+ij}$, where $\bar m_i$ and $\bar m_j$ are the effective multiplicative bias of the shear sample $i$ and shear sample $j$, respectively.  
Then
\begin{equation}
\hat \omega_{ij}(\theta)/\hat \xi_{+ij}(\theta)=2[\alpha(m_{lim})-1]/(1+\bar{m_j}).
\label{ratio}
\end{equation}
Thus the ratio of Eqn.(\ref{ratio}) provides a means to calibrate $m$ in a cosmology-independent way using the observational data themselves if $\alpha$ can be determined accurately. 

In the following, we apply the cross-correlation analyses to the HSC-SSP data. 

\section{Observational data}
\label{Obsdata}
HSC-SSP is a large imaging survey with weak lensing cosmology as a major science driver \citep{HSC2018}.   
In our analyses, we use data from first and second data release PDR1 and PDR2 \citep{HSCPDR1, HSCPDR2}, and the first-year shear catalog (S16A) \citep{HSCshear}. 

\begin{figure*}
\includegraphics[width=2.0\columnwidth, height=0.76\columnwidth]{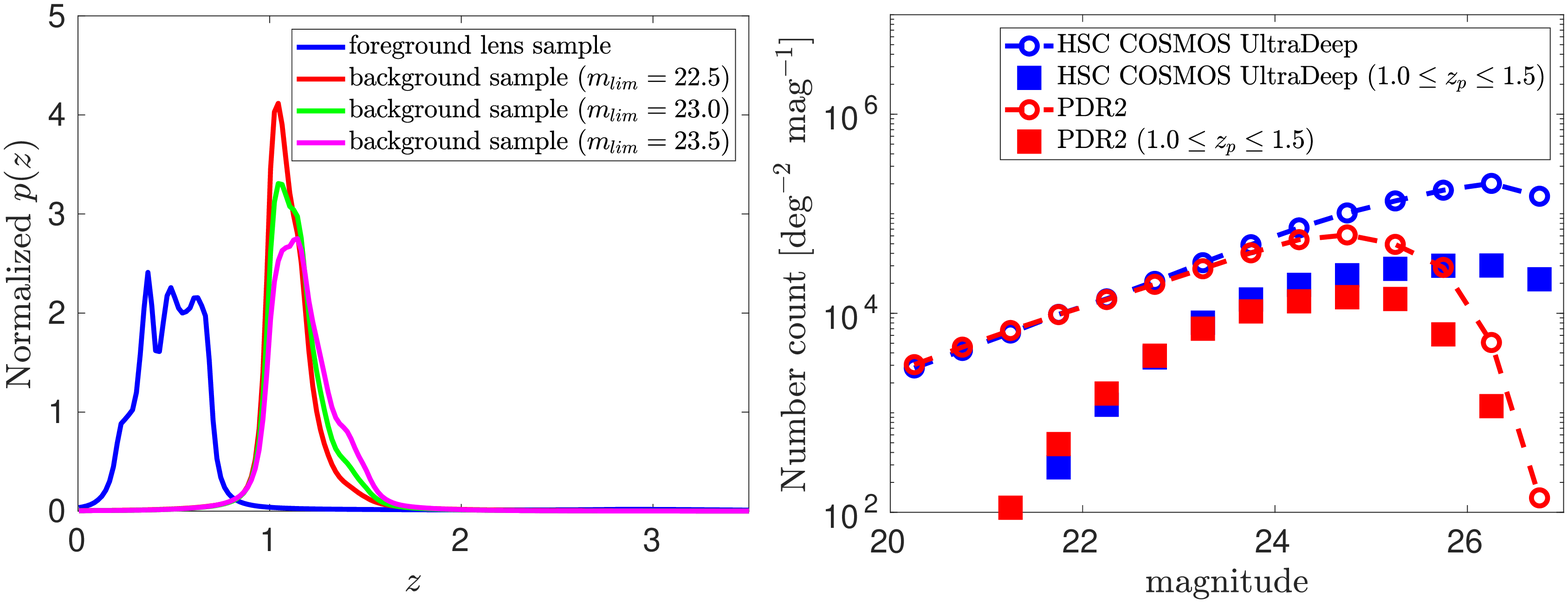}
\caption{Left: Normalized redshift distributions of the foreground galaxy sample and background PDR2 galaxy subsamples with different magnitude cuts.
Right: $i$-band differential galaxy number density of PDR2 wide (red) in comparison with that of
HSC PDR1 COSMOS UltraDeep catalog (blue). Circles and squares are for the full and high redshift samples, respectively.}
\label{fig:pz_mag_plot}
\end{figure*}

To implement the cross correlation analyses, we divide galaxies into low- and high-redshift bins, with $0.2\le z_p\le 0.7$ and $1.0\le z_p\le 1.5$, respectively, where
$z_p$ is the best-fit value of the photometric redshift  (photo-z) of a galaxy \citep{HSCpz1, HSCpz2}.
For the low-redshift bin, we use the S16A shear catalog to construct the
convergence field, adopting the selection criteria listed in Table 4 of \citet{HSCshear}. The sample's redshift distribution  
is shown in blue in the left panel of Figure \ref{fig:pz_mag_plot} by stacking the photo-z distribution of individual galaxies. 

For the high-redshift bin, we take the photometric data from PDR2.
To ensure the data quality and avoid introducing significant photometric bias, 
besides the common flag criteria in \citet{HSCPDR1}, two extra cuts are applied: the $i$-band flux signal-to-noise (S/N) ratio $\ge 5$ and iblendness\_abs\_flux, a
measure of how strongly an object is blended, $\le 10^{-0.375}$. We further take i\_{extendedness}\_{value} = 1 at $i < 24.5$ to remove the impact of point sources \citep{HSCPDR1}. 
In addition, we make a cross match in position between S16A and PDR2, and exclude those galaxies with 
a small position difference ($\le 0.5$ arcsec) but a large magnitude difference ($\ge 0.5$), which amount to $\sim 3\%$ of the whole sample.
For magnification detections, we calculate the magnitude of each galaxy by ${mag}=\mathrm{i\_cmodel\_mag}-a_i$ with $a_i$ the 
Galactic extinction correction in $i$-band given in the PDR2 catalog. The right panel of Figure \ref{fig:pz_mag_plot} shows the magnitude distributions.
By comparing with the HSC COSMOS UltraDeep data \citep{HSCPDR1}, we find that the high-redshift sample is about 80\% and 
65\% complete up to $mag\approx 23.5$ and $24.5$, respectively.  
For magnification analyses, we then construct different subsamples with ${mag} \le 22.5$, $\le 23.0$, and $\le 23.5$, respectively. The impact of the incompleteness is tested with our mock simulations to be described later in the paper.
The redshift distributions of these subsamples are shown in different colors in the left panel of Figure \ref{fig:pz_mag_plot}.

It is seen that the low- and high-redshift samples have only a small fraction of redshift overlap. 
We numerically check the contamination from the galaxy-convergence intrinsic correlation in the overlapped region. It is about $1-2\%$ to $\omega_{ij}/\{2[\alpha(m_{lim})-1]\}$, negligible in the analyses here. 
For future high precision studies, we should investigate this contamination carefully.

To perform the analyses, we select 52 overlapped fields between S16A and PDR2, with 40 having an area of $1.5\times1.5\deg^2$ each, and 
12 with $1.0\times 1.0\deg^2$ each cropped from partially-overlapped tracts. The total area is $\sim 100\deg^2$.  
For the low-redshift shear sample, the weighted number density is $n_g\sim 8.0~\mathrm{arcmin}^{-2}$.
For the high-redshift PDR2 data, $n_g\sim 0.28$, $0.77$, and $1.69~\mathrm{arcmin}^{-2}$, for $m_{lim}=22.5$, $23.0$, and $23.5$, respectively. 
We estimate the slope $\alpha$ by bootstrapping the 52 fields to generate 1000 sets of data. The average $\alpha$ 
for the three $m_{lim}$ is $2.48$, $1.97$, and $1.48$ calculated with the interval of $\Delta m=0.01$ around $m_{lim}$. 
The corresponding $1\sigma$ uncertainty is $0.052$, $0.030$, and $0.018$. We test by using $\Delta m=0.05$, and the $\alpha$ value differences are less than $1\%$.

For the corresponding shear-shear correlation analyses, S16A is used for both low- and high-redshift samples. We apply the same $m_{lim}$  
to construct high-redshift shear subsamples. Their redshift distributions 
are nearly identical to the ones shown in the left panel of Figure \ref{fig:pz_mag_plot}. 

\section{Cross-correlation analyses}
\label{CCanalyses}
We first reconstruct the convergence fields from the low-redshift shear sample with the same procedures 
detailed in \citet{Liu2015}, which are summarized here. 
(1) The smoothed shear fields are calculated following \citet{Oguri2018}. Specifically,  
\begin{equation}
\begin{aligned}
\langle\bm{\epsilon}\rangle(\bm{\theta})
&=\frac{\sum_k w_{k}W_{\theta_\mathrm{G}}(\bm{\theta_k-\theta})\bm{e}_k}{2\sum_k w_{k}(1-e_{\mathrm{rms},k}^2)(1+m_k)W_{\theta_\mathrm{G}}(\bm{\theta_k-\theta})}\\
&-\frac{\sum_k w_{k}W_{\theta_\mathrm{G}}(\bm{\theta_k-\theta})\bm{c}_k}{\sum_k w_{k}(1+m_k)W_{\theta_\mathrm{G}}(\bm{\theta_k-\theta})}
\end{aligned}
\label{smoothing}
\end{equation}
where $\bm{e}_k$ is the two-component ellipticity of galaxy $k$, and $w_{k}$, $\bm{c}_k$, $m_k$ and $e_{\mathrm{rms},k}$ are the corresponding weight, additive and 
multiplicative biases of shear measurements, and the intrinsic ellipticity dispersion per component. We take the Gaussian smoothing kernel with 
$W_{\theta_{\rm G}}(\boldsymbol{\theta})=1/({\pi\theta_{\rm G}^2})\exp{(-{|\boldsymbol{\theta}|^2}/{\theta_{\rm G}^2})}$ and $\theta_{\rm G}=1.5$ arcmin.  
(2) From $\langle\bm{\epsilon}\rangle(\bm{\theta})$, we reconstruct the convergence field $K_{\rm GRID}$ sampled on $1024\times 1024$ pixels for each of the 52 fields. 
(3) We generate smoothed filling-factor maps $f$ from the galaxy spatial distributions of the shear sample,
and exclude regions with $f<0.6$ \citep{Liu2015}. The outermost $5\theta_\mathrm{G}$ pixels in each of the four sides of a map are also removed to reduce the boundary effects. 
(4) We rotate each of the galaxies randomly, and reconstruct the noise part $N_\mathrm{GRID}$ 
of the convergence fields following steps (1) and (2). In total, 30 sets of $N_\mathrm{GRID}$ are generated for each field. 

For the high-redshift galaxies from PDR2, we apply the same mask and boundary exclusion criteria as those of the foreground convergence fields. 
The remaining galaxies make up the background subsamples $G_\mathrm{L}$ with different $m_{lim}$.  
We also generate 30 sets of random samples $G_\mathrm{R}$ from the galaxies in each field by randomly populating them spatially.

\begin{figure*}
\includegraphics[width=2.0\columnwidth, height=0.63\columnwidth]{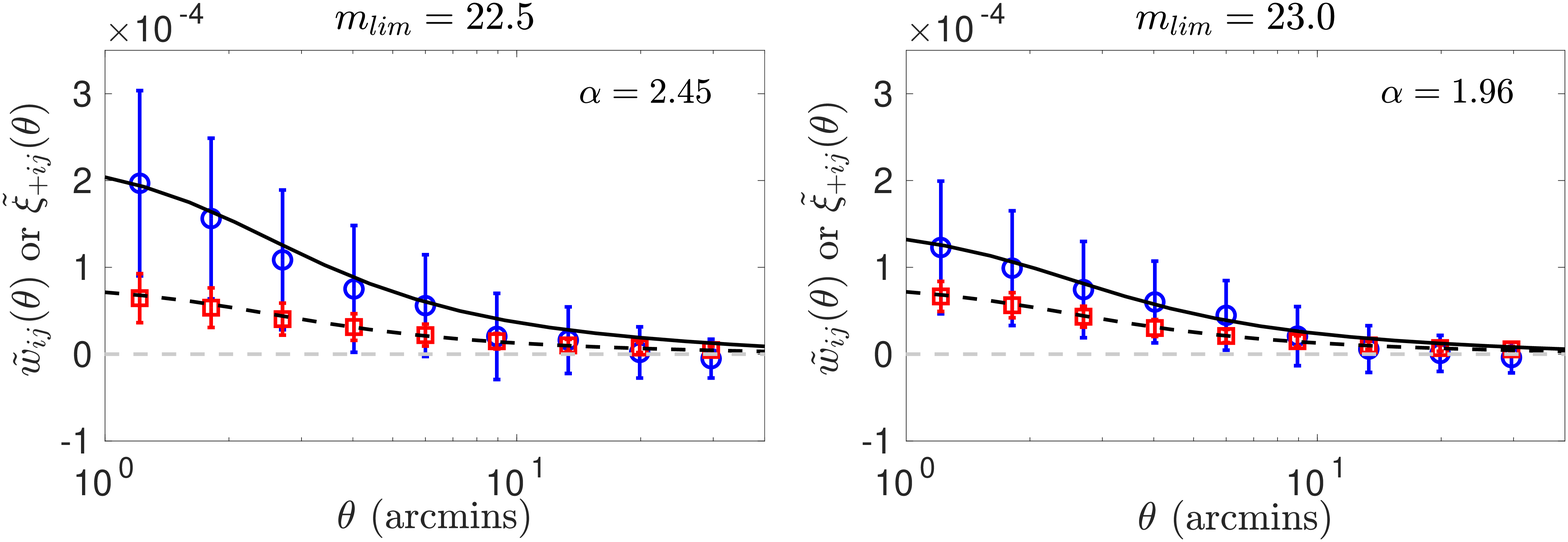}
\includegraphics[width=2.0\columnwidth, height=0.63\columnwidth]{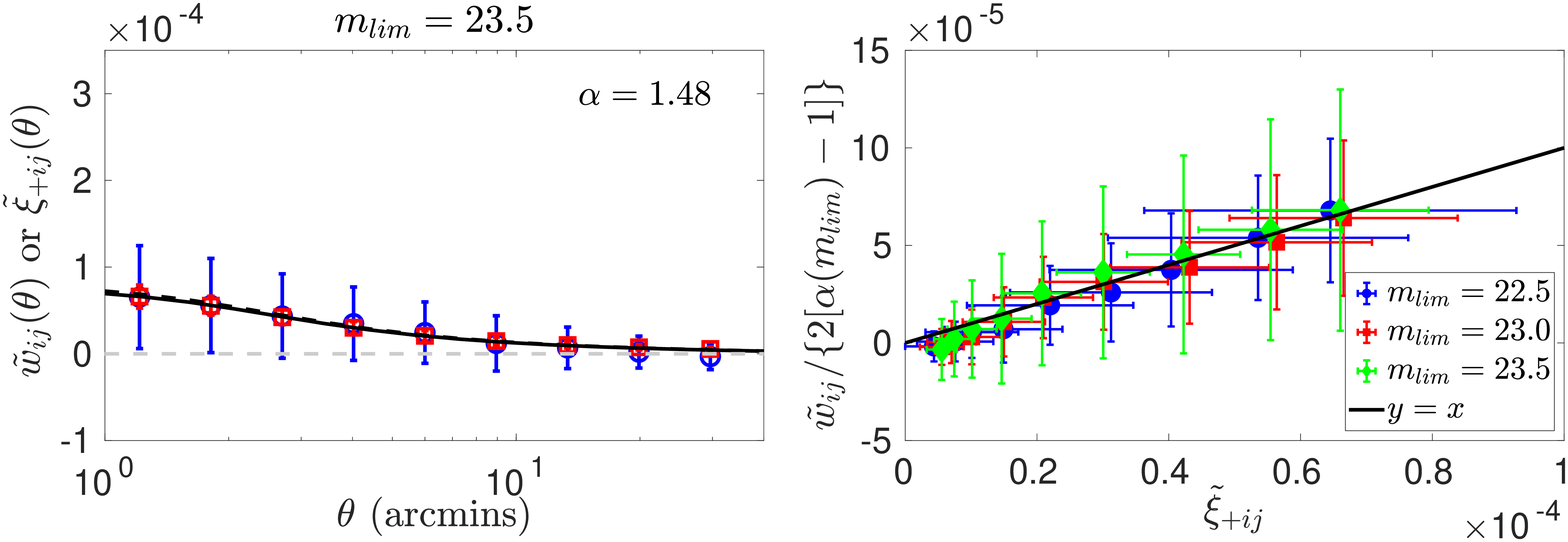}
\caption{\label{fig:mockvalidate} Results from the simulated mocks with $\tilde \omega_{ij}$ and $\tilde \xi_{+ij}$ shown in blue and red, respectively. 
The black solid and dashed lines are the corresponding theoretical predictions. 
The bottom right panel shows $\tilde \xi_{+ij}$ vs. $\tilde \omega_{ij}/\{2[\alpha(m_{lim})-1]\}$ with the theoretical expectation of 1:1 shown as the solid line. 
}
\end{figure*}

An estimator for the shear-number density cross-correlation is then given by 
\begin{equation}
\begin{aligned}
\tilde \omega_{ij}(\theta)&
=(\langle K_\mathrm{GRID} G_\mathrm{L}\rangle-\langle K_\mathrm{GRID} G_\mathrm{R}\rangle)\\
&-(\langle N_\mathrm{GRID} G_\mathrm{L}\rangle-\langle N_\mathrm{GRID} G_\mathrm{R}\rangle).
\end{aligned}
\label{wijfinal2}
\end{equation}
The calculations are done using \textbf{Athena}\footnote{http://www.cosmostat.org/software/athena} \citep{Sch2002}. 
Considering the size of each selected field, a total of 9 bins with logarithmic bin-width of $\Delta \mathrm{log_{10}} \theta\sim0.173$ are chosen with 
the central $\theta$ ranging from 1.21 to 29.65 arcmin. Note that in model calculations, the kernel $W_G$ applied to the foreground needs to be included  
in Eqn.(\ref{2ptwij_mid}) and also in Eqn.(\ref{2ptfinal}).

To estimate the shear-shear correlation between low- and high-redshift bins, we construct low-redshift shear fields $\Gamma_{\mathrm{GRID}}$ on pixels 
directly from the smoothed shear fields $\langle\bm{\epsilon}\rangle$.  
For high redshift part, we build shear subsamples $\bm{\gamma_{j}}$ with $m_{lim}=22.5, 23.0$, and $23.5$ from S16A. 
The cosmic shear two-point correlations are then calculated 
over all preserved pixel-galaxy pairs ($ab$) within some  bin  around $\theta$ using $\tilde \xi_{+ij}(\theta)=\langle\Gamma_{t\mathrm{GRID}}\gamma_{tj}+\Gamma_{\times \mathrm{GRID}}\gamma_{\times j}\rangle=\sum_{ab}{w_b[\Gamma_{1\mathrm{GRID}}(\bm{\theta_a})\gamma_{1j}(\bm{\theta_b})+\Gamma_{2\mathrm{GRID}}(\bm{\theta_a})\gamma_{2j}(\bm{\theta_b})]}/\sum_{ab}w_b$, where $\bm{\gamma_j}(\bm{\theta_b})=\frac{1}{(1+\hat{m})}[\frac{\bm{e_b}}{2\mathcal{R}}-\bm{c_b}]$, with an ensemble correction of multiplicative bias $\hat{m}$ and reponsivity factor $\mathcal{R}$ applied following \citep{HSCshear}.

\begin{figure*}
\includegraphics[width=2.0\columnwidth, height=0.63\columnwidth]{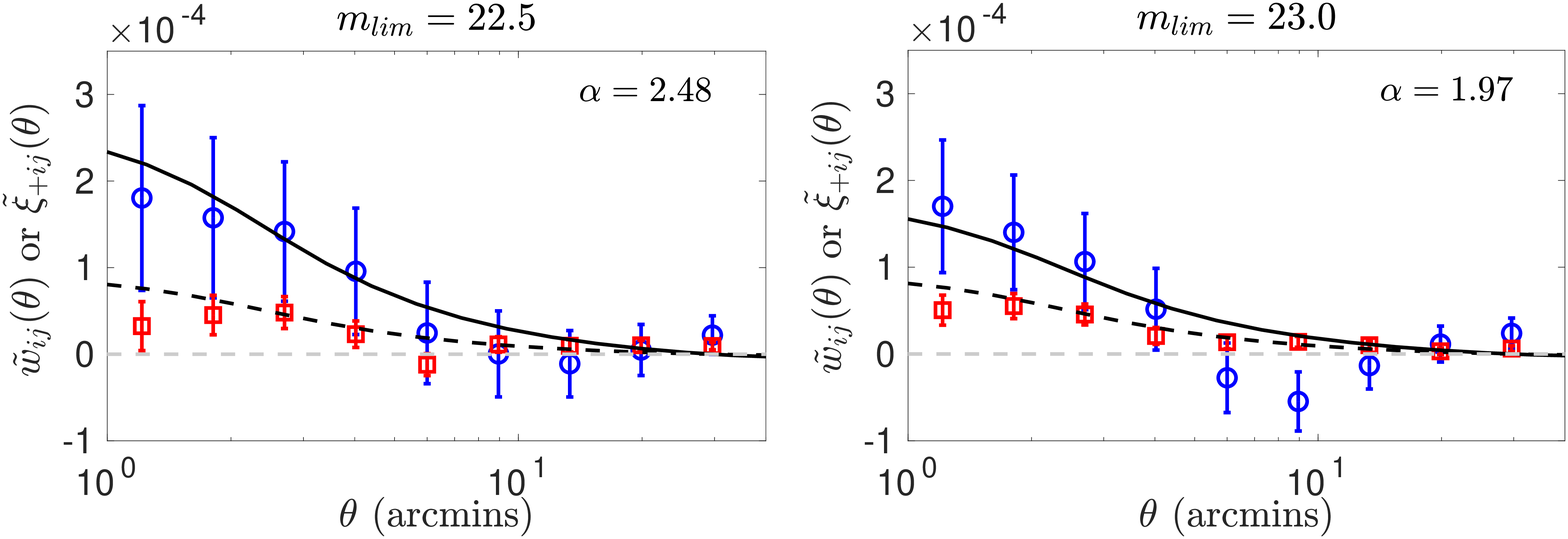}
\includegraphics[width=2.0\columnwidth, height=0.63\columnwidth]{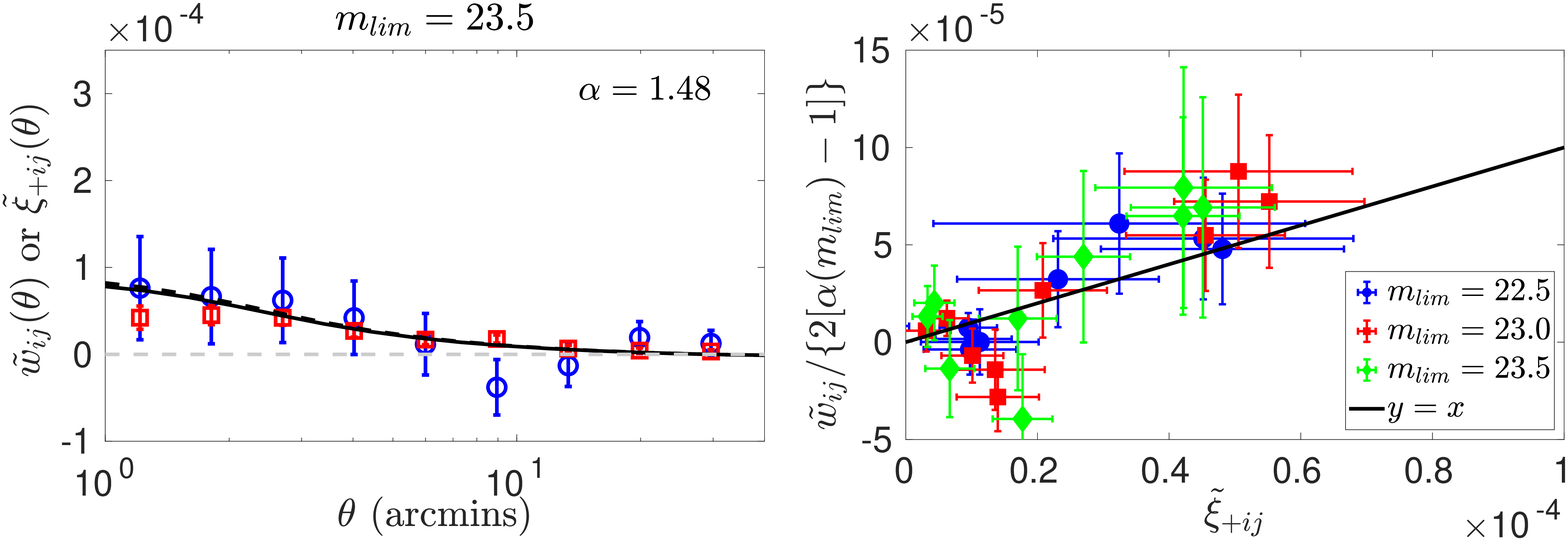}
\caption{\label{fig:obs_results} 
Correlation results from HSC-SSP observational data.
}
\end{figure*}

\section{HSC mock validation and observational results}
\label{results}
Before applying to data, we validate our analyses by building HSC-SSP mocks from ray-tracing simulations described in \citet{Liu2014}. 
We have 59 lensing-plane outputs between $z=3$ and $z=0$, and each consists of 96 maps of $3.5\times 3.5\deg^2$ each.

With these lensing maps, we create HSC mocks as follows. (1) For high-redshift PDR2 mocks, the galaxy positions and magnitudes from PDR2 of the 52 fields are adopted. Because of the independence between the original HSC galaxy positions/magnitudes 
and our simulated lensing signals, we regard them as the unlensed quantities. The redshift $z_p$ is also kept. 
(2) For shear mocks, we preserve the position, magnitude, redshift $z_p$, and shear related quantities of each galaxy from S16A. 
We randomly rotate them to eliminate the original shear signals in the data. Otherwise the signals embedded in the HSC data will contaminate the mock simulation results of both shear-number density and shear-shear correlations.
(3) We pad the mock galaxies into our simulated maps, and calculate the lensing signals at their positions  
by interpolating (in redshift and angular position) the grid signals on lensing maps. For high redshift PDR2 mocks, we obtain the deflection angle and magnification $\mu$ for each galaxy, and adjust its position and magnitude (adding a factor of 2.5$\log_{10}(1/\mu)$) accordingly.  
For shear mocks, we follow the procedures of \citet{Oguri2018} including the shear bias from S16A to our simulated lensing signal of each galaxy.
(4) We do different paddings of (3), and generate 20 mocks for each of the 52 fields. (5) We randomly select one from 
the 20 for each field and compose one set of HSC mock with 52 fields. In total 200 sets of mock data are generated.  
(6) The same convergence reconstruction and correlation estimates are done as for the observational data. 
(7) The covariance matrices are estimated from the 200 mocks. The results are about the same as those calculated from 400 mocks. 

In Figure \ref{fig:mockvalidate}, we show the results averaged over the 200 mocks. The blue and red symbols are for the shear-number density and shear-shear
correlations, respectively. The solid and dashed lines are the corresponding model predictions with the listed $\alpha(m_{lim})$ values. 
The last panel shows $\tilde \xi_{+ij}$ vs. $\tilde \omega_{ij}/\{2[\alpha(m_{lim})-1]\}$ with the theoretical expectation of 1:1 black line noting that we correct for the shear bias in calculating the correlations as shown in Eqns.(\ref{smoothing}) and (\ref{wijfinal2}) and $\tilde \xi_{+ij}(\theta)$.  
The mock results agree well with that of the model, validating our analyese.   

Our mock settings implicitly assume the completeness of the high-redshift PDR2 sample, which is not true for the real data. 
Ideally, to evaluate the effect of incompleteness, we should generate mock data from deeper surveys, and do analyses  
for the selected incomplete samples. This is infeasible here because of the small coverage of the HSC UltraDeep survey. 
We thus assess the impact by assuming an artificial incompleteness function to select galaxies from our high-redshift PDR2 mocks.
Our results show, consistent with other studies, that the magnification signals from incomplete samples follow the predictions with $\alpha$ calculated from the underlying 
complete sample\citep{Hilde2016}. For our specific mocks, with the completeness fraction of $\sim 80\%$ and $\sim 70\%$, 
the $\alpha$ value estimated from the incomplete samples is lower by $\sim 4\%$ and $\sim 8\%$, respectively, than that from the complete sample. 
The effect of such changes on the shear-number density correlation is well within the statistical uncertainties here. In follow-up studies, 
we will investigate in detail the impact of incompleteness on future high precision analyses.

The results from the observational data are shown in Figure \ref{fig:obs_results}
with symbols and lines the same as those in Figure \ref{fig:mockvalidate}, except that the model is $\Omega_{\rm m}=0.346$ and $\sigma_8=0.749$ \citep{Hamana2020}.    

The observed shear-number density correlations follow the expected trend of varying $\alpha$, and are consistent
with the theoretical predictions. With 27 non-independent data points and the full covariance,  
$\chi_T^2=24.0$ and $\chi_0^2=34.1$ with respect to the model and to the null, respectively. The corresponding
p-value of $p(\ge\chi_T^2)$ is $0.63$, four times higher than $p(\ge\chi_0^2)=0.16$. 
Assuming Gaussian likelihood and equal probability of the considered cosmological model and the null, the Bayes factor
is $\log_{10}B_{T0}=\log_{10}\{\exp[-0.5(\chi^2_T-\chi^2_0)]\}\approx 2.2$, showing a solid detection of the signals \citep{Raftery1995}.  

\section{Cosmological potentials}
\label{forecast}
With the shear multiplicative bias $m$ corrected according to that given in S16A, we see from Figure \ref{fig:obs_results} that   
$\tilde \omega_{ij}/\{2[\alpha(m_{lim})-1]\}$ and $\tilde \xi_{+ij}$ follow each other well showing the consistency of the HSC shear and magnification signals,
and the residual shear bias after correction is insignificant within the error bars.

Without the bias information (or it is incorrect),
the ratio of the two correlations is given in Eqn.(\ref{ratio}) containing the factor of $(1+\bar m_j)$. If $\alpha(m_{lim})$ can be well determined,
the combination of the two can be used to calibrate $\bar m_j$ using observational data themselves, independent of cosmology.
To demonstrate the idea, we perform a forecast of cosmological constraints by combining the shear-number density and shear-shear correlations in comparison with
that using the shear-shear correlation alone.

To avoid the instability problem, we do not directly calculate the ratio. Instead, we perform analyses by combining the data vectors of $\tilde \omega_{ij}$ and $\tilde \xi_{+ij}$. 
Specifically, we consider the same two redshift bins as we did for the HSC analyses, but scale the effective survey area from $\sim 60 \deg^2$ to $15000 \deg^2$. 
For the high redshift bin, we use the subsample with $m_{lim}=23$. How to combine different samples with different $m_{lim}$ to enhance the constraining power will be our future tasks. 
The fiducial values of the correlation signals are taken from the mock results, and the covariance matrices are scaled by the considered area. The number densities of different galaxy samples are assumed to be the same as those of HSC analyses.   

We consider three parameters $(\Omega_{\rm m}, \sigma_8, \bar m)$ assuming the same $\bar m$ for low- and high-redshift shear samples. 
The value of $\alpha(m_{lim})=1.96$ from our mocks, and its uncertainty is estimated from the HSC data by bootstrap different fields and then scaled to
15000 $\deg^2$, which gives $\sigma_\alpha\approx 0.002$. 

\begin{figure}
\includegraphics[width=1.0\columnwidth,
height=0.75\columnwidth]{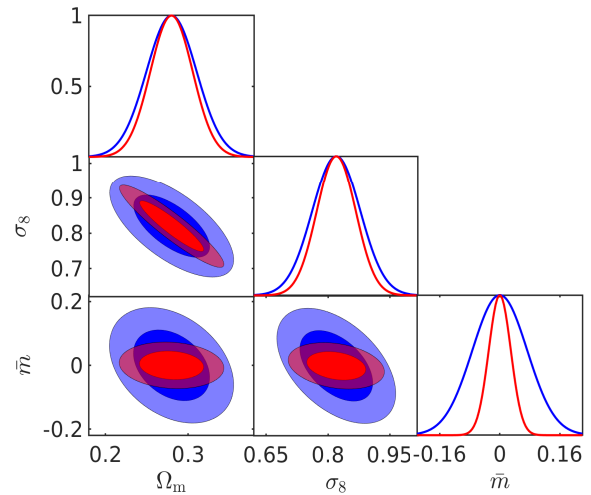}
\caption{\label{fig:fisher} 
Fisher forecast for the constraints on ($\Omega_\mathrm{m}$, $\sigma_8$, $\bar{m}$) with the survey area of $15000\deg^2$ and $m_{lim}= 23.0$ for the high-redshift sample. 
Blue and red are, respectively, the results from $\tilde \xi_{+ij}$ only and that of combining $\tilde \xi_{+ij}$ and $\tilde \omega_{ij}$.}
\end{figure}

The Fisher forecast results are shown in Figure \ref{fig:fisher}, where the blue and red parts are for the case with $\tilde \xi_{+ij}$ only and that of combining $\tilde \xi_{+ij}$ and $\tilde \omega_{ij}$, 
respectively. By including $\tilde \omega_{ij}$, the constraint on $\bar m$ do improve significantly, and the degeneracies with $\Omega_{\rm m}$ and $\sigma_8$ 
are largely eliminated. Consequently, the constraints on the cosmological parameters are tightened, as presented in Table \ref{tab:constraint}.     
With SS and SP for shear-shear and shear-number density correlations and $S_8=\sigma_8(\Omega_{\rm m}/0.3)^{0.5}$, 
the improvement on $\bar m$ and $S_8$ is $2.3$ and $2.1$ times, respectively.

The combination of the two correlations can also potentially reveal scale-dependent shear measurement bias or other systematics because  
then the ratio of the two should have a scale dependence. In the analyses here, because of the relatively large error bars, such a dependence is not seen clearly for HSC data.
We will explore this potential in our follow-up studies with respect to future observations.

\begin{table}
\caption{Forecast constraints for a $15000\deg^2$ survey}
\label{tab:constraint}
\begin{center}
  \leavevmode
    \begin{tabular}{c c c} \hline
                           &  SS only & SS \& SP\\
                             \hline
                $\Omega_{\mathrm{m}}$ ($68\%$CL) & 0.032 & 0.027\\
                $\sigma_8$ ($68\%$CL) & 0.059 & 0.051\\
                $\bar{m}$ ($68\%$CL) & 0.075 & 0.032\\
                $S_8$ ($68\%$CL) & 0.043  & 0.020\\
                             \hline
     \end{tabular}
    \end{center}
\end{table}

\section{Summary and Discussion}
\label{final}
In this study, we propose the shear-number density correlation analyses and apply to HSC data. The cosmic magnification signals are clearly detected with no galaxy bias involved. 
Combining with the shear-shear correlation allows us to check the data consistency internally.  
Our results show that the cosmic shear and magnification signals from HSC are consistent with each other within error bars. 
We further explore the potential of including shear-number density correlations in cosmological studies, revealing significant improvements of the constraints on $\bar m$ and $S_8$.
Thus for future large weak lensing surveys, we advocate to include this information  
in addition to the shear-shear, galaxy clustering and galaxy-galaxy lensing analyses to enhance the cosmological gains. For that, this paper provides a proof of concept.
We will carry out more detailed studies about different effects, such as sample incompleteness, dust effect, photo-z errors and intrinsic alignments. We will also investigate 
analyses methods to increase the signal-to-noise ratio of shear-number density correlations and how to inlcude them in multi-bin tomographic studies. 

\begin{acknowledgments}
The Hyper Suprime-Cam (HSC) collaboration includes the astronomical communities of Japan and Taiwan, and Princeton University. The HSC instrumentation and software were developed by the National Astronomical Observatory of Japan (NAOJ), the Kavli Institute for the Physics and Mathematics of the Universe (Kavli IPMU), the University of Tokyo, the High Energy Accelerator Research Organization (KEK), the Academia Sinica Institute for Astronomy and Astrophysics in Taiwan (ASIAA), and Princeton University. Funding was contributed by the FIRST program from Japanese Cabinet Office, the Ministry of Education, Culture, Sports, Science and Technology (MEXT), the Japan Society for the Promotion of Science (JSPS), Japan Science and Technology Agency (JST), the Toray Science Foundation, NAOJ, Kavli IPMU, KEK, ASIAA, and Princeton University. 

This paper makes use of software developed for the Large Synoptic Survey Telescope. We thank the LSST Project for making their code available as free software at  http://dm.lsst.org.

The Pan-STARRS1 Surveys (PS1) have been made possible through contributions of the Institute for Astronomy, the University of Hawaii, the Pan-STARRS Project Office, the Max-Planck Society and its participating institutes, the Max Planck Institute for Astronomy, Heidelberg and the Max Planck Institute for Extraterrestrial Physics, Garching, The Johns Hopkins University, Durham University, the University of Edinburgh, Queen’s University Belfast, the Harvard-Smithsonian Center for Astrophysics, the Las Cumbres Observatory Global Telescope Network Incorporated, the National Central University of Taiwan, the Space Telescope Science Institute, the National Aeronautics and Space Administration under Grant No. NNX08AR22G issued through the Planetary Science Division of the NASA Science Mission Directorate, the National Science Foundation under Grant No. AST-1238877, the University of Maryland, and Eotvos Lorand University (ELTE) and the Los Alamos National Laboratory.

The analyses of this paper are based on data collected at the Subaru Telescope and retrieved from the HSC data archive system, which is operated by Subaru Telescope and Astronomy Data Center at National Astronomical Observatory of Japan.

We thank the stimulating discussions with Pengjie Zhang, Jun Zhang and Martin Kilbinger.
The calculations of this study are partly done on the Yunnan University Astronomy Supercomputer.
X.K.L. acknowledges the supports from NSFC of China under grant 11803028 and YNU grant C176220100008. Z.H.F. is supported by NSFC of China under grants 11933002, U1931210 and 11653001.  Z.H.F. and X.K.L. are also supported by a grant of CAS Interdisciplinary Innovation Team. D.Z.L. acknowledges the support from the China Postdoctoral Science Foundation under the
grant 2019M663582. C.L.W. acknowledges the support from NSFC of China under grant 11903082. G.L.L. is supported by NSFC of China under grants 11673065, U1931210, and 11273061. 
LPF acknowledges the support from NSFC grants 11722326, 11673018 and 11933002, STCSM grant 18590780100 and 188014066, and the Dawn Program 19SG41 and
the Innovation Program 2019-01-07-00-02-E00032 of SMEC.
T.F. is supported by Grant-in-Aids for Scientific Research from JSPS (No.20 K03937), and also acknowledges the SWIFAR visiting fellow program. 
We also acknowledge the support from ISSI/ISSI-BJ International Team Program - Weak gravitational lensing studies from space missions. 
\end{acknowledgments}

\bibliography{ms}

\begin{thebibliography}{35}%
\makeatletter
\providecommand \@ifxundefined [1]{%
 \@ifx{#1\undefined}
}%
\providecommand \@ifnum [1]{%
 \ifnum #1\expandafter \@firstoftwo
 \else \expandafter \@secondoftwo
 \fi
}%
\providecommand \@ifx [1]{%
 \ifx #1\expandafter \@firstoftwo
 \else \expandafter \@secondoftwo
 \fi
}%
\providecommand \natexlab [1]{#1}%
\providecommand \enquote  [1]{``#1''}%
\providecommand \bibnamefont  [1]{#1}%
\providecommand \bibfnamefont [1]{#1}%
\providecommand \citenamefont [1]{#1}%
\providecommand \href@noop [0]{\@secondoftwo}%
\providecommand \href [0]{\begingroup \@sanitize@url \@href}%
\providecommand \@href[1]{\@@startlink{#1}\@@href}%
\providecommand \@@href[1]{\endgroup#1\@@endlink}%
\providecommand \@sanitize@url [0]{\catcode `\\12\catcode `\$12\catcode
  `\&12\catcode `\#12\catcode `\^12\catcode `\_12\catcode `\%12\relax}%
\providecommand \@@startlink[1]{}%
\providecommand \@@endlink[0]{}%
\providecommand \url  [0]{\begingroup\@sanitize@url \@url }%
\providecommand \@url [1]{\endgroup\@href {#1}{\urlprefix }}%
\providecommand \urlprefix  [0]{URL }%
\providecommand \Eprint [0]{\href }%
\providecommand \doibase [0]{https://doi.org/}%
\providecommand \selectlanguage [0]{\@gobble}%
\providecommand \bibinfo  [0]{\@secondoftwo}%
\providecommand \bibfield  [0]{\@secondoftwo}%
\providecommand \translation [1]{[#1]}%
\providecommand \BibitemOpen [0]{}%
\providecommand \bibitemStop [0]{}%
\providecommand \bibitemNoStop [0]{.\EOS\space}%
\providecommand \EOS [0]{\spacefactor3000\relax}%
\providecommand \BibitemShut  [1]{\csname bibitem#1\endcsname}%
\let\auto@bib@innerbib\@empty
\bibitem [{\citenamefont {{Hamana}}\ \emph {et~al.}(2020)\citenamefont
  {{Hamana}}, \citenamefont {{Shirasaki}}, \citenamefont {{Miyazaki}},
  \citenamefont {{Hikage}}, \citenamefont {{Oguri}}, \citenamefont {{More}},
  \citenamefont {{Armstrong}}, \citenamefont {{Leauthaud}}, \citenamefont
  {{Mandelbaum}}, \citenamefont {{Miyatake}}, \citenamefont {{Nishizawa}},
  \citenamefont {{Simet}}, \citenamefont {{Takada}}, \citenamefont {{Aihara}},
  \citenamefont {{Bosch}}, \citenamefont {{Komiyama}}, \citenamefont
  {{Lupton}}, \citenamefont {{Murayama}}, \citenamefont {{Strauss}},\ and\
  \citenamefont {{Tanaka}}}]{Hamana2020}%
  \BibitemOpen
  \bibfield  {author} {\bibinfo {author} {\bibfnamefont {T.}~\bibnamefont
  {{Hamana}}}, \bibinfo {author} {\bibfnamefont {M.}~\bibnamefont
  {{Shirasaki}}}, \bibinfo {author} {\bibfnamefont {S.}~\bibnamefont
  {{Miyazaki}}}, \bibinfo {author} {\bibfnamefont {C.}~\bibnamefont
  {{Hikage}}}, \bibinfo {author} {\bibfnamefont {M.}~\bibnamefont {{Oguri}}},
  \bibinfo {author} {\bibfnamefont {S.}~\bibnamefont {{More}}}, \bibinfo
  {author} {\bibfnamefont {R.}~\bibnamefont {{Armstrong}}}, \bibinfo {author}
  {\bibfnamefont {A.}~\bibnamefont {{Leauthaud}}}, \bibinfo {author}
  {\bibfnamefont {R.}~\bibnamefont {{Mandelbaum}}}, \bibinfo {author}
  {\bibfnamefont {H.}~\bibnamefont {{Miyatake}}}, \bibinfo {author}
  {\bibfnamefont {A.~J.}\ \bibnamefont {{Nishizawa}}}, \bibinfo {author}
  {\bibfnamefont {M.}~\bibnamefont {{Simet}}}, \bibinfo {author} {\bibfnamefont
  {M.}~\bibnamefont {{Takada}}}, \bibinfo {author} {\bibfnamefont
  {H.}~\bibnamefont {{Aihara}}}, \bibinfo {author} {\bibfnamefont
  {J.}~\bibnamefont {{Bosch}}}, \bibinfo {author} {\bibfnamefont
  {Y.}~\bibnamefont {{Komiyama}}}, \bibinfo {author} {\bibfnamefont
  {R.}~\bibnamefont {{Lupton}}}, \bibinfo {author} {\bibfnamefont
  {H.}~\bibnamefont {{Murayama}}}, \bibinfo {author} {\bibfnamefont {M.~A.}\
  \bibnamefont {{Strauss}}},\ and\ \bibinfo {author} {\bibfnamefont
  {M.}~\bibnamefont {{Tanaka}}},\ }\bibfield  {title} {\bibinfo {title}
  {{Cosmological constraints from cosmic shear two-point correlation functions
  with HSC survey first-year data}},\ }\href
  {https://doi.org/10.1093/pasj/psz138} {\bibfield  {journal} {\bibinfo
  {journal} {\pasj}\ }\textbf {\bibinfo {volume} {72}},\ \bibinfo {eid} {16}
  (\bibinfo {year} {2020})},\ \Eprint {https://arxiv.org/abs/1906.06041}
  {arXiv:1906.06041 [astro-ph.CO]} \BibitemShut {NoStop}%
\bibitem [{\citenamefont {{Bartelmann}}\ and\ \citenamefont
  {{Schneider}}(2001)}]{BS2001}%
  \BibitemOpen
  \bibfield  {author} {\bibinfo {author} {\bibfnamefont {M.}~\bibnamefont
  {{Bartelmann}}}\ and\ \bibinfo {author} {\bibfnamefont {P.}~\bibnamefont
  {{Schneider}}},\ }\bibfield  {title} {\bibinfo {title} {{Weak gravitational
  lensing}},\ }\href {https://doi.org/10.1016/S0370-1573(00)00082-X} {\bibfield
   {journal} {\bibinfo  {journal} {\physrep}\ }\textbf {\bibinfo {volume}
  {340}},\ \bibinfo {pages} {291} (\bibinfo {year} {2001})},\ \Eprint
  {https://arxiv.org/abs/astro-ph/9912508} {arXiv:astro-ph/9912508 [astro-ph]}
  \BibitemShut {NoStop}%
\bibitem [{\citenamefont {{Hoekstra}}\ and\ \citenamefont
  {{Jain}}(2008)}]{HJ2008}%
  \BibitemOpen
  \bibfield  {author} {\bibinfo {author} {\bibfnamefont {H.}~\bibnamefont
  {{Hoekstra}}}\ and\ \bibinfo {author} {\bibfnamefont {B.}~\bibnamefont
  {{Jain}}},\ }\bibfield  {title} {\bibinfo {title} {{Weak Gravitational
  Lensing and Its Cosmological Applications}},\ }\href
  {https://doi.org/10.1146/annurev.nucl.58.110707.171151} {\bibfield  {journal}
  {\bibinfo  {journal} {Annual Review of Nuclear and Particle Science}\
  }\textbf {\bibinfo {volume} {58}},\ \bibinfo {pages} {99} (\bibinfo {year}
  {2008})},\ \Eprint {https://arxiv.org/abs/0805.0139} {arXiv:0805.0139
  [astro-ph]} \BibitemShut {NoStop}%
\bibitem [{\citenamefont {{Van Waerbeke}}\ \emph {et~al.}(2010)\citenamefont
  {{Van Waerbeke}}, \citenamefont {{Hildebrandt}}, \citenamefont {{Ford}},\
  and\ \citenamefont {{Milkeraitis}}}]{van2010a}%
  \BibitemOpen
  \bibfield  {author} {\bibinfo {author} {\bibfnamefont {L.}~\bibnamefont {{Van
  Waerbeke}}}, \bibinfo {author} {\bibfnamefont {H.}~\bibnamefont
  {{Hildebrandt}}}, \bibinfo {author} {\bibfnamefont {J.}~\bibnamefont
  {{Ford}}},\ and\ \bibinfo {author} {\bibfnamefont {M.}~\bibnamefont
  {{Milkeraitis}}},\ }\bibfield  {title} {\bibinfo {title} {{Magnification as a
  Probe of Dark Matter Halos at High Redshifts}},\ }\href
  {https://doi.org/10.1088/2041-8205/723/1/L13} {\bibfield  {journal} {\bibinfo
   {journal} {\apjl}\ }\textbf {\bibinfo {volume} {723}},\ \bibinfo {pages}
  {L13} (\bibinfo {year} {2010})},\ \Eprint {https://arxiv.org/abs/1004.3793}
  {arXiv:1004.3793 [astro-ph.CO]} \BibitemShut {NoStop}%
\bibitem [{\citenamefont {{Fu}}\ and\ \citenamefont {{Fan}}(2014)}]{FuFan2014}%
  \BibitemOpen
  \bibfield  {author} {\bibinfo {author} {\bibfnamefont {L.-P.}\ \bibnamefont
  {{Fu}}}\ and\ \bibinfo {author} {\bibfnamefont {Z.-H.}\ \bibnamefont
  {{Fan}}},\ }\bibfield  {title} {\bibinfo {title} {{Probing the dark side of
  the Universe with weak gravitational lensing effects}},\ }\href
  {https://doi.org/10.1088/1674-4527/14/9/002} {\bibfield  {journal} {\bibinfo
  {journal} {Research in Astronomy and Astrophysics}\ }\textbf {\bibinfo
  {volume} {14}},\ \bibinfo {eid} {1061-1120} (\bibinfo {year}
  {2014})}\BibitemShut {NoStop}%
\bibitem [{\citenamefont {{Kilbinger}}(2015)}]{Kil2015}%
  \BibitemOpen
  \bibfield  {author} {\bibinfo {author} {\bibfnamefont {M.}~\bibnamefont
  {{Kilbinger}}},\ }\bibfield  {title} {\bibinfo {title} {{Cosmology with
  cosmic shear observations: a review}},\ }\href
  {https://doi.org/10.1088/0034-4885/78/8/086901} {\bibfield  {journal}
  {\bibinfo  {journal} {Reports on Progress in Physics}\ }\textbf {\bibinfo
  {volume} {78}},\ \bibinfo {eid} {086901} (\bibinfo {year} {2015})},\ \Eprint
  {https://arxiv.org/abs/1411.0115} {arXiv:1411.0115 [astro-ph.CO]}
  \BibitemShut {NoStop}%
\bibitem [{\citenamefont {{Dark Energy Survey Collaboration}}\ \emph
  {et~al.}(2016)\citenamefont {{Dark Energy Survey Collaboration}},
  \citenamefont {{Abbott}}, \citenamefont {{Abdalla}}, \citenamefont
  {{Aleksi{\'c}}}, \citenamefont {{Allam}}, \citenamefont {{Amara}},
  \citenamefont {{Bacon}}, \citenamefont {{Balbinot}}, \citenamefont
  {{Banerji}}, \citenamefont {{Bechtol}}, \citenamefont {{Benoit-L{\'e}vy}},
  \citenamefont {{Bernstein}}, \citenamefont {{Bertin}}, \citenamefont
  {{Blazek}}, \citenamefont {{Bonnett}}, \citenamefont {{Bridle}},
  \citenamefont {{Brooks}}, \citenamefont {{Brunner}}, \citenamefont
  {{Buckley-Geer}}, \citenamefont {{Burke}}, \citenamefont {{Caminha}},
  \citenamefont {{Capozzi}}, \citenamefont {{Carlsen}}, \citenamefont
  {{Carnero-Rosell}}, \citenamefont {{Carollo}}, \citenamefont
  {{Carrasco-Kind}}, \citenamefont {{Carretero}}, \citenamefont {{Castander}},
  \citenamefont {{Clerkin}}, \citenamefont {{Collett}}, \citenamefont
  {{Conselice}}, \citenamefont {{Crocce}}, \citenamefont {{Cunha}},
  \citenamefont {{D'Andrea}}, \citenamefont {{da Costa}}, \citenamefont
  {{Davis}}, \citenamefont {{Desai}}, \citenamefont {{Diehl}}, \citenamefont
  {{Dietrich}}, \citenamefont {{Dodelson}}, \citenamefont {{Doel}},
  \citenamefont {{Drlica-Wagner}}, \citenamefont {{Estrada}}, \citenamefont
  {{Etherington}}, \citenamefont {{Evrard}}, \citenamefont {{Fabbri}},
  \citenamefont {{Finley}}, \citenamefont {{Flaugher}}, \citenamefont
  {{Foley}}, \citenamefont {{Fosalba}}, \citenamefont {{Frieman}},
  \citenamefont {{Garc{\'\i}a-Bellido}}, \citenamefont {{Gaztanaga}},
  \citenamefont {{Gerdes}}, \citenamefont {{Giannantonio}}, \citenamefont
  {{Goldstein}}, \citenamefont {{Gruen}}, \citenamefont {{Gruendl}},
  \citenamefont {{Guarnieri}}, \citenamefont {{Gutierrez}}, \citenamefont
  {{Hartley}}, \citenamefont {{Honscheid}}, \citenamefont {{Jain}},
  \citenamefont {{James}}, \citenamefont {{Jeltema}}, \citenamefont {{Jouvel}},
  \citenamefont {{Kessler}}, \citenamefont {{King}}, \citenamefont {{Kirk}},
  \citenamefont {{Kron}}, \citenamefont {{Kuehn}}, \citenamefont
  {{Kuropatkin}}, \citenamefont {{Lahav}}, \citenamefont {{Li}}, \citenamefont
  {{Lima}}, \citenamefont {{Lin}}, \citenamefont {{Maia}}, \citenamefont
  {{Makler}}, \citenamefont {{Manera}}, \citenamefont {{Maraston}},
  \citenamefont {{Marshall}}, \citenamefont {{Martini}}, \citenamefont
  {{McMahon}}, \citenamefont {{Melchior}}, \citenamefont {{Merson}},
  \citenamefont {{Miller}}, \citenamefont {{Miquel}}, \citenamefont {{Mohr}},
  \citenamefont {{Morice-Atkinson}}, \citenamefont {{Naidoo}}, \citenamefont
  {{Neilsen}}, \citenamefont {{Nichol}}, \citenamefont {{Nord}}, \citenamefont
  {{Ogando}}, \citenamefont {{Ostrovski}}, \citenamefont {{Palmese}},
  \citenamefont {{Papadopoulos}}, \citenamefont {{Peiris}}, \citenamefont
  {{Peoples}}, \citenamefont {{Percival}}, \citenamefont {{Plazas}},
  \citenamefont {{Reed}}, \citenamefont {{Refregier}}, \citenamefont {{Romer}},
  \citenamefont {{Roodman}}, \citenamefont {{Ross}}, \citenamefont {{Rozo}},
  \citenamefont {{Rykoff}}, \citenamefont {{Sadeh}}, \citenamefont {{Sako}},
  \citenamefont {{S{\'a}nchez}}, \citenamefont {{Sanchez}}, \citenamefont
  {{Santiago}}, \citenamefont {{Scarpine}}, \citenamefont {{Schubnell}},
  \citenamefont {{Sevilla-Noarbe}}, \citenamefont {{Sheldon}}, \citenamefont
  {{Smith}}, \citenamefont {{Smith}}, \citenamefont {{Soares-Santos}},
  \citenamefont {{Sobreira}}, \citenamefont {{Soumagnac}}, \citenamefont
  {{Suchyta}}, \citenamefont {{Sullivan}}, \citenamefont {{Swanson}},
  \citenamefont {{Tarle}}, \citenamefont {{Thaler}}, \citenamefont {{Thomas}},
  \citenamefont {{Thomas}}, \citenamefont {{Tucker}}, \citenamefont {{Vieira}},
  \citenamefont {{Vikram}}, \citenamefont {{Walker}}, \citenamefont
  {{Wechsler}}, \citenamefont {{Weller}}, \citenamefont {{Wester}},
  \citenamefont {{Whiteway}}, \citenamefont {{Wilcox}}, \citenamefont
  {{Yanny}}, \citenamefont {{Zhang}},\ and\ \citenamefont {{Zuntz}}}]{DES2016}%
  \BibitemOpen
  \bibfield  {author} {\bibinfo {author} {\bibnamefont {{Dark Energy Survey
  Collaboration}}}, \bibinfo {author} {\bibfnamefont {T.}~\bibnamefont
  {{Abbott}}}, \bibinfo {author} {\bibfnamefont {F.~B.}\ \bibnamefont
  {{Abdalla}}}, \bibinfo {author} {\bibfnamefont {J.}~\bibnamefont
  {{Aleksi{\'c}}}}, \bibinfo {author} {\bibfnamefont {S.}~\bibnamefont
  {{Allam}}}, \bibinfo {author} {\bibfnamefont {A.}~\bibnamefont {{Amara}}},
  \bibinfo {author} {\bibfnamefont {D.}~\bibnamefont {{Bacon}}}, \bibinfo
  {author} {\bibfnamefont {E.}~\bibnamefont {{Balbinot}}}, \bibinfo {author}
  {\bibfnamefont {M.}~\bibnamefont {{Banerji}}}, \bibinfo {author}
  {\bibfnamefont {K.}~\bibnamefont {{Bechtol}}}, \bibinfo {author}
  {\bibfnamefont {A.}~\bibnamefont {{Benoit-L{\'e}vy}}}, \bibinfo {author}
  {\bibfnamefont {G.~M.}\ \bibnamefont {{Bernstein}}}, \bibinfo {author}
  {\bibfnamefont {E.}~\bibnamefont {{Bertin}}}, \bibinfo {author}
  {\bibfnamefont {J.}~\bibnamefont {{Blazek}}}, \bibinfo {author}
  {\bibfnamefont {C.}~\bibnamefont {{Bonnett}}}, \bibinfo {author}
  {\bibfnamefont {S.}~\bibnamefont {{Bridle}}}, \bibinfo {author}
  {\bibfnamefont {D.}~\bibnamefont {{Brooks}}}, \bibinfo {author}
  {\bibfnamefont {R.~J.}\ \bibnamefont {{Brunner}}}, \bibinfo {author}
  {\bibfnamefont {E.}~\bibnamefont {{Buckley-Geer}}}, \bibinfo {author}
  {\bibfnamefont {D.~L.}\ \bibnamefont {{Burke}}}, \bibinfo {author}
  {\bibfnamefont {G.~B.}\ \bibnamefont {{Caminha}}}, \bibinfo {author}
  {\bibfnamefont {D.}~\bibnamefont {{Capozzi}}}, \bibinfo {author}
  {\bibfnamefont {J.}~\bibnamefont {{Carlsen}}}, \bibinfo {author}
  {\bibfnamefont {A.}~\bibnamefont {{Carnero-Rosell}}}, \bibinfo {author}
  {\bibfnamefont {M.}~\bibnamefont {{Carollo}}}, \bibinfo {author}
  {\bibfnamefont {M.}~\bibnamefont {{Carrasco-Kind}}}, \bibinfo {author}
  {\bibfnamefont {J.}~\bibnamefont {{Carretero}}}, \bibinfo {author}
  {\bibfnamefont {F.~J.}\ \bibnamefont {{Castander}}}, \bibinfo {author}
  {\bibfnamefont {L.}~\bibnamefont {{Clerkin}}}, \bibinfo {author}
  {\bibfnamefont {T.}~\bibnamefont {{Collett}}}, \bibinfo {author}
  {\bibfnamefont {C.}~\bibnamefont {{Conselice}}}, \bibinfo {author}
  {\bibfnamefont {M.}~\bibnamefont {{Crocce}}}, \bibinfo {author}
  {\bibfnamefont {C.~E.}\ \bibnamefont {{Cunha}}}, \bibinfo {author}
  {\bibfnamefont {C.~B.}\ \bibnamefont {{D'Andrea}}}, \bibinfo {author}
  {\bibfnamefont {L.~N.}\ \bibnamefont {{da Costa}}}, \bibinfo {author}
  {\bibfnamefont {T.~M.}\ \bibnamefont {{Davis}}}, \bibinfo {author}
  {\bibfnamefont {S.}~\bibnamefont {{Desai}}}, \bibinfo {author} {\bibfnamefont
  {H.~T.}\ \bibnamefont {{Diehl}}}, \bibinfo {author} {\bibfnamefont {J.~P.}\
  \bibnamefont {{Dietrich}}}, \bibinfo {author} {\bibfnamefont
  {S.}~\bibnamefont {{Dodelson}}}, \bibinfo {author} {\bibfnamefont
  {P.}~\bibnamefont {{Doel}}}, \bibinfo {author} {\bibfnamefont
  {A.}~\bibnamefont {{Drlica-Wagner}}}, \bibinfo {author} {\bibfnamefont
  {J.}~\bibnamefont {{Estrada}}}, \bibinfo {author} {\bibfnamefont
  {J.}~\bibnamefont {{Etherington}}}, \bibinfo {author} {\bibfnamefont {A.~E.}\
  \bibnamefont {{Evrard}}}, \bibinfo {author} {\bibfnamefont {J.}~\bibnamefont
  {{Fabbri}}}, \bibinfo {author} {\bibfnamefont {D.~A.}\ \bibnamefont
  {{Finley}}}, \bibinfo {author} {\bibfnamefont {B.}~\bibnamefont
  {{Flaugher}}}, \bibinfo {author} {\bibfnamefont {R.~J.}\ \bibnamefont
  {{Foley}}}, \bibinfo {author} {\bibfnamefont {P.}~\bibnamefont {{Fosalba}}},
  \bibinfo {author} {\bibfnamefont {J.}~\bibnamefont {{Frieman}}}, \bibinfo
  {author} {\bibfnamefont {J.}~\bibnamefont {{Garc{\'\i}a-Bellido}}}, \bibinfo
  {author} {\bibfnamefont {E.}~\bibnamefont {{Gaztanaga}}}, \bibinfo {author}
  {\bibfnamefont {D.~W.}\ \bibnamefont {{Gerdes}}}, \bibinfo {author}
  {\bibfnamefont {T.}~\bibnamefont {{Giannantonio}}}, \bibinfo {author}
  {\bibfnamefont {D.~A.}\ \bibnamefont {{Goldstein}}}, \bibinfo {author}
  {\bibfnamefont {D.}~\bibnamefont {{Gruen}}}, \bibinfo {author} {\bibfnamefont
  {R.~A.}\ \bibnamefont {{Gruendl}}}, \bibinfo {author} {\bibfnamefont
  {P.}~\bibnamefont {{Guarnieri}}}, \bibinfo {author} {\bibfnamefont
  {G.}~\bibnamefont {{Gutierrez}}}, \bibinfo {author} {\bibfnamefont
  {W.}~\bibnamefont {{Hartley}}}, \bibinfo {author} {\bibfnamefont
  {K.}~\bibnamefont {{Honscheid}}}, \bibinfo {author} {\bibfnamefont
  {B.}~\bibnamefont {{Jain}}}, \bibinfo {author} {\bibfnamefont {D.~J.}\
  \bibnamefont {{James}}}, \bibinfo {author} {\bibfnamefont {T.}~\bibnamefont
  {{Jeltema}}}, \bibinfo {author} {\bibfnamefont {S.}~\bibnamefont {{Jouvel}}},
  \bibinfo {author} {\bibfnamefont {R.}~\bibnamefont {{Kessler}}}, \bibinfo
  {author} {\bibfnamefont {A.}~\bibnamefont {{King}}}, \bibinfo {author}
  {\bibfnamefont {D.}~\bibnamefont {{Kirk}}}, \bibinfo {author} {\bibfnamefont
  {R.}~\bibnamefont {{Kron}}}, \bibinfo {author} {\bibfnamefont
  {K.}~\bibnamefont {{Kuehn}}}, \bibinfo {author} {\bibfnamefont
  {N.}~\bibnamefont {{Kuropatkin}}}, \bibinfo {author} {\bibfnamefont
  {O.}~\bibnamefont {{Lahav}}}, \bibinfo {author} {\bibfnamefont {T.~S.}\
  \bibnamefont {{Li}}}, \bibinfo {author} {\bibfnamefont {M.}~\bibnamefont
  {{Lima}}}, \bibinfo {author} {\bibfnamefont {H.}~\bibnamefont {{Lin}}},
  \bibinfo {author} {\bibfnamefont {M.~A.~G.}\ \bibnamefont {{Maia}}}, \bibinfo
  {author} {\bibfnamefont {M.}~\bibnamefont {{Makler}}}, \bibinfo {author}
  {\bibfnamefont {M.}~\bibnamefont {{Manera}}}, \bibinfo {author}
  {\bibfnamefont {C.}~\bibnamefont {{Maraston}}}, \bibinfo {author}
  {\bibfnamefont {J.~L.}\ \bibnamefont {{Marshall}}}, \bibinfo {author}
  {\bibfnamefont {P.}~\bibnamefont {{Martini}}}, \bibinfo {author}
  {\bibfnamefont {R.~G.}\ \bibnamefont {{McMahon}}}, \bibinfo {author}
  {\bibfnamefont {P.}~\bibnamefont {{Melchior}}}, \bibinfo {author}
  {\bibfnamefont {A.}~\bibnamefont {{Merson}}}, \bibinfo {author}
  {\bibfnamefont {C.~J.}\ \bibnamefont {{Miller}}}, \bibinfo {author}
  {\bibfnamefont {R.}~\bibnamefont {{Miquel}}}, \bibinfo {author}
  {\bibfnamefont {J.~J.}\ \bibnamefont {{Mohr}}}, \bibinfo {author}
  {\bibfnamefont {X.}~\bibnamefont {{Morice-Atkinson}}}, \bibinfo {author}
  {\bibfnamefont {K.}~\bibnamefont {{Naidoo}}}, \bibinfo {author}
  {\bibfnamefont {E.}~\bibnamefont {{Neilsen}}}, \bibinfo {author}
  {\bibfnamefont {R.~C.}\ \bibnamefont {{Nichol}}}, \bibinfo {author}
  {\bibfnamefont {B.}~\bibnamefont {{Nord}}}, \bibinfo {author} {\bibfnamefont
  {R.}~\bibnamefont {{Ogando}}}, \bibinfo {author} {\bibfnamefont
  {F.}~\bibnamefont {{Ostrovski}}}, \bibinfo {author} {\bibfnamefont
  {A.}~\bibnamefont {{Palmese}}}, \bibinfo {author} {\bibfnamefont
  {A.}~\bibnamefont {{Papadopoulos}}}, \bibinfo {author} {\bibfnamefont
  {H.~V.}\ \bibnamefont {{Peiris}}}, \bibinfo {author} {\bibfnamefont
  {J.}~\bibnamefont {{Peoples}}}, \bibinfo {author} {\bibfnamefont {W.~J.}\
  \bibnamefont {{Percival}}}, \bibinfo {author} {\bibfnamefont {A.~A.}\
  \bibnamefont {{Plazas}}}, \bibinfo {author} {\bibfnamefont {S.~L.}\
  \bibnamefont {{Reed}}}, \bibinfo {author} {\bibfnamefont {A.}~\bibnamefont
  {{Refregier}}}, \bibinfo {author} {\bibfnamefont {A.~K.}\ \bibnamefont
  {{Romer}}}, \bibinfo {author} {\bibfnamefont {A.}~\bibnamefont {{Roodman}}},
  \bibinfo {author} {\bibfnamefont {A.}~\bibnamefont {{Ross}}}, \bibinfo
  {author} {\bibfnamefont {E.}~\bibnamefont {{Rozo}}}, \bibinfo {author}
  {\bibfnamefont {E.~S.}\ \bibnamefont {{Rykoff}}}, \bibinfo {author}
  {\bibfnamefont {I.}~\bibnamefont {{Sadeh}}}, \bibinfo {author} {\bibfnamefont
  {M.}~\bibnamefont {{Sako}}}, \bibinfo {author} {\bibfnamefont
  {C.}~\bibnamefont {{S{\'a}nchez}}}, \bibinfo {author} {\bibfnamefont
  {E.}~\bibnamefont {{Sanchez}}}, \bibinfo {author} {\bibfnamefont
  {B.}~\bibnamefont {{Santiago}}}, \bibinfo {author} {\bibfnamefont
  {V.}~\bibnamefont {{Scarpine}}}, \bibinfo {author} {\bibfnamefont
  {M.}~\bibnamefont {{Schubnell}}}, \bibinfo {author} {\bibfnamefont
  {I.}~\bibnamefont {{Sevilla-Noarbe}}}, \bibinfo {author} {\bibfnamefont
  {E.}~\bibnamefont {{Sheldon}}}, \bibinfo {author} {\bibfnamefont
  {M.}~\bibnamefont {{Smith}}}, \bibinfo {author} {\bibfnamefont {R.~C.}\
  \bibnamefont {{Smith}}}, \bibinfo {author} {\bibfnamefont {M.}~\bibnamefont
  {{Soares-Santos}}}, \bibinfo {author} {\bibfnamefont {F.}~\bibnamefont
  {{Sobreira}}}, \bibinfo {author} {\bibfnamefont {M.}~\bibnamefont
  {{Soumagnac}}}, \bibinfo {author} {\bibfnamefont {E.}~\bibnamefont
  {{Suchyta}}}, \bibinfo {author} {\bibfnamefont {M.}~\bibnamefont
  {{Sullivan}}}, \bibinfo {author} {\bibfnamefont {M.}~\bibnamefont
  {{Swanson}}}, \bibinfo {author} {\bibfnamefont {G.}~\bibnamefont {{Tarle}}},
  \bibinfo {author} {\bibfnamefont {J.}~\bibnamefont {{Thaler}}}, \bibinfo
  {author} {\bibfnamefont {D.}~\bibnamefont {{Thomas}}}, \bibinfo {author}
  {\bibfnamefont {R.~C.}\ \bibnamefont {{Thomas}}}, \bibinfo {author}
  {\bibfnamefont {D.}~\bibnamefont {{Tucker}}}, \bibinfo {author}
  {\bibfnamefont {J.~D.}\ \bibnamefont {{Vieira}}}, \bibinfo {author}
  {\bibfnamefont {V.}~\bibnamefont {{Vikram}}}, \bibinfo {author}
  {\bibfnamefont {A.~R.}\ \bibnamefont {{Walker}}}, \bibinfo {author}
  {\bibfnamefont {R.~H.}\ \bibnamefont {{Wechsler}}}, \bibinfo {author}
  {\bibfnamefont {J.}~\bibnamefont {{Weller}}}, \bibinfo {author}
  {\bibfnamefont {W.}~\bibnamefont {{Wester}}}, \bibinfo {author}
  {\bibfnamefont {L.}~\bibnamefont {{Whiteway}}}, \bibinfo {author}
  {\bibfnamefont {H.}~\bibnamefont {{Wilcox}}}, \bibinfo {author}
  {\bibfnamefont {B.}~\bibnamefont {{Yanny}}}, \bibinfo {author} {\bibfnamefont
  {Y.}~\bibnamefont {{Zhang}}},\ and\ \bibinfo {author} {\bibfnamefont
  {J.}~\bibnamefont {{Zuntz}}},\ }\bibfield  {title} {\bibinfo {title} {{The
  Dark Energy Survey: more than dark energy - an overview}},\ }\href
  {https://doi.org/10.1093/mnras/stw641} {\bibfield  {journal} {\bibinfo
  {journal} {\mnras}\ }\textbf {\bibinfo {volume} {460}},\ \bibinfo {pages}
  {1270} (\bibinfo {year} {2016})},\ \Eprint {https://arxiv.org/abs/1601.00329}
  {arXiv:1601.00329 [astro-ph.CO]} \BibitemShut {NoStop}%
\bibitem [{\citenamefont {{de Jong}}\ \emph {et~al.}(2013)\citenamefont {{de
  Jong}}, \citenamefont {{Verdoes Kleijn}}, \citenamefont {{Kuijken}},\ and\
  \citenamefont {{Valentijn}}}]{KiDS2013}%
  \BibitemOpen
  \bibfield  {author} {\bibinfo {author} {\bibfnamefont {J.~T.~A.}\
  \bibnamefont {{de Jong}}}, \bibinfo {author} {\bibfnamefont {G.~A.}\
  \bibnamefont {{Verdoes Kleijn}}}, \bibinfo {author} {\bibfnamefont {K.~H.}\
  \bibnamefont {{Kuijken}}},\ and\ \bibinfo {author} {\bibfnamefont {E.~A.}\
  \bibnamefont {{Valentijn}}},\ }\bibfield  {title} {\bibinfo {title} {{The
  Kilo-Degree Survey}},\ }\href {https://doi.org/10.1007/s10686-012-9306-1}
  {\bibfield  {journal} {\bibinfo  {journal} {Experimental Astronomy}\ }\textbf
  {\bibinfo {volume} {35}},\ \bibinfo {pages} {25} (\bibinfo {year} {2013})},\
  \Eprint {https://arxiv.org/abs/1206.1254} {arXiv:1206.1254 [astro-ph.CO]}
  \BibitemShut {NoStop}%
\bibitem [{\citenamefont {{Aihara}}\ \emph
  {et~al.}(2018{\natexlab{a}})\citenamefont {{Aihara}}, \citenamefont
  {{Arimoto}}, \citenamefont {{Armstrong}}, \citenamefont {{Arnouts}},
  \citenamefont {{Bahcall}}, \citenamefont {{Bickerton}}, \citenamefont
  {{Bosch}}, \citenamefont {{Bundy}}, \citenamefont {{Capak}}, \citenamefont
  {{Chan}}, \citenamefont {{Chiba}}, \citenamefont {{Coupon}}, \citenamefont
  {{Egami}}, \citenamefont {{Enoki}}, \citenamefont {{Finet}}, \citenamefont
  {{Fujimori}}, \citenamefont {{Fujimoto}}, \citenamefont {{Furusawa}},
  \citenamefont {{Furusawa}}, \citenamefont {{Goto}}, \citenamefont
  {{Goulding}}, \citenamefont {{Greco}}, \citenamefont {{Greene}},
  \citenamefont {{Gunn}}, \citenamefont {{Hamana}}, \citenamefont {{Harikane}},
  \citenamefont {{Hashimoto}}, \citenamefont {{Hattori}}, \citenamefont
  {{Hayashi}}, \citenamefont {{Hayashi}}, \citenamefont {{He{\l}miniak}},
  \citenamefont {{Higuchi}}, \citenamefont {{Hikage}}, \citenamefont {{Ho}},
  \citenamefont {{Hsieh}}, \citenamefont {{Huang}}, \citenamefont {{Huang}},
  \citenamefont {{Ikeda}}, \citenamefont {{Imanishi}}, \citenamefont {{Inoue}},
  \citenamefont {{Iwasawa}}, \citenamefont {{Iwata}}, \citenamefont
  {{Jaelani}}, \citenamefont {{Jian}}, \citenamefont {{Kamata}}, \citenamefont
  {{Karoji}}, \citenamefont {{Kashikawa}}, \citenamefont {{Katayama}},
  \citenamefont {{Kawanomoto}}, \citenamefont {{Kayo}}, \citenamefont {{Koda}},
  \citenamefont {{Koike}}, \citenamefont {{Kojima}}, \citenamefont
  {{Komiyama}}, \citenamefont {{Konno}}, \citenamefont {{Koshida}},
  \citenamefont {{Koyama}}, \citenamefont {{Kusakabe}}, \citenamefont
  {{Leauthaud}}, \citenamefont {{Lee}}, \citenamefont {{Lin}}, \citenamefont
  {{Lin}}, \citenamefont {{Lupton}}, \citenamefont {{Mandelbaum}},
  \citenamefont {{Matsuoka}}, \citenamefont {{Medezinski}}, \citenamefont
  {{Mineo}}, \citenamefont {{Miyama}}, \citenamefont {{Miyatake}},
  \citenamefont {{Miyazaki}}, \citenamefont {{Momose}}, \citenamefont {{More}},
  \citenamefont {{More}}, \citenamefont {{Moritani}}, \citenamefont {{Moriya}},
  \citenamefont {{Morokuma}}, \citenamefont {{Mukae}}, \citenamefont
  {{Murata}}, \citenamefont {{Murayama}}, \citenamefont {{Nagao}},
  \citenamefont {{Nakata}}, \citenamefont {{Niida}}, \citenamefont {{Niikura}},
  \citenamefont {{Nishizawa}}, \citenamefont {{Obuchi}}, \citenamefont
  {{Oguri}}, \citenamefont {{Oishi}}, \citenamefont {{Okabe}}, \citenamefont
  {{Okamoto}}, \citenamefont {{Okura}}, \citenamefont {{Ono}}, \citenamefont
  {{Onodera}}, \citenamefont {{Onoue}}, \citenamefont {{Osato}}, \citenamefont
  {{Ouchi}}, \citenamefont {{Price}}, \citenamefont {{Pyo}}, \citenamefont
  {{Sako}}, \citenamefont {{Sawicki}}, \citenamefont {{Shibuya}}, \citenamefont
  {{Shimasaku}}, \citenamefont {{Shimono}}, \citenamefont {{Shirasaki}},
  \citenamefont {{Silverman}}, \citenamefont {{Simet}}, \citenamefont
  {{Speagle}}, \citenamefont {{Spergel}}, \citenamefont {{Strauss}},
  \citenamefont {{Sugahara}}, \citenamefont {{Sugiyama}}, \citenamefont
  {{Suto}}, \citenamefont {{Suyu}}, \citenamefont {{Suzuki}}, \citenamefont
  {{Tait}}, \citenamefont {{Takada}}, \citenamefont {{Takata}}, \citenamefont
  {{Tamura}}, \citenamefont {{Tanaka}}, \citenamefont {{Tanaka}}, \citenamefont
  {{Tanaka}}, \citenamefont {{Tanaka}}, \citenamefont {{Terai}}, \citenamefont
  {{Terashima}}, \citenamefont {{Toba}}, \citenamefont {{Tominaga}},
  \citenamefont {{Toshikawa}}, \citenamefont {{Turner}}, \citenamefont
  {{Uchida}}, \citenamefont {{Uchiyama}}, \citenamefont {{Umetsu}},
  \citenamefont {{Uraguchi}}, \citenamefont {{Urata}}, \citenamefont {{Usuda}},
  \citenamefont {{Utsumi}}, \citenamefont {{Wang}}, \citenamefont {{Wang}},
  \citenamefont {{Wong}}, \citenamefont {{Yabe}}, \citenamefont {{Yamada}},
  \citenamefont {{Yamanoi}}, \citenamefont {{Yasuda}}, \citenamefont {{Yeh}},
  \citenamefont {{Yonehara}},\ and\ \citenamefont {{Yuma}}}]{HSC2018}%
  \BibitemOpen
  \bibfield  {author} {\bibinfo {author} {\bibfnamefont {H.}~\bibnamefont
  {{Aihara}}}, \bibinfo {author} {\bibfnamefont {N.}~\bibnamefont {{Arimoto}}},
  \bibinfo {author} {\bibfnamefont {R.}~\bibnamefont {{Armstrong}}}, \bibinfo
  {author} {\bibfnamefont {S.}~\bibnamefont {{Arnouts}}}, \bibinfo {author}
  {\bibfnamefont {N.~A.}\ \bibnamefont {{Bahcall}}}, \bibinfo {author}
  {\bibfnamefont {S.}~\bibnamefont {{Bickerton}}}, \bibinfo {author}
  {\bibfnamefont {J.}~\bibnamefont {{Bosch}}}, \bibinfo {author} {\bibfnamefont
  {K.}~\bibnamefont {{Bundy}}}, \bibinfo {author} {\bibfnamefont {P.~L.}\
  \bibnamefont {{Capak}}}, \bibinfo {author} {\bibfnamefont {J.~H.~H.}\
  \bibnamefont {{Chan}}}, \bibinfo {author} {\bibfnamefont {M.}~\bibnamefont
  {{Chiba}}}, \bibinfo {author} {\bibfnamefont {J.}~\bibnamefont {{Coupon}}},
  \bibinfo {author} {\bibfnamefont {E.}~\bibnamefont {{Egami}}}, \bibinfo
  {author} {\bibfnamefont {M.}~\bibnamefont {{Enoki}}}, \bibinfo {author}
  {\bibfnamefont {F.}~\bibnamefont {{Finet}}}, \bibinfo {author} {\bibfnamefont
  {H.}~\bibnamefont {{Fujimori}}}, \bibinfo {author} {\bibfnamefont
  {S.}~\bibnamefont {{Fujimoto}}}, \bibinfo {author} {\bibfnamefont
  {H.}~\bibnamefont {{Furusawa}}}, \bibinfo {author} {\bibfnamefont
  {J.}~\bibnamefont {{Furusawa}}}, \bibinfo {author} {\bibfnamefont
  {T.}~\bibnamefont {{Goto}}}, \bibinfo {author} {\bibfnamefont
  {A.}~\bibnamefont {{Goulding}}}, \bibinfo {author} {\bibfnamefont {J.~P.}\
  \bibnamefont {{Greco}}}, \bibinfo {author} {\bibfnamefont {J.~E.}\
  \bibnamefont {{Greene}}}, \bibinfo {author} {\bibfnamefont {J.~E.}\
  \bibnamefont {{Gunn}}}, \bibinfo {author} {\bibfnamefont {T.}~\bibnamefont
  {{Hamana}}}, \bibinfo {author} {\bibfnamefont {Y.}~\bibnamefont
  {{Harikane}}}, \bibinfo {author} {\bibfnamefont {Y.}~\bibnamefont
  {{Hashimoto}}}, \bibinfo {author} {\bibfnamefont {T.}~\bibnamefont
  {{Hattori}}}, \bibinfo {author} {\bibfnamefont {M.}~\bibnamefont
  {{Hayashi}}}, \bibinfo {author} {\bibfnamefont {Y.}~\bibnamefont
  {{Hayashi}}}, \bibinfo {author} {\bibfnamefont {K.~G.}\ \bibnamefont
  {{He{\l}miniak}}}, \bibinfo {author} {\bibfnamefont {R.}~\bibnamefont
  {{Higuchi}}}, \bibinfo {author} {\bibfnamefont {C.}~\bibnamefont {{Hikage}}},
  \bibinfo {author} {\bibfnamefont {P.~T.~P.}\ \bibnamefont {{Ho}}}, \bibinfo
  {author} {\bibfnamefont {B.-C.}\ \bibnamefont {{Hsieh}}}, \bibinfo {author}
  {\bibfnamefont {K.}~\bibnamefont {{Huang}}}, \bibinfo {author} {\bibfnamefont
  {S.}~\bibnamefont {{Huang}}}, \bibinfo {author} {\bibfnamefont
  {H.}~\bibnamefont {{Ikeda}}}, \bibinfo {author} {\bibfnamefont
  {M.}~\bibnamefont {{Imanishi}}}, \bibinfo {author} {\bibfnamefont {A.~K.}\
  \bibnamefont {{Inoue}}}, \bibinfo {author} {\bibfnamefont {K.}~\bibnamefont
  {{Iwasawa}}}, \bibinfo {author} {\bibfnamefont {I.}~\bibnamefont {{Iwata}}},
  \bibinfo {author} {\bibfnamefont {A.~T.}\ \bibnamefont {{Jaelani}}}, \bibinfo
  {author} {\bibfnamefont {H.-Y.}\ \bibnamefont {{Jian}}}, \bibinfo {author}
  {\bibfnamefont {Y.}~\bibnamefont {{Kamata}}}, \bibinfo {author}
  {\bibfnamefont {H.}~\bibnamefont {{Karoji}}}, \bibinfo {author}
  {\bibfnamefont {N.}~\bibnamefont {{Kashikawa}}}, \bibinfo {author}
  {\bibfnamefont {N.}~\bibnamefont {{Katayama}}}, \bibinfo {author}
  {\bibfnamefont {S.}~\bibnamefont {{Kawanomoto}}}, \bibinfo {author}
  {\bibfnamefont {I.}~\bibnamefont {{Kayo}}}, \bibinfo {author} {\bibfnamefont
  {J.}~\bibnamefont {{Koda}}}, \bibinfo {author} {\bibfnamefont
  {M.}~\bibnamefont {{Koike}}}, \bibinfo {author} {\bibfnamefont
  {T.}~\bibnamefont {{Kojima}}}, \bibinfo {author} {\bibfnamefont
  {Y.}~\bibnamefont {{Komiyama}}}, \bibinfo {author} {\bibfnamefont
  {A.}~\bibnamefont {{Konno}}}, \bibinfo {author} {\bibfnamefont
  {S.}~\bibnamefont {{Koshida}}}, \bibinfo {author} {\bibfnamefont
  {Y.}~\bibnamefont {{Koyama}}}, \bibinfo {author} {\bibfnamefont
  {H.}~\bibnamefont {{Kusakabe}}}, \bibinfo {author} {\bibfnamefont
  {A.}~\bibnamefont {{Leauthaud}}}, \bibinfo {author} {\bibfnamefont {C.-H.}\
  \bibnamefont {{Lee}}}, \bibinfo {author} {\bibfnamefont {L.}~\bibnamefont
  {{Lin}}}, \bibinfo {author} {\bibfnamefont {Y.-T.}\ \bibnamefont {{Lin}}},
  \bibinfo {author} {\bibfnamefont {R.~H.}\ \bibnamefont {{Lupton}}}, \bibinfo
  {author} {\bibfnamefont {R.}~\bibnamefont {{Mandelbaum}}}, \bibinfo {author}
  {\bibfnamefont {Y.}~\bibnamefont {{Matsuoka}}}, \bibinfo {author}
  {\bibfnamefont {E.}~\bibnamefont {{Medezinski}}}, \bibinfo {author}
  {\bibfnamefont {S.}~\bibnamefont {{Mineo}}}, \bibinfo {author} {\bibfnamefont
  {S.}~\bibnamefont {{Miyama}}}, \bibinfo {author} {\bibfnamefont
  {H.}~\bibnamefont {{Miyatake}}}, \bibinfo {author} {\bibfnamefont
  {S.}~\bibnamefont {{Miyazaki}}}, \bibinfo {author} {\bibfnamefont
  {R.}~\bibnamefont {{Momose}}}, \bibinfo {author} {\bibfnamefont
  {A.}~\bibnamefont {{More}}}, \bibinfo {author} {\bibfnamefont
  {S.}~\bibnamefont {{More}}}, \bibinfo {author} {\bibfnamefont
  {Y.}~\bibnamefont {{Moritani}}}, \bibinfo {author} {\bibfnamefont {T.~J.}\
  \bibnamefont {{Moriya}}}, \bibinfo {author} {\bibfnamefont {T.}~\bibnamefont
  {{Morokuma}}}, \bibinfo {author} {\bibfnamefont {S.}~\bibnamefont {{Mukae}}},
  \bibinfo {author} {\bibfnamefont {R.}~\bibnamefont {{Murata}}}, \bibinfo
  {author} {\bibfnamefont {H.}~\bibnamefont {{Murayama}}}, \bibinfo {author}
  {\bibfnamefont {T.}~\bibnamefont {{Nagao}}}, \bibinfo {author} {\bibfnamefont
  {F.}~\bibnamefont {{Nakata}}}, \bibinfo {author} {\bibfnamefont
  {M.}~\bibnamefont {{Niida}}}, \bibinfo {author} {\bibfnamefont
  {H.}~\bibnamefont {{Niikura}}}, \bibinfo {author} {\bibfnamefont {A.~J.}\
  \bibnamefont {{Nishizawa}}}, \bibinfo {author} {\bibfnamefont
  {Y.}~\bibnamefont {{Obuchi}}}, \bibinfo {author} {\bibfnamefont
  {M.}~\bibnamefont {{Oguri}}}, \bibinfo {author} {\bibfnamefont
  {Y.}~\bibnamefont {{Oishi}}}, \bibinfo {author} {\bibfnamefont
  {N.}~\bibnamefont {{Okabe}}}, \bibinfo {author} {\bibfnamefont
  {S.}~\bibnamefont {{Okamoto}}}, \bibinfo {author} {\bibfnamefont
  {Y.}~\bibnamefont {{Okura}}}, \bibinfo {author} {\bibfnamefont
  {Y.}~\bibnamefont {{Ono}}}, \bibinfo {author} {\bibfnamefont
  {M.}~\bibnamefont {{Onodera}}}, \bibinfo {author} {\bibfnamefont
  {M.}~\bibnamefont {{Onoue}}}, \bibinfo {author} {\bibfnamefont
  {K.}~\bibnamefont {{Osato}}}, \bibinfo {author} {\bibfnamefont
  {M.}~\bibnamefont {{Ouchi}}}, \bibinfo {author} {\bibfnamefont {P.~A.}\
  \bibnamefont {{Price}}}, \bibinfo {author} {\bibfnamefont {T.-S.}\
  \bibnamefont {{Pyo}}}, \bibinfo {author} {\bibfnamefont {M.}~\bibnamefont
  {{Sako}}}, \bibinfo {author} {\bibfnamefont {M.}~\bibnamefont {{Sawicki}}},
  \bibinfo {author} {\bibfnamefont {T.}~\bibnamefont {{Shibuya}}}, \bibinfo
  {author} {\bibfnamefont {K.}~\bibnamefont {{Shimasaku}}}, \bibinfo {author}
  {\bibfnamefont {A.}~\bibnamefont {{Shimono}}}, \bibinfo {author}
  {\bibfnamefont {M.}~\bibnamefont {{Shirasaki}}}, \bibinfo {author}
  {\bibfnamefont {J.~D.}\ \bibnamefont {{Silverman}}}, \bibinfo {author}
  {\bibfnamefont {M.}~\bibnamefont {{Simet}}}, \bibinfo {author} {\bibfnamefont
  {J.}~\bibnamefont {{Speagle}}}, \bibinfo {author} {\bibfnamefont {D.~N.}\
  \bibnamefont {{Spergel}}}, \bibinfo {author} {\bibfnamefont {M.~A.}\
  \bibnamefont {{Strauss}}}, \bibinfo {author} {\bibfnamefont {Y.}~\bibnamefont
  {{Sugahara}}}, \bibinfo {author} {\bibfnamefont {N.}~\bibnamefont
  {{Sugiyama}}}, \bibinfo {author} {\bibfnamefont {Y.}~\bibnamefont {{Suto}}},
  \bibinfo {author} {\bibfnamefont {S.~H.}\ \bibnamefont {{Suyu}}}, \bibinfo
  {author} {\bibfnamefont {N.}~\bibnamefont {{Suzuki}}}, \bibinfo {author}
  {\bibfnamefont {P.~J.}\ \bibnamefont {{Tait}}}, \bibinfo {author}
  {\bibfnamefont {M.}~\bibnamefont {{Takada}}}, \bibinfo {author}
  {\bibfnamefont {T.}~\bibnamefont {{Takata}}}, \bibinfo {author}
  {\bibfnamefont {N.}~\bibnamefont {{Tamura}}}, \bibinfo {author}
  {\bibfnamefont {M.~M.}\ \bibnamefont {{Tanaka}}}, \bibinfo {author}
  {\bibfnamefont {M.}~\bibnamefont {{Tanaka}}}, \bibinfo {author}
  {\bibfnamefont {M.}~\bibnamefont {{Tanaka}}}, \bibinfo {author}
  {\bibfnamefont {Y.}~\bibnamefont {{Tanaka}}}, \bibinfo {author}
  {\bibfnamefont {T.}~\bibnamefont {{Terai}}}, \bibinfo {author} {\bibfnamefont
  {Y.}~\bibnamefont {{Terashima}}}, \bibinfo {author} {\bibfnamefont
  {Y.}~\bibnamefont {{Toba}}}, \bibinfo {author} {\bibfnamefont
  {N.}~\bibnamefont {{Tominaga}}}, \bibinfo {author} {\bibfnamefont
  {J.}~\bibnamefont {{Toshikawa}}}, \bibinfo {author} {\bibfnamefont {E.~L.}\
  \bibnamefont {{Turner}}}, \bibinfo {author} {\bibfnamefont {T.}~\bibnamefont
  {{Uchida}}}, \bibinfo {author} {\bibfnamefont {H.}~\bibnamefont
  {{Uchiyama}}}, \bibinfo {author} {\bibfnamefont {K.}~\bibnamefont
  {{Umetsu}}}, \bibinfo {author} {\bibfnamefont {F.}~\bibnamefont
  {{Uraguchi}}}, \bibinfo {author} {\bibfnamefont {Y.}~\bibnamefont {{Urata}}},
  \bibinfo {author} {\bibfnamefont {T.}~\bibnamefont {{Usuda}}}, \bibinfo
  {author} {\bibfnamefont {Y.}~\bibnamefont {{Utsumi}}}, \bibinfo {author}
  {\bibfnamefont {S.-Y.}\ \bibnamefont {{Wang}}}, \bibinfo {author}
  {\bibfnamefont {W.-H.}\ \bibnamefont {{Wang}}}, \bibinfo {author}
  {\bibfnamefont {K.~C.}\ \bibnamefont {{Wong}}}, \bibinfo {author}
  {\bibfnamefont {K.}~\bibnamefont {{Yabe}}}, \bibinfo {author} {\bibfnamefont
  {Y.}~\bibnamefont {{Yamada}}}, \bibinfo {author} {\bibfnamefont
  {H.}~\bibnamefont {{Yamanoi}}}, \bibinfo {author} {\bibfnamefont
  {N.}~\bibnamefont {{Yasuda}}}, \bibinfo {author} {\bibfnamefont
  {S.}~\bibnamefont {{Yeh}}}, \bibinfo {author} {\bibfnamefont
  {A.}~\bibnamefont {{Yonehara}}},\ and\ \bibinfo {author} {\bibfnamefont
  {S.}~\bibnamefont {{Yuma}}},\ }\bibfield  {title} {\bibinfo {title} {{The
  Hyper Suprime-Cam SSP Survey: Overview and survey design}},\ }\href
  {https://doi.org/10.1093/pasj/psx066} {\bibfield  {journal} {\bibinfo
  {journal} {\pasj}\ }\textbf {\bibinfo {volume} {70}},\ \bibinfo {eid} {S4}
  (\bibinfo {year} {2018}{\natexlab{a}})},\ \Eprint
  {https://arxiv.org/abs/1704.05858} {arXiv:1704.05858 [astro-ph.IM]}
  \BibitemShut {NoStop}%
\bibitem [{\citenamefont {{Kilbinger}}\ \emph {et~al.}(2013)\citenamefont
  {{Kilbinger}}, \citenamefont {{Fu}}, \citenamefont {{Heymans}}, \citenamefont
  {{Simpson}}, \citenamefont {{Benjamin}}, \citenamefont {{Erben}},
  \citenamefont {{Harnois-D{\'e}raps}}, \citenamefont {{Hoekstra}},
  \citenamefont {{Hildebrandt}}, \citenamefont {{Kitching}}, \citenamefont
  {{Mellier}}, \citenamefont {{Miller}}, \citenamefont {{Van Waerbeke}},
  \citenamefont {{Benabed}}, \citenamefont {{Bonnett}}, \citenamefont
  {{Coupon}}, \citenamefont {{Hudson}}, \citenamefont {{Kuijken}},
  \citenamefont {{Rowe}}, \citenamefont {{Schrabback}}, \citenamefont
  {{Semboloni}}, \citenamefont {{Vafaei}},\ and\ \citenamefont
  {{Velander}}}]{Kil2013}%
  \BibitemOpen
  \bibfield  {author} {\bibinfo {author} {\bibfnamefont {M.}~\bibnamefont
  {{Kilbinger}}}, \bibinfo {author} {\bibfnamefont {L.}~\bibnamefont {{Fu}}},
  \bibinfo {author} {\bibfnamefont {C.}~\bibnamefont {{Heymans}}}, \bibinfo
  {author} {\bibfnamefont {F.}~\bibnamefont {{Simpson}}}, \bibinfo {author}
  {\bibfnamefont {J.}~\bibnamefont {{Benjamin}}}, \bibinfo {author}
  {\bibfnamefont {T.}~\bibnamefont {{Erben}}}, \bibinfo {author} {\bibfnamefont
  {J.}~\bibnamefont {{Harnois-D{\'e}raps}}}, \bibinfo {author} {\bibfnamefont
  {H.}~\bibnamefont {{Hoekstra}}}, \bibinfo {author} {\bibfnamefont
  {H.}~\bibnamefont {{Hildebrandt}}}, \bibinfo {author} {\bibfnamefont {T.~D.}\
  \bibnamefont {{Kitching}}}, \bibinfo {author} {\bibfnamefont
  {Y.}~\bibnamefont {{Mellier}}}, \bibinfo {author} {\bibfnamefont
  {L.}~\bibnamefont {{Miller}}}, \bibinfo {author} {\bibfnamefont
  {L.}~\bibnamefont {{Van Waerbeke}}}, \bibinfo {author} {\bibfnamefont
  {K.}~\bibnamefont {{Benabed}}}, \bibinfo {author} {\bibfnamefont
  {C.}~\bibnamefont {{Bonnett}}}, \bibinfo {author} {\bibfnamefont
  {J.}~\bibnamefont {{Coupon}}}, \bibinfo {author} {\bibfnamefont {M.~J.}\
  \bibnamefont {{Hudson}}}, \bibinfo {author} {\bibfnamefont {K.}~\bibnamefont
  {{Kuijken}}}, \bibinfo {author} {\bibfnamefont {B.}~\bibnamefont {{Rowe}}},
  \bibinfo {author} {\bibfnamefont {T.}~\bibnamefont {{Schrabback}}}, \bibinfo
  {author} {\bibfnamefont {E.}~\bibnamefont {{Semboloni}}}, \bibinfo {author}
  {\bibfnamefont {S.}~\bibnamefont {{Vafaei}}},\ and\ \bibinfo {author}
  {\bibfnamefont {M.}~\bibnamefont {{Velander}}},\ }\bibfield  {title}
  {\bibinfo {title} {{CFHTLenS: combined probe cosmological model comparison
  using 2D weak gravitational lensing}},\ }\href
  {https://doi.org/10.1093/mnras/stt041} {\bibfield  {journal} {\bibinfo
  {journal} {\mnras}\ }\textbf {\bibinfo {volume} {430}},\ \bibinfo {pages}
  {2200} (\bibinfo {year} {2013})},\ \Eprint {https://arxiv.org/abs/1212.3338}
  {arXiv:1212.3338 [astro-ph.CO]} \BibitemShut {NoStop}%
\bibitem [{\citenamefont {{Hildebrandt}}\ \emph {et~al.}(2017)\citenamefont
  {{Hildebrandt}}, \citenamefont {{Viola}}, \citenamefont {{Heymans}},
  \citenamefont {{Joudaki}}, \citenamefont {{Kuijken}}, \citenamefont
  {{Blake}}, \citenamefont {{Erben}}, \citenamefont {{Joachimi}}, \citenamefont
  {{Klaes}}, \citenamefont {{Miller}}, \citenamefont {{Morrison}},
  \citenamefont {{Nakajima}}, \citenamefont {{Verdoes Kleijn}}, \citenamefont
  {{Amon}}, \citenamefont {{Choi}}, \citenamefont {{Covone}}, \citenamefont
  {{de Jong}}, \citenamefont {{Dvornik}}, \citenamefont {{Fenech Conti}},
  \citenamefont {{Grado}}, \citenamefont {{Harnois-D{\'e}raps}}, \citenamefont
  {{Herbonnet}}, \citenamefont {{Hoekstra}}, \citenamefont {{K{\"o}hlinger}},
  \citenamefont {{McFarland}}, \citenamefont {{Mead}}, \citenamefont
  {{Merten}}, \citenamefont {{Napolitano}}, \citenamefont {{Peacock}},
  \citenamefont {{Radovich}}, \citenamefont {{Schneider}}, \citenamefont
  {{Simon}}, \citenamefont {{Valentijn}}, \citenamefont {{van den Busch}},
  \citenamefont {{van Uitert}},\ and\ \citenamefont {{Van
  Waerbeke}}}]{Hilde2017}%
  \BibitemOpen
  \bibfield  {author} {\bibinfo {author} {\bibfnamefont {H.}~\bibnamefont
  {{Hildebrandt}}}, \bibinfo {author} {\bibfnamefont {M.}~\bibnamefont
  {{Viola}}}, \bibinfo {author} {\bibfnamefont {C.}~\bibnamefont {{Heymans}}},
  \bibinfo {author} {\bibfnamefont {S.}~\bibnamefont {{Joudaki}}}, \bibinfo
  {author} {\bibfnamefont {K.}~\bibnamefont {{Kuijken}}}, \bibinfo {author}
  {\bibfnamefont {C.}~\bibnamefont {{Blake}}}, \bibinfo {author} {\bibfnamefont
  {T.}~\bibnamefont {{Erben}}}, \bibinfo {author} {\bibfnamefont
  {B.}~\bibnamefont {{Joachimi}}}, \bibinfo {author} {\bibfnamefont
  {D.}~\bibnamefont {{Klaes}}}, \bibinfo {author} {\bibfnamefont
  {L.}~\bibnamefont {{Miller}}}, \bibinfo {author} {\bibfnamefont {C.~B.}\
  \bibnamefont {{Morrison}}}, \bibinfo {author} {\bibfnamefont
  {R.}~\bibnamefont {{Nakajima}}}, \bibinfo {author} {\bibfnamefont
  {G.}~\bibnamefont {{Verdoes Kleijn}}}, \bibinfo {author} {\bibfnamefont
  {A.}~\bibnamefont {{Amon}}}, \bibinfo {author} {\bibfnamefont
  {A.}~\bibnamefont {{Choi}}}, \bibinfo {author} {\bibfnamefont
  {G.}~\bibnamefont {{Covone}}}, \bibinfo {author} {\bibfnamefont {J.~T.~A.}\
  \bibnamefont {{de Jong}}}, \bibinfo {author} {\bibfnamefont {A.}~\bibnamefont
  {{Dvornik}}}, \bibinfo {author} {\bibfnamefont {I.}~\bibnamefont {{Fenech
  Conti}}}, \bibinfo {author} {\bibfnamefont {A.}~\bibnamefont {{Grado}}},
  \bibinfo {author} {\bibfnamefont {J.}~\bibnamefont {{Harnois-D{\'e}raps}}},
  \bibinfo {author} {\bibfnamefont {R.}~\bibnamefont {{Herbonnet}}}, \bibinfo
  {author} {\bibfnamefont {H.}~\bibnamefont {{Hoekstra}}}, \bibinfo {author}
  {\bibfnamefont {F.}~\bibnamefont {{K{\"o}hlinger}}}, \bibinfo {author}
  {\bibfnamefont {J.}~\bibnamefont {{McFarland}}}, \bibinfo {author}
  {\bibfnamefont {A.}~\bibnamefont {{Mead}}}, \bibinfo {author} {\bibfnamefont
  {J.}~\bibnamefont {{Merten}}}, \bibinfo {author} {\bibfnamefont
  {N.}~\bibnamefont {{Napolitano}}}, \bibinfo {author} {\bibfnamefont {J.~A.}\
  \bibnamefont {{Peacock}}}, \bibinfo {author} {\bibfnamefont {M.}~\bibnamefont
  {{Radovich}}}, \bibinfo {author} {\bibfnamefont {P.}~\bibnamefont
  {{Schneider}}}, \bibinfo {author} {\bibfnamefont {P.}~\bibnamefont
  {{Simon}}}, \bibinfo {author} {\bibfnamefont {E.~A.}\ \bibnamefont
  {{Valentijn}}}, \bibinfo {author} {\bibfnamefont {J.~L.}\ \bibnamefont {{van
  den Busch}}}, \bibinfo {author} {\bibfnamefont {E.}~\bibnamefont {{van
  Uitert}}},\ and\ \bibinfo {author} {\bibfnamefont {L.}~\bibnamefont {{Van
  Waerbeke}}},\ }\bibfield  {title} {\bibinfo {title} {{KiDS-450: cosmological
  parameter constraints from tomographic weak gravitational lensing}},\ }\href
  {https://doi.org/10.1093/mnras/stw2805} {\bibfield  {journal} {\bibinfo
  {journal} {\mnras}\ }\textbf {\bibinfo {volume} {465}},\ \bibinfo {pages}
  {1454} (\bibinfo {year} {2017})},\ \Eprint {https://arxiv.org/abs/1606.05338}
  {arXiv:1606.05338 [astro-ph.CO]} \BibitemShut {NoStop}%
\bibitem [{\citenamefont {{Abbott}}\ \emph {et~al.}(2018)\citenamefont
  {{Abbott}}, \citenamefont {{Abdalla}}, \citenamefont {{Alarcon}},
  \citenamefont {{Aleksi{\'c}}}, \citenamefont {{Allam}}, \citenamefont
  {{Allen}}, \citenamefont {{Amara}}, \citenamefont {{Annis}}, \citenamefont
  {{Asorey}}, \citenamefont {{Avila}}, \citenamefont {{Bacon}}, \citenamefont
  {{Balbinot}}, \citenamefont {{Banerji}}, \citenamefont {{Banik}},
  \citenamefont {{Barkhouse}}, \citenamefont {{Baumer}}, \citenamefont
  {{Baxter}}, \citenamefont {{Bechtol}}, \citenamefont {{Becker}},
  \citenamefont {{Benoit-L{\'e}vy}}, \citenamefont {{Benson}}, \citenamefont
  {{Bernstein}}, \citenamefont {{Bertin}}, \citenamefont {{Blazek}},
  \citenamefont {{Bridle}}, \citenamefont {{Brooks}}, \citenamefont {{Brout}},
  \citenamefont {{Buckley-Geer}}, \citenamefont {{Burke}}, \citenamefont
  {{Busha}}, \citenamefont {{Campos}}, \citenamefont {{Capozzi}}, \citenamefont
  {{Carnero Rosell}}, \citenamefont {{Carrasco Kind}}, \citenamefont
  {{Carretero}}, \citenamefont {{Castander}}, \citenamefont {{Cawthon}},
  \citenamefont {{Chang}}, \citenamefont {{Chen}}, \citenamefont {{Childress}},
  \citenamefont {{Choi}}, \citenamefont {{Conselice}}, \citenamefont
  {{Crittenden}}, \citenamefont {{Crocce}}, \citenamefont {{Cunha}},
  \citenamefont {{D'Andrea}}, \citenamefont {{da Costa}}, \citenamefont
  {{Das}}, \citenamefont {{Davis}}, \citenamefont {{Davis}}, \citenamefont {{De
  Vicente}}, \citenamefont {{DePoy}}, \citenamefont {{DeRose}}, \citenamefont
  {{Desai}}, \citenamefont {{Diehl}}, \citenamefont {{Dietrich}}, \citenamefont
  {{Dodelson}}, \citenamefont {{Doel}}, \citenamefont {{Drlica-Wagner}},
  \citenamefont {{Eifler}}, \citenamefont {{Elliott}}, \citenamefont
  {{Elsner}}, \citenamefont {{Elvin-Poole}}, \citenamefont {{Estrada}},
  \citenamefont {{Evrard}}, \citenamefont {{Fang}}, \citenamefont
  {{Fernandez}}, \citenamefont {{Fert{\'e}}}, \citenamefont {{Finley}},
  \citenamefont {{Flaugher}}, \citenamefont {{Fosalba}}, \citenamefont
  {{Friedrich}}, \citenamefont {{Frieman}}, \citenamefont
  {{Garc{\'\i}a-Bellido}}, \citenamefont {{Garcia-Fernandez}}, \citenamefont
  {{Gatti}}, \citenamefont {{Gaztanaga}}, \citenamefont {{Gerdes}},
  \citenamefont {{Giannantonio}}, \citenamefont {{Gill}}, \citenamefont
  {{Glazebrook}}, \citenamefont {{Goldstein}}, \citenamefont {{Gruen}},
  \citenamefont {{Gruendl}}, \citenamefont {{Gschwend}}, \citenamefont
  {{Gutierrez}}, \citenamefont {{Hamilton}}, \citenamefont {{Hartley}},
  \citenamefont {{Hinton}}, \citenamefont {{Honscheid}}, \citenamefont
  {{Hoyle}}, \citenamefont {{Huterer}}, \citenamefont {{Jain}}, \citenamefont
  {{James}}, \citenamefont {{Jarvis}}, \citenamefont {{Jeltema}}, \citenamefont
  {{Johnson}}, \citenamefont {{Johnson}}, \citenamefont {{Kacprzak}},
  \citenamefont {{Kent}}, \citenamefont {{Kim}}, \citenamefont {{King}},
  \citenamefont {{Kirk}}, \citenamefont {{Kokron}}, \citenamefont {{Kovacs}},
  \citenamefont {{Krause}}, \citenamefont {{Krawiec}}, \citenamefont
  {{Kremin}}, \citenamefont {{Kuehn}}, \citenamefont {{Kuhlmann}},
  \citenamefont {{Kuropatkin}}, \citenamefont {{Lacasa}}, \citenamefont
  {{Lahav}}, \citenamefont {{Li}}, \citenamefont {{Liddle}}, \citenamefont
  {{Lidman}}, \citenamefont {{Lima}}, \citenamefont {{Lin}}, \citenamefont
  {{MacCrann}}, \citenamefont {{Maia}}, \citenamefont {{Makler}}, \citenamefont
  {{Manera}}, \citenamefont {{March}}, \citenamefont {{Marshall}},
  \citenamefont {{Martini}}, \citenamefont {{McMahon}}, \citenamefont
  {{Melchior}}, \citenamefont {{Menanteau}}, \citenamefont {{Miquel}},
  \citenamefont {{Miranda}}, \citenamefont {{Mudd}}, \citenamefont {{Muir}},
  \citenamefont {{M{\"o}ller}}, \citenamefont {{Neilsen}}, \citenamefont
  {{Nichol}}, \citenamefont {{Nord}}, \citenamefont {{Nugent}}, \citenamefont
  {{Ogando}}, \citenamefont {{Palmese}}, \citenamefont {{Peacock}},
  \citenamefont {{Peiris}}, \citenamefont {{Peoples}}, \citenamefont
  {{Percival}}, \citenamefont {{Petravick}}, \citenamefont {{Plazas}},
  \citenamefont {{Porredon}}, \citenamefont {{Prat}}, \citenamefont {{Pujol}},
  \citenamefont {{Rau}}, \citenamefont {{Refregier}}, \citenamefont {{Ricker}},
  \citenamefont {{Roe}}, \citenamefont {{Rollins}}, \citenamefont {{Romer}},
  \citenamefont {{Roodman}}, \citenamefont {{Rosenfeld}}, \citenamefont
  {{Ross}}, \citenamefont {{Rozo}}, \citenamefont {{Rykoff}}, \citenamefont
  {{Sako}}, \citenamefont {{Salvador}}, \citenamefont {{Samuroff}},
  \citenamefont {{S{\'a}nchez}}, \citenamefont {{Sanchez}}, \citenamefont
  {{Santiago}}, \citenamefont {{Scarpine}}, \citenamefont {{Schindler}},
  \citenamefont {{Scolnic}}, \citenamefont {{Secco}}, \citenamefont
  {{Serrano}}, \citenamefont {{Sevilla-Noarbe}}, \citenamefont {{Sheldon}},
  \citenamefont {{Smith}}, \citenamefont {{Smith}}, \citenamefont {{Smith}},
  \citenamefont {{Soares-Santos}}, \citenamefont {{Sobreira}}, \citenamefont
  {{Suchyta}}, \citenamefont {{Tarle}}, \citenamefont {{Thomas}}, \citenamefont
  {{Troxel}}, \citenamefont {{Tucker}}, \citenamefont {{Tucker}}, \citenamefont
  {{Uddin}}, \citenamefont {{Varga}}, \citenamefont {{Vielzeuf}}, \citenamefont
  {{Vikram}}, \citenamefont {{Vivas}}, \citenamefont {{Walker}}, \citenamefont
  {{Wang}}, \citenamefont {{Wechsler}}, \citenamefont {{Weller}}, \citenamefont
  {{Wester}}, \citenamefont {{Wolf}}, \citenamefont {{Yanny}}, \citenamefont
  {{Yuan}}, \citenamefont {{Zenteno}}, \citenamefont {{Zhang}}, \citenamefont
  {{Zhang}}, \citenamefont {{Zuntz}},\ and\ \citenamefont {{Dark Energy Survey
  Collaboration}}}]{Abbott2018}%
  \BibitemOpen
  \bibfield  {author} {\bibinfo {author} {\bibfnamefont {T.~M.~C.}\
  \bibnamefont {{Abbott}}}, \bibinfo {author} {\bibfnamefont {F.~B.}\
  \bibnamefont {{Abdalla}}}, \bibinfo {author} {\bibfnamefont {A.}~\bibnamefont
  {{Alarcon}}}, \bibinfo {author} {\bibfnamefont {J.}~\bibnamefont
  {{Aleksi{\'c}}}}, \bibinfo {author} {\bibfnamefont {S.}~\bibnamefont
  {{Allam}}}, \bibinfo {author} {\bibfnamefont {S.}~\bibnamefont {{Allen}}},
  \bibinfo {author} {\bibfnamefont {A.}~\bibnamefont {{Amara}}}, \bibinfo
  {author} {\bibfnamefont {J.}~\bibnamefont {{Annis}}}, \bibinfo {author}
  {\bibfnamefont {J.}~\bibnamefont {{Asorey}}}, \bibinfo {author}
  {\bibfnamefont {S.}~\bibnamefont {{Avila}}}, \bibinfo {author} {\bibfnamefont
  {D.}~\bibnamefont {{Bacon}}}, \bibinfo {author} {\bibfnamefont
  {E.}~\bibnamefont {{Balbinot}}}, \bibinfo {author} {\bibfnamefont
  {M.}~\bibnamefont {{Banerji}}}, \bibinfo {author} {\bibfnamefont
  {N.}~\bibnamefont {{Banik}}}, \bibinfo {author} {\bibfnamefont
  {W.}~\bibnamefont {{Barkhouse}}}, \bibinfo {author} {\bibfnamefont
  {M.}~\bibnamefont {{Baumer}}}, \bibinfo {author} {\bibfnamefont
  {E.}~\bibnamefont {{Baxter}}}, \bibinfo {author} {\bibfnamefont
  {K.}~\bibnamefont {{Bechtol}}}, \bibinfo {author} {\bibfnamefont {M.~R.}\
  \bibnamefont {{Becker}}}, \bibinfo {author} {\bibfnamefont {A.}~\bibnamefont
  {{Benoit-L{\'e}vy}}}, \bibinfo {author} {\bibfnamefont {B.~A.}\ \bibnamefont
  {{Benson}}}, \bibinfo {author} {\bibfnamefont {G.~M.}\ \bibnamefont
  {{Bernstein}}}, \bibinfo {author} {\bibfnamefont {E.}~\bibnamefont
  {{Bertin}}}, \bibinfo {author} {\bibfnamefont {J.}~\bibnamefont {{Blazek}}},
  \bibinfo {author} {\bibfnamefont {S.~L.}\ \bibnamefont {{Bridle}}}, \bibinfo
  {author} {\bibfnamefont {D.}~\bibnamefont {{Brooks}}}, \bibinfo {author}
  {\bibfnamefont {D.}~\bibnamefont {{Brout}}}, \bibinfo {author} {\bibfnamefont
  {E.}~\bibnamefont {{Buckley-Geer}}}, \bibinfo {author} {\bibfnamefont
  {D.~L.}\ \bibnamefont {{Burke}}}, \bibinfo {author} {\bibfnamefont {M.~T.}\
  \bibnamefont {{Busha}}}, \bibinfo {author} {\bibfnamefont {A.}~\bibnamefont
  {{Campos}}}, \bibinfo {author} {\bibfnamefont {D.}~\bibnamefont {{Capozzi}}},
  \bibinfo {author} {\bibfnamefont {A.}~\bibnamefont {{Carnero Rosell}}},
  \bibinfo {author} {\bibfnamefont {M.}~\bibnamefont {{Carrasco Kind}}},
  \bibinfo {author} {\bibfnamefont {J.}~\bibnamefont {{Carretero}}}, \bibinfo
  {author} {\bibfnamefont {F.~J.}\ \bibnamefont {{Castander}}}, \bibinfo
  {author} {\bibfnamefont {R.}~\bibnamefont {{Cawthon}}}, \bibinfo {author}
  {\bibfnamefont {C.}~\bibnamefont {{Chang}}}, \bibinfo {author} {\bibfnamefont
  {N.}~\bibnamefont {{Chen}}}, \bibinfo {author} {\bibfnamefont
  {M.}~\bibnamefont {{Childress}}}, \bibinfo {author} {\bibfnamefont
  {A.}~\bibnamefont {{Choi}}}, \bibinfo {author} {\bibfnamefont
  {C.}~\bibnamefont {{Conselice}}}, \bibinfo {author} {\bibfnamefont
  {R.}~\bibnamefont {{Crittenden}}}, \bibinfo {author} {\bibfnamefont
  {M.}~\bibnamefont {{Crocce}}}, \bibinfo {author} {\bibfnamefont {C.~E.}\
  \bibnamefont {{Cunha}}}, \bibinfo {author} {\bibfnamefont {C.~B.}\
  \bibnamefont {{D'Andrea}}}, \bibinfo {author} {\bibfnamefont {L.~N.}\
  \bibnamefont {{da Costa}}}, \bibinfo {author} {\bibfnamefont
  {R.}~\bibnamefont {{Das}}}, \bibinfo {author} {\bibfnamefont {T.~M.}\
  \bibnamefont {{Davis}}}, \bibinfo {author} {\bibfnamefont {C.}~\bibnamefont
  {{Davis}}}, \bibinfo {author} {\bibfnamefont {J.}~\bibnamefont {{De
  Vicente}}}, \bibinfo {author} {\bibfnamefont {D.~L.}\ \bibnamefont
  {{DePoy}}}, \bibinfo {author} {\bibfnamefont {J.}~\bibnamefont {{DeRose}}},
  \bibinfo {author} {\bibfnamefont {S.}~\bibnamefont {{Desai}}}, \bibinfo
  {author} {\bibfnamefont {H.~T.}\ \bibnamefont {{Diehl}}}, \bibinfo {author}
  {\bibfnamefont {J.~P.}\ \bibnamefont {{Dietrich}}}, \bibinfo {author}
  {\bibfnamefont {S.}~\bibnamefont {{Dodelson}}}, \bibinfo {author}
  {\bibfnamefont {P.}~\bibnamefont {{Doel}}}, \bibinfo {author} {\bibfnamefont
  {A.}~\bibnamefont {{Drlica-Wagner}}}, \bibinfo {author} {\bibfnamefont
  {T.~F.}\ \bibnamefont {{Eifler}}}, \bibinfo {author} {\bibfnamefont {A.~E.}\
  \bibnamefont {{Elliott}}}, \bibinfo {author} {\bibfnamefont {F.}~\bibnamefont
  {{Elsner}}}, \bibinfo {author} {\bibfnamefont {J.}~\bibnamefont
  {{Elvin-Poole}}}, \bibinfo {author} {\bibfnamefont {J.}~\bibnamefont
  {{Estrada}}}, \bibinfo {author} {\bibfnamefont {A.~E.}\ \bibnamefont
  {{Evrard}}}, \bibinfo {author} {\bibfnamefont {Y.}~\bibnamefont {{Fang}}},
  \bibinfo {author} {\bibfnamefont {E.}~\bibnamefont {{Fernandez}}}, \bibinfo
  {author} {\bibfnamefont {A.}~\bibnamefont {{Fert{\'e}}}}, \bibinfo {author}
  {\bibfnamefont {D.~A.}\ \bibnamefont {{Finley}}}, \bibinfo {author}
  {\bibfnamefont {B.}~\bibnamefont {{Flaugher}}}, \bibinfo {author}
  {\bibfnamefont {P.}~\bibnamefont {{Fosalba}}}, \bibinfo {author}
  {\bibfnamefont {O.}~\bibnamefont {{Friedrich}}}, \bibinfo {author}
  {\bibfnamefont {J.}~\bibnamefont {{Frieman}}}, \bibinfo {author}
  {\bibfnamefont {J.}~\bibnamefont {{Garc{\'\i}a-Bellido}}}, \bibinfo {author}
  {\bibfnamefont {M.}~\bibnamefont {{Garcia-Fernandez}}}, \bibinfo {author}
  {\bibfnamefont {M.}~\bibnamefont {{Gatti}}}, \bibinfo {author} {\bibfnamefont
  {E.}~\bibnamefont {{Gaztanaga}}}, \bibinfo {author} {\bibfnamefont {D.~W.}\
  \bibnamefont {{Gerdes}}}, \bibinfo {author} {\bibfnamefont {T.}~\bibnamefont
  {{Giannantonio}}}, \bibinfo {author} {\bibfnamefont {M.~S.~S.}\ \bibnamefont
  {{Gill}}}, \bibinfo {author} {\bibfnamefont {K.}~\bibnamefont
  {{Glazebrook}}}, \bibinfo {author} {\bibfnamefont {D.~A.}\ \bibnamefont
  {{Goldstein}}}, \bibinfo {author} {\bibfnamefont {D.}~\bibnamefont
  {{Gruen}}}, \bibinfo {author} {\bibfnamefont {R.~A.}\ \bibnamefont
  {{Gruendl}}}, \bibinfo {author} {\bibfnamefont {J.}~\bibnamefont
  {{Gschwend}}}, \bibinfo {author} {\bibfnamefont {G.}~\bibnamefont
  {{Gutierrez}}}, \bibinfo {author} {\bibfnamefont {S.}~\bibnamefont
  {{Hamilton}}}, \bibinfo {author} {\bibfnamefont {W.~G.}\ \bibnamefont
  {{Hartley}}}, \bibinfo {author} {\bibfnamefont {S.~R.}\ \bibnamefont
  {{Hinton}}}, \bibinfo {author} {\bibfnamefont {K.}~\bibnamefont
  {{Honscheid}}}, \bibinfo {author} {\bibfnamefont {B.}~\bibnamefont
  {{Hoyle}}}, \bibinfo {author} {\bibfnamefont {D.}~\bibnamefont {{Huterer}}},
  \bibinfo {author} {\bibfnamefont {B.}~\bibnamefont {{Jain}}}, \bibinfo
  {author} {\bibfnamefont {D.~J.}\ \bibnamefont {{James}}}, \bibinfo {author}
  {\bibfnamefont {M.}~\bibnamefont {{Jarvis}}}, \bibinfo {author}
  {\bibfnamefont {T.}~\bibnamefont {{Jeltema}}}, \bibinfo {author}
  {\bibfnamefont {M.~D.}\ \bibnamefont {{Johnson}}}, \bibinfo {author}
  {\bibfnamefont {M.~W.~G.}\ \bibnamefont {{Johnson}}}, \bibinfo {author}
  {\bibfnamefont {T.}~\bibnamefont {{Kacprzak}}}, \bibinfo {author}
  {\bibfnamefont {S.}~\bibnamefont {{Kent}}}, \bibinfo {author} {\bibfnamefont
  {A.~G.}\ \bibnamefont {{Kim}}}, \bibinfo {author} {\bibfnamefont
  {A.}~\bibnamefont {{King}}}, \bibinfo {author} {\bibfnamefont
  {D.}~\bibnamefont {{Kirk}}}, \bibinfo {author} {\bibfnamefont
  {N.}~\bibnamefont {{Kokron}}}, \bibinfo {author} {\bibfnamefont
  {A.}~\bibnamefont {{Kovacs}}}, \bibinfo {author} {\bibfnamefont
  {E.}~\bibnamefont {{Krause}}}, \bibinfo {author} {\bibfnamefont
  {C.}~\bibnamefont {{Krawiec}}}, \bibinfo {author} {\bibfnamefont
  {A.}~\bibnamefont {{Kremin}}}, \bibinfo {author} {\bibfnamefont
  {K.}~\bibnamefont {{Kuehn}}}, \bibinfo {author} {\bibfnamefont
  {S.}~\bibnamefont {{Kuhlmann}}}, \bibinfo {author} {\bibfnamefont
  {N.}~\bibnamefont {{Kuropatkin}}}, \bibinfo {author} {\bibfnamefont
  {F.}~\bibnamefont {{Lacasa}}}, \bibinfo {author} {\bibfnamefont
  {O.}~\bibnamefont {{Lahav}}}, \bibinfo {author} {\bibfnamefont {T.~S.}\
  \bibnamefont {{Li}}}, \bibinfo {author} {\bibfnamefont {A.~R.}\ \bibnamefont
  {{Liddle}}}, \bibinfo {author} {\bibfnamefont {C.}~\bibnamefont {{Lidman}}},
  \bibinfo {author} {\bibfnamefont {M.}~\bibnamefont {{Lima}}}, \bibinfo
  {author} {\bibfnamefont {H.}~\bibnamefont {{Lin}}}, \bibinfo {author}
  {\bibfnamefont {N.}~\bibnamefont {{MacCrann}}}, \bibinfo {author}
  {\bibfnamefont {M.~A.~G.}\ \bibnamefont {{Maia}}}, \bibinfo {author}
  {\bibfnamefont {M.}~\bibnamefont {{Makler}}}, \bibinfo {author}
  {\bibfnamefont {M.}~\bibnamefont {{Manera}}}, \bibinfo {author}
  {\bibfnamefont {M.}~\bibnamefont {{March}}}, \bibinfo {author} {\bibfnamefont
  {J.~L.}\ \bibnamefont {{Marshall}}}, \bibinfo {author} {\bibfnamefont
  {P.}~\bibnamefont {{Martini}}}, \bibinfo {author} {\bibfnamefont {R.~G.}\
  \bibnamefont {{McMahon}}}, \bibinfo {author} {\bibfnamefont {P.}~\bibnamefont
  {{Melchior}}}, \bibinfo {author} {\bibfnamefont {F.}~\bibnamefont
  {{Menanteau}}}, \bibinfo {author} {\bibfnamefont {R.}~\bibnamefont
  {{Miquel}}}, \bibinfo {author} {\bibfnamefont {V.}~\bibnamefont {{Miranda}}},
  \bibinfo {author} {\bibfnamefont {D.}~\bibnamefont {{Mudd}}}, \bibinfo
  {author} {\bibfnamefont {J.}~\bibnamefont {{Muir}}}, \bibinfo {author}
  {\bibfnamefont {A.}~\bibnamefont {{M{\"o}ller}}}, \bibinfo {author}
  {\bibfnamefont {E.}~\bibnamefont {{Neilsen}}}, \bibinfo {author}
  {\bibfnamefont {R.~C.}\ \bibnamefont {{Nichol}}}, \bibinfo {author}
  {\bibfnamefont {B.}~\bibnamefont {{Nord}}}, \bibinfo {author} {\bibfnamefont
  {P.}~\bibnamefont {{Nugent}}}, \bibinfo {author} {\bibfnamefont {R.~L.~C.}\
  \bibnamefont {{Ogando}}}, \bibinfo {author} {\bibfnamefont {A.}~\bibnamefont
  {{Palmese}}}, \bibinfo {author} {\bibfnamefont {J.}~\bibnamefont
  {{Peacock}}}, \bibinfo {author} {\bibfnamefont {H.~V.}\ \bibnamefont
  {{Peiris}}}, \bibinfo {author} {\bibfnamefont {J.}~\bibnamefont {{Peoples}}},
  \bibinfo {author} {\bibfnamefont {W.~J.}\ \bibnamefont {{Percival}}},
  \bibinfo {author} {\bibfnamefont {D.}~\bibnamefont {{Petravick}}}, \bibinfo
  {author} {\bibfnamefont {A.~A.}\ \bibnamefont {{Plazas}}}, \bibinfo {author}
  {\bibfnamefont {A.}~\bibnamefont {{Porredon}}}, \bibinfo {author}
  {\bibfnamefont {J.}~\bibnamefont {{Prat}}}, \bibinfo {author} {\bibfnamefont
  {A.}~\bibnamefont {{Pujol}}}, \bibinfo {author} {\bibfnamefont {M.~M.}\
  \bibnamefont {{Rau}}}, \bibinfo {author} {\bibfnamefont {A.}~\bibnamefont
  {{Refregier}}}, \bibinfo {author} {\bibfnamefont {P.~M.}\ \bibnamefont
  {{Ricker}}}, \bibinfo {author} {\bibfnamefont {N.}~\bibnamefont {{Roe}}},
  \bibinfo {author} {\bibfnamefont {R.~P.}\ \bibnamefont {{Rollins}}}, \bibinfo
  {author} {\bibfnamefont {A.~K.}\ \bibnamefont {{Romer}}}, \bibinfo {author}
  {\bibfnamefont {A.}~\bibnamefont {{Roodman}}}, \bibinfo {author}
  {\bibfnamefont {R.}~\bibnamefont {{Rosenfeld}}}, \bibinfo {author}
  {\bibfnamefont {A.~J.}\ \bibnamefont {{Ross}}}, \bibinfo {author}
  {\bibfnamefont {E.}~\bibnamefont {{Rozo}}}, \bibinfo {author} {\bibfnamefont
  {E.~S.}\ \bibnamefont {{Rykoff}}}, \bibinfo {author} {\bibfnamefont
  {M.}~\bibnamefont {{Sako}}}, \bibinfo {author} {\bibfnamefont {A.~I.}\
  \bibnamefont {{Salvador}}}, \bibinfo {author} {\bibfnamefont
  {S.}~\bibnamefont {{Samuroff}}}, \bibinfo {author} {\bibfnamefont
  {C.}~\bibnamefont {{S{\'a}nchez}}}, \bibinfo {author} {\bibfnamefont
  {E.}~\bibnamefont {{Sanchez}}}, \bibinfo {author} {\bibfnamefont
  {B.}~\bibnamefont {{Santiago}}}, \bibinfo {author} {\bibfnamefont
  {V.}~\bibnamefont {{Scarpine}}}, \bibinfo {author} {\bibfnamefont
  {R.}~\bibnamefont {{Schindler}}}, \bibinfo {author} {\bibfnamefont
  {D.}~\bibnamefont {{Scolnic}}}, \bibinfo {author} {\bibfnamefont {L.~F.}\
  \bibnamefont {{Secco}}}, \bibinfo {author} {\bibfnamefont {S.}~\bibnamefont
  {{Serrano}}}, \bibinfo {author} {\bibfnamefont {I.}~\bibnamefont
  {{Sevilla-Noarbe}}}, \bibinfo {author} {\bibfnamefont {E.}~\bibnamefont
  {{Sheldon}}}, \bibinfo {author} {\bibfnamefont {R.~C.}\ \bibnamefont
  {{Smith}}}, \bibinfo {author} {\bibfnamefont {M.}~\bibnamefont {{Smith}}},
  \bibinfo {author} {\bibfnamefont {J.}~\bibnamefont {{Smith}}}, \bibinfo
  {author} {\bibfnamefont {M.}~\bibnamefont {{Soares-Santos}}}, \bibinfo
  {author} {\bibfnamefont {F.}~\bibnamefont {{Sobreira}}}, \bibinfo {author}
  {\bibfnamefont {E.}~\bibnamefont {{Suchyta}}}, \bibinfo {author}
  {\bibfnamefont {G.}~\bibnamefont {{Tarle}}}, \bibinfo {author} {\bibfnamefont
  {D.}~\bibnamefont {{Thomas}}}, \bibinfo {author} {\bibfnamefont {M.~A.}\
  \bibnamefont {{Troxel}}}, \bibinfo {author} {\bibfnamefont {D.~L.}\
  \bibnamefont {{Tucker}}}, \bibinfo {author} {\bibfnamefont {B.~E.}\
  \bibnamefont {{Tucker}}}, \bibinfo {author} {\bibfnamefont {S.~A.}\
  \bibnamefont {{Uddin}}}, \bibinfo {author} {\bibfnamefont {T.~N.}\
  \bibnamefont {{Varga}}}, \bibinfo {author} {\bibfnamefont {P.}~\bibnamefont
  {{Vielzeuf}}}, \bibinfo {author} {\bibfnamefont {V.}~\bibnamefont
  {{Vikram}}}, \bibinfo {author} {\bibfnamefont {A.~K.}\ \bibnamefont
  {{Vivas}}}, \bibinfo {author} {\bibfnamefont {A.~R.}\ \bibnamefont
  {{Walker}}}, \bibinfo {author} {\bibfnamefont {M.}~\bibnamefont {{Wang}}},
  \bibinfo {author} {\bibfnamefont {R.~H.}\ \bibnamefont {{Wechsler}}},
  \bibinfo {author} {\bibfnamefont {J.}~\bibnamefont {{Weller}}}, \bibinfo
  {author} {\bibfnamefont {W.}~\bibnamefont {{Wester}}}, \bibinfo {author}
  {\bibfnamefont {R.~C.}\ \bibnamefont {{Wolf}}}, \bibinfo {author}
  {\bibfnamefont {B.}~\bibnamefont {{Yanny}}}, \bibinfo {author} {\bibfnamefont
  {F.}~\bibnamefont {{Yuan}}}, \bibinfo {author} {\bibfnamefont
  {A.}~\bibnamefont {{Zenteno}}}, \bibinfo {author} {\bibfnamefont
  {B.}~\bibnamefont {{Zhang}}}, \bibinfo {author} {\bibfnamefont
  {Y.}~\bibnamefont {{Zhang}}}, \bibinfo {author} {\bibfnamefont
  {J.}~\bibnamefont {{Zuntz}}},\ and\ \bibinfo {author} {\bibnamefont {{Dark
  Energy Survey Collaboration}}},\ }\bibfield  {title} {\bibinfo {title} {{Dark
  Energy Survey year 1 results: Cosmological constraints from galaxy clustering
  and weak lensing}},\ }\href {https://doi.org/10.1103/PhysRevD.98.043526}
  {\bibfield  {journal} {\bibinfo  {journal} {\prd}\ }\textbf {\bibinfo
  {volume} {98}},\ \bibinfo {eid} {043526} (\bibinfo {year} {2018})},\ \Eprint
  {https://arxiv.org/abs/1708.01530} {arXiv:1708.01530 [astro-ph.CO]}
  \BibitemShut {NoStop}%
\bibitem [{\citenamefont {{Blandford}}\ and\ \citenamefont
  {{Narayan}}(1992)}]{BN1992}%
  \BibitemOpen
  \bibfield  {author} {\bibinfo {author} {\bibfnamefont {R.~D.}\ \bibnamefont
  {{Blandford}}}\ and\ \bibinfo {author} {\bibfnamefont {R.}~\bibnamefont
  {{Narayan}}},\ }\bibfield  {title} {\bibinfo {title} {{Cosmological
  applications of gravitational lensing.}},\ }\href
  {https://doi.org/10.1146/annurev.astro.30.1.311} {\bibfield  {journal}
  {\bibinfo  {journal} {\araa}\ }\textbf {\bibinfo {volume} {30}},\ \bibinfo
  {pages} {311} (\bibinfo {year} {1992})}\BibitemShut {NoStop}%
\bibitem [{\citenamefont {{Bartelmann}}\ and\ \citenamefont
  {{Narayan}}(1995)}]{BN1995}%
  \BibitemOpen
  \bibfield  {author} {\bibinfo {author} {\bibfnamefont {M.}~\bibnamefont
  {{Bartelmann}}}\ and\ \bibinfo {author} {\bibfnamefont {R.}~\bibnamefont
  {{Narayan}}},\ }\bibfield  {title} {\bibinfo {title} {{Gravitational lensing
  and the mass distribution of clusters}},\ }in\ \href
  {https://doi.org/10.1063/1.48350} {\emph {\bibinfo {booktitle} {Dark
  Matter}}},\ \bibinfo {series} {American Institute of Physics Conference
  Series}, Vol.\ \bibinfo {volume} {336},\ \bibinfo {editor} {edited by\
  \bibinfo {editor} {\bibfnamefont {S.~S.}\ \bibnamefont {{Holt}}}\ and\
  \bibinfo {editor} {\bibfnamefont {C.~L.}\ \bibnamefont {{Bennett}}}}\
  (\bibinfo {year} {1995})\ pp.\ \bibinfo {pages} {307--319},\ \Eprint
  {https://arxiv.org/abs/astro-ph/9411033} {arXiv:astro-ph/9411033 [astro-ph]}
  \BibitemShut {NoStop}%
\bibitem [{\citenamefont {{Menard}}\ and\ \citenamefont
  {{Bartelmann}}(2002)}]{Menard2002}%
  \BibitemOpen
  \bibfield  {author} {\bibinfo {author} {\bibfnamefont {B.}~\bibnamefont
  {{Menard}}}\ and\ \bibinfo {author} {\bibfnamefont {M.}~\bibnamefont
  {{Bartelmann}}},\ }\bibfield  {title} {\bibinfo {title} {{Cosmological
  information from quasar-galaxy correlations induced by weak lensing}},\
  }\href {https://doi.org/10.1051/0004-6361:20020274} {\bibfield  {journal}
  {\bibinfo  {journal} {\aap}\ }\textbf {\bibinfo {volume} {386}},\ \bibinfo
  {eid} {784} (\bibinfo {year} {2002})},\ \Eprint
  {https://arxiv.org/abs/0203163} {astro-ph:0203163 [astro-ph.CO]} \BibitemShut
  {NoStop}%
\bibitem [{\citenamefont {{Menard}}\ \emph {et~al.}(2003)\citenamefont
  {{Menard}}, \citenamefont {{Bartelmann}},\ and\ \citenamefont
  {{Mellier}}}]{Menard2003}%
  \BibitemOpen
  \bibfield  {author} {\bibinfo {author} {\bibfnamefont {B.}~\bibnamefont
  {{Menard}}}, \bibinfo {author} {\bibfnamefont {M.}~\bibnamefont
  {{Bartelmann}}},\ and\ \bibinfo {author} {\bibfnamefont {Y.}~\bibnamefont
  {{Mellier}}},\ }\bibfield  {title} {\bibinfo {title} {{Measuring Omega0 with
  higher-order quasar-galaxy correlations induced by weak lensing}},\ }\href
  {https://doi.org/10.1051/0004-6361:20031095} {\bibfield  {journal} {\bibinfo
  {journal} {\aap}\ }\textbf {\bibinfo {volume} {409}},\ \bibinfo {eid} {411}
  (\bibinfo {year} {2003})}\BibitemShut {NoStop}%
\bibitem [{\citenamefont {{Menard}}\ \emph {et~al.}(2010)\citenamefont
  {{Menard}}, \citenamefont {{Scranton}}, \citenamefont {{Fukugita}},\ and\
  \citenamefont {{Richards}}}]{Menard2010}%
  \BibitemOpen
  \bibfield  {author} {\bibinfo {author} {\bibfnamefont {B.}~\bibnamefont
  {{Menard}}}, \bibinfo {author} {\bibfnamefont {R.}~\bibnamefont
  {{Scranton}}}, \bibinfo {author} {\bibfnamefont {M.}~\bibnamefont
  {{Fukugita}}},\ and\ \bibinfo {author} {\bibfnamefont {G.}~\bibnamefont
  {{Richards}}},\ }\bibfield  {title} {\bibinfo {title} {{Measuring the
  galaxy-mass and galaxy-dust correlations through magnification and
  reddeing}},\ }\href {https://doi.org/10.1111/j.1365-2966.2010.16486.x}
  {\bibfield  {journal} {\bibinfo  {journal} {\mnras}\ }\textbf {\bibinfo
  {volume} {405}},\ \bibinfo {eid} {1025} (\bibinfo {year} {2010})},\ \Eprint
  {https://arxiv.org/abs/0902.4240} {arXiv:0902.4240 [astro-ph.CO]}
  \BibitemShut {NoStop}%
\bibitem [{\citenamefont {{van Waerbeke}}(2010)}]{van2010b}%
  \BibitemOpen
  \bibfield  {author} {\bibinfo {author} {\bibfnamefont {L.}~\bibnamefont {{van
  Waerbeke}}},\ }\bibfield  {title} {\bibinfo {title} {{Shear and
  magnification: cosmic complementarity}},\ }\href
  {https://doi.org/10.1111/j.1365-2966.2009.15809.x} {\bibfield  {journal}
  {\bibinfo  {journal} {\mnras}\ }\textbf {\bibinfo {volume} {401}},\ \bibinfo
  {pages} {2093} (\bibinfo {year} {2010})},\ \Eprint
  {https://arxiv.org/abs/0906.1583} {arXiv:0906.1583 [astro-ph.CO]}
  \BibitemShut {NoStop}%
\bibitem [{\citenamefont {{Yang}}\ \emph {et~al.}(2017)\citenamefont {{Yang}},
  \citenamefont {{Zhang}}, \citenamefont {{Yu}},\ and\ \citenamefont
  {{Zhang}}}]{Yangxj2017}%
  \BibitemOpen
  \bibfield  {author} {\bibinfo {author} {\bibfnamefont {X.}~\bibnamefont
  {{Yang}}}, \bibinfo {author} {\bibfnamefont {J.}~\bibnamefont {{Zhang}}},
  \bibinfo {author} {\bibfnamefont {Y.}~\bibnamefont {{Yu}}},\ and\ \bibinfo
  {author} {\bibfnamefont {P.}~\bibnamefont {{Zhang}}},\ }\bibfield  {title}
  {\bibinfo {title} {{Weak-lensing Power Spectrum Reconstruction by Counting
  Galaxies. I. The ABS Method}},\ }\href
  {https://doi.org/10.3847/1538-4357/aa7ed4} {\bibfield  {journal} {\bibinfo
  {journal} {\apj}\ }\textbf {\bibinfo {volume} {845}},\ \bibinfo {eid} {174}
  (\bibinfo {year} {2017})},\ \Eprint {https://arxiv.org/abs/1703.01575}
  {arXiv:1703.01575 [astro-ph.CO]} \BibitemShut {NoStop}%
\bibitem [{\citenamefont {{Zhang}}\ \emph {et~al.}(2018)\citenamefont
  {{Zhang}}, \citenamefont {{Yang}}, \citenamefont {{Zhang}},\ and\
  \citenamefont {{Yu}}}]{Zhangpj2018}%
  \BibitemOpen
  \bibfield  {author} {\bibinfo {author} {\bibfnamefont {P.}~\bibnamefont
  {{Zhang}}}, \bibinfo {author} {\bibfnamefont {X.}~\bibnamefont {{Yang}}},
  \bibinfo {author} {\bibfnamefont {J.}~\bibnamefont {{Zhang}}},\ and\ \bibinfo
  {author} {\bibfnamefont {Y.}~\bibnamefont {{Yu}}},\ }\bibfield  {title}
  {\bibinfo {title} {{Weak-lensing Power Spectrum Reconstruction by Counting
  Galaxies. II. Improving the ABS Method with the Shift Parameter}},\ }\href
  {https://doi.org/10.3847/1538-4357/aad0f1} {\bibfield  {journal} {\bibinfo
  {journal} {\apj}\ }\textbf {\bibinfo {volume} {864}},\ \bibinfo {eid} {10}
  (\bibinfo {year} {2018})},\ \Eprint {https://arxiv.org/abs/1807.00443}
  {arXiv:1807.00443 [astro-ph.CO]} \BibitemShut {NoStop}%
\bibitem [{\citenamefont {{Morrison}}\ \emph {et~al.}(2012)\citenamefont
  {{Morrison}}, \citenamefont {{Scranton}}, \citenamefont {{M{\'e}nard}},
  \citenamefont {{Schmidt}}, \citenamefont {{Tyson}}, \citenamefont {{Ryan}},
  \citenamefont {{Choi}},\ and\ \citenamefont {{Wittman}}}]{MS2012}%
  \BibitemOpen
  \bibfield  {author} {\bibinfo {author} {\bibfnamefont {C.~B.}\ \bibnamefont
  {{Morrison}}}, \bibinfo {author} {\bibfnamefont {R.}~\bibnamefont
  {{Scranton}}}, \bibinfo {author} {\bibfnamefont {B.}~\bibnamefont
  {{M{\'e}nard}}}, \bibinfo {author} {\bibfnamefont {S.~J.}\ \bibnamefont
  {{Schmidt}}}, \bibinfo {author} {\bibfnamefont {J.~A.}\ \bibnamefont
  {{Tyson}}}, \bibinfo {author} {\bibfnamefont {R.}~\bibnamefont {{Ryan}}},
  \bibinfo {author} {\bibfnamefont {A.}~\bibnamefont {{Choi}}},\ and\ \bibinfo
  {author} {\bibfnamefont {D.~M.}\ \bibnamefont {{Wittman}}},\ }\bibfield
  {title} {\bibinfo {title} {{Tomographic magnification of Lyman-break galaxies
  in the Deep Lens Survey}},\ }\href
  {https://doi.org/10.1111/j.1365-2966.2012.21826.x} {\bibfield  {journal}
  {\bibinfo  {journal} {\mnras}\ }\textbf {\bibinfo {volume} {426}},\ \bibinfo
  {pages} {2489} (\bibinfo {year} {2012})},\ \Eprint
  {https://arxiv.org/abs/1204.2830} {arXiv:1204.2830 [astro-ph.CO]}
  \BibitemShut {NoStop}%
\bibitem [{\citenamefont {{Ford}}\ \emph {et~al.}(2014)\citenamefont {{Ford}},
  \citenamefont {{Hildebrandt}}, \citenamefont {{Van Waerbeke}}, \citenamefont
  {{Erben}}, \citenamefont {{Laigle}}, \citenamefont {{Milkeraitis}},\ and\
  \citenamefont {{Morrison}}}]{FH2014}%
  \BibitemOpen
  \bibfield  {author} {\bibinfo {author} {\bibfnamefont {J.}~\bibnamefont
  {{Ford}}}, \bibinfo {author} {\bibfnamefont {H.}~\bibnamefont
  {{Hildebrandt}}}, \bibinfo {author} {\bibfnamefont {L.}~\bibnamefont {{Van
  Waerbeke}}}, \bibinfo {author} {\bibfnamefont {T.}~\bibnamefont {{Erben}}},
  \bibinfo {author} {\bibfnamefont {C.}~\bibnamefont {{Laigle}}}, \bibinfo
  {author} {\bibfnamefont {M.}~\bibnamefont {{Milkeraitis}}},\ and\ \bibinfo
  {author} {\bibfnamefont {C.~B.}\ \bibnamefont {{Morrison}}},\ }\bibfield
  {title} {\bibinfo {title} {{Cluster magnification and the mass-richness
  relation in CFHTLenS}},\ }\href {https://doi.org/10.1093/mnras/stu225}
  {\bibfield  {journal} {\bibinfo  {journal} {\mnras}\ }\textbf {\bibinfo
  {volume} {439}},\ \bibinfo {pages} {3755} (\bibinfo {year} {2014})},\ \Eprint
  {https://arxiv.org/abs/1310.2295} {arXiv:1310.2295 [astro-ph.CO]}
  \BibitemShut {NoStop}%
\bibitem [{\citenamefont {{Garcia-Fernandez}}\ \emph
  {et~al.}(2018)\citenamefont {{Garcia-Fernandez}}, \citenamefont {{Sanchez}},
  \citenamefont {{Sevilla-Noarbe}}, \citenamefont {{Suchyta}}, \citenamefont
  {{Huff}}, \citenamefont {{Gaztanaga}}, \citenamefont {{Aleksi{\'c}}},
  \citenamefont {{}}, \citenamefont {{Ponce}}, \citenamefont {{Castander}},
  \citenamefont {{Hoyle}}, \citenamefont {{Abbott}}, \citenamefont {{Abdalla}},
  \citenamefont {{Allam}}, \citenamefont {{Annis}}, \citenamefont
  {{Benoit-L{\'e}vy}}, \citenamefont {{Bernstein}}, \citenamefont {{Bertin}},
  \citenamefont {{Brooks}}, \citenamefont {{Buckley-Geer}}, \citenamefont
  {{Burke}}, \citenamefont {{Carnero Rosell}}, \citenamefont {{Carrasco Kind}},
  \citenamefont {{Carretero}}, \citenamefont {{Crocce}}, \citenamefont
  {{Cunha}}, \citenamefont {{D'Andrea}}, \citenamefont {{da Costa}},
  \citenamefont {{DePoy}}, \citenamefont {{Desai}}, \citenamefont {{Diehl}},
  \citenamefont {{Eifler}}, \citenamefont {{Evrard}}, \citenamefont
  {{Fernandez}}, \citenamefont {{Flaugher}}, \citenamefont {{Fosalba}},
  \citenamefont {{Frieman}}, \citenamefont {{Garc{\'\i}a-Bellido}},
  \citenamefont {{Gerdes}}, \citenamefont {{Giannantonio}}, \citenamefont
  {{Gruen}}, \citenamefont {{Gruendl}}, \citenamefont {{Gschwend}},
  \citenamefont {{Gutierrez}}, \citenamefont {{James}}, \citenamefont
  {{Jarvis}}, \citenamefont {{Kirk}}, \citenamefont {{Krause}}, \citenamefont
  {{Kuehn}}, \citenamefont {{Kuropatkin}}, \citenamefont {{Lahav}},
  \citenamefont {{Lima}}, \citenamefont {{MacCrann}}, \citenamefont {{Maia}},
  \citenamefont {{March}}, \citenamefont {{Marshall}}, \citenamefont
  {{Melchior}}, \citenamefont {{Miquel}}, \citenamefont {{Mohr}}, \citenamefont
  {{Plazas}}, \citenamefont {{Romer}}, \citenamefont {{Roodman}}, \citenamefont
  {{Rykoff}}, \citenamefont {{Scarpine}}, \citenamefont {{Schubnell}},
  \citenamefont {{Smith}}, \citenamefont {{Soares-Santos}}, \citenamefont
  {{Sobreira}}, \citenamefont {{Tarle}}, \citenamefont {{Thomas}},
  \citenamefont {{Walker}}, \citenamefont {{Wester}},\ and\ \citenamefont {{DES
  Collaboration}}}]{GS2018}%
  \BibitemOpen
  \bibfield  {author} {\bibinfo {author} {\bibfnamefont {M.}~\bibnamefont
  {{Garcia-Fernandez}}}, \bibinfo {author} {\bibfnamefont {E.}~\bibnamefont
  {{Sanchez}}}, \bibinfo {author} {\bibfnamefont {I.}~\bibnamefont
  {{Sevilla-Noarbe}}}, \bibinfo {author} {\bibfnamefont {E.}~\bibnamefont
  {{Suchyta}}}, \bibinfo {author} {\bibfnamefont {E.~M.}\ \bibnamefont
  {{Huff}}}, \bibinfo {author} {\bibfnamefont {E.}~\bibnamefont {{Gaztanaga}}},
  \bibinfo {author} {\bibnamefont {{Aleksi{\'c}}}}, \bibinfo {author}
  {\bibfnamefont {J.}~\bibnamefont {{}}}, \bibinfo {author} {\bibfnamefont
  {R.}~\bibnamefont {{Ponce}}}, \bibinfo {author} {\bibfnamefont {F.~J.}\
  \bibnamefont {{Castander}}}, \bibinfo {author} {\bibfnamefont
  {B.}~\bibnamefont {{Hoyle}}}, \bibinfo {author} {\bibfnamefont {T.~M.~C.}\
  \bibnamefont {{Abbott}}}, \bibinfo {author} {\bibfnamefont {F.~B.}\
  \bibnamefont {{Abdalla}}}, \bibinfo {author} {\bibfnamefont {S.}~\bibnamefont
  {{Allam}}}, \bibinfo {author} {\bibfnamefont {J.}~\bibnamefont {{Annis}}},
  \bibinfo {author} {\bibfnamefont {A.}~\bibnamefont {{Benoit-L{\'e}vy}}},
  \bibinfo {author} {\bibfnamefont {G.~M.}\ \bibnamefont {{Bernstein}}},
  \bibinfo {author} {\bibfnamefont {E.}~\bibnamefont {{Bertin}}}, \bibinfo
  {author} {\bibfnamefont {D.}~\bibnamefont {{Brooks}}}, \bibinfo {author}
  {\bibfnamefont {E.}~\bibnamefont {{Buckley-Geer}}}, \bibinfo {author}
  {\bibfnamefont {D.~L.}\ \bibnamefont {{Burke}}}, \bibinfo {author}
  {\bibfnamefont {A.}~\bibnamefont {{Carnero Rosell}}}, \bibinfo {author}
  {\bibfnamefont {M.}~\bibnamefont {{Carrasco Kind}}}, \bibinfo {author}
  {\bibfnamefont {J.}~\bibnamefont {{Carretero}}}, \bibinfo {author}
  {\bibfnamefont {M.}~\bibnamefont {{Crocce}}}, \bibinfo {author}
  {\bibfnamefont {C.~E.}\ \bibnamefont {{Cunha}}}, \bibinfo {author}
  {\bibfnamefont {C.~B.}\ \bibnamefont {{D'Andrea}}}, \bibinfo {author}
  {\bibfnamefont {L.~N.}\ \bibnamefont {{da Costa}}}, \bibinfo {author}
  {\bibfnamefont {D.~L.}\ \bibnamefont {{DePoy}}}, \bibinfo {author}
  {\bibfnamefont {S.}~\bibnamefont {{Desai}}}, \bibinfo {author} {\bibfnamefont
  {H.~T.}\ \bibnamefont {{Diehl}}}, \bibinfo {author} {\bibfnamefont {T.~F.}\
  \bibnamefont {{Eifler}}}, \bibinfo {author} {\bibfnamefont {A.~E.}\
  \bibnamefont {{Evrard}}}, \bibinfo {author} {\bibfnamefont {E.}~\bibnamefont
  {{Fernandez}}}, \bibinfo {author} {\bibfnamefont {B.}~\bibnamefont
  {{Flaugher}}}, \bibinfo {author} {\bibfnamefont {P.}~\bibnamefont
  {{Fosalba}}}, \bibinfo {author} {\bibfnamefont {J.}~\bibnamefont
  {{Frieman}}}, \bibinfo {author} {\bibfnamefont {J.}~\bibnamefont
  {{Garc{\'\i}a-Bellido}}}, \bibinfo {author} {\bibfnamefont {D.~W.}\
  \bibnamefont {{Gerdes}}}, \bibinfo {author} {\bibfnamefont {T.}~\bibnamefont
  {{Giannantonio}}}, \bibinfo {author} {\bibfnamefont {D.}~\bibnamefont
  {{Gruen}}}, \bibinfo {author} {\bibfnamefont {R.~A.}\ \bibnamefont
  {{Gruendl}}}, \bibinfo {author} {\bibfnamefont {J.}~\bibnamefont
  {{Gschwend}}}, \bibinfo {author} {\bibfnamefont {G.}~\bibnamefont
  {{Gutierrez}}}, \bibinfo {author} {\bibfnamefont {D.~J.}\ \bibnamefont
  {{James}}}, \bibinfo {author} {\bibfnamefont {M.}~\bibnamefont {{Jarvis}}},
  \bibinfo {author} {\bibfnamefont {D.}~\bibnamefont {{Kirk}}}, \bibinfo
  {author} {\bibfnamefont {E.}~\bibnamefont {{Krause}}}, \bibinfo {author}
  {\bibfnamefont {K.}~\bibnamefont {{Kuehn}}}, \bibinfo {author} {\bibfnamefont
  {N.}~\bibnamefont {{Kuropatkin}}}, \bibinfo {author} {\bibfnamefont
  {O.}~\bibnamefont {{Lahav}}}, \bibinfo {author} {\bibfnamefont
  {M.}~\bibnamefont {{Lima}}}, \bibinfo {author} {\bibfnamefont
  {N.}~\bibnamefont {{MacCrann}}}, \bibinfo {author} {\bibfnamefont {M.~A.~G.}\
  \bibnamefont {{Maia}}}, \bibinfo {author} {\bibfnamefont {M.}~\bibnamefont
  {{March}}}, \bibinfo {author} {\bibfnamefont {J.~L.}\ \bibnamefont
  {{Marshall}}}, \bibinfo {author} {\bibfnamefont {P.}~\bibnamefont
  {{Melchior}}}, \bibinfo {author} {\bibfnamefont {R.}~\bibnamefont
  {{Miquel}}}, \bibinfo {author} {\bibfnamefont {J.~J.}\ \bibnamefont
  {{Mohr}}}, \bibinfo {author} {\bibfnamefont {A.~A.}\ \bibnamefont
  {{Plazas}}}, \bibinfo {author} {\bibfnamefont {A.~K.}\ \bibnamefont
  {{Romer}}}, \bibinfo {author} {\bibfnamefont {A.}~\bibnamefont {{Roodman}}},
  \bibinfo {author} {\bibfnamefont {E.~S.}\ \bibnamefont {{Rykoff}}}, \bibinfo
  {author} {\bibfnamefont {V.}~\bibnamefont {{Scarpine}}}, \bibinfo {author}
  {\bibfnamefont {M.}~\bibnamefont {{Schubnell}}}, \bibinfo {author}
  {\bibfnamefont {R.~C.}\ \bibnamefont {{Smith}}}, \bibinfo {author}
  {\bibfnamefont {M.}~\bibnamefont {{Soares-Santos}}}, \bibinfo {author}
  {\bibfnamefont {F.}~\bibnamefont {{Sobreira}}}, \bibinfo {author}
  {\bibfnamefont {G.}~\bibnamefont {{Tarle}}}, \bibinfo {author} {\bibfnamefont
  {D.}~\bibnamefont {{Thomas}}}, \bibinfo {author} {\bibfnamefont {A.~R.}\
  \bibnamefont {{Walker}}}, \bibinfo {author} {\bibfnamefont {W.}~\bibnamefont
  {{Wester}}},\ and\ \bibinfo {author} {\bibnamefont {{DES Collaboration}}},\
  }\bibfield  {title} {\bibinfo {title} {{Weak lensing magnification in the
  Dark Energy Survey Science Verification data}},\ }\href
  {https://doi.org/10.1093/mnras/sty282} {\bibfield  {journal} {\bibinfo
  {journal} {\mnras}\ }\textbf {\bibinfo {volume} {476}},\ \bibinfo {pages}
  {1071} (\bibinfo {year} {2018})}\BibitemShut {NoStop}%
\bibitem [{\citenamefont {{Umetsu}}\ \emph {et~al.}(2014)\citenamefont
  {{Umetsu}}, \citenamefont {{Medezinski}}, \citenamefont {{Nonino}},
  \citenamefont {{Merten}}, \citenamefont {{Postman}}, \citenamefont
  {{Meneghetti}}, \citenamefont {{Donahue}}, \citenamefont {{Czakon}},
  \citenamefont {{Molino}}, \citenamefont {{Seitz}}, \citenamefont {{Gruen}},
  \citenamefont {{Lemze}}, \citenamefont {{Balestra}}, \citenamefont
  {{Ben{\'\i}tez}}, \citenamefont {{Biviano}}, \citenamefont {{Broadhurst}},
  \citenamefont {{Ford}}, \citenamefont {{Grillo}}, \citenamefont
  {{Koekemoer}}, \citenamefont {{Melchior}}, \citenamefont {{Mercurio}},
  \citenamefont {{Moustakas}}, \citenamefont {{Rosati}},\ and\ \citenamefont
  {{Zitrin}}}]{Umetsu2014}%
  \BibitemOpen
  \bibfield  {author} {\bibinfo {author} {\bibfnamefont {K.}~\bibnamefont
  {{Umetsu}}}, \bibinfo {author} {\bibfnamefont {E.}~\bibnamefont
  {{Medezinski}}}, \bibinfo {author} {\bibfnamefont {M.}~\bibnamefont
  {{Nonino}}}, \bibinfo {author} {\bibfnamefont {J.}~\bibnamefont {{Merten}}},
  \bibinfo {author} {\bibfnamefont {M.}~\bibnamefont {{Postman}}}, \bibinfo
  {author} {\bibfnamefont {M.}~\bibnamefont {{Meneghetti}}}, \bibinfo {author}
  {\bibfnamefont {M.}~\bibnamefont {{Donahue}}}, \bibinfo {author}
  {\bibfnamefont {N.}~\bibnamefont {{Czakon}}}, \bibinfo {author}
  {\bibfnamefont {A.}~\bibnamefont {{Molino}}}, \bibinfo {author}
  {\bibfnamefont {S.}~\bibnamefont {{Seitz}}}, \bibinfo {author} {\bibfnamefont
  {D.}~\bibnamefont {{Gruen}}}, \bibinfo {author} {\bibfnamefont
  {D.}~\bibnamefont {{Lemze}}}, \bibinfo {author} {\bibfnamefont
  {I.}~\bibnamefont {{Balestra}}}, \bibinfo {author} {\bibfnamefont
  {N.}~\bibnamefont {{Ben{\'\i}tez}}}, \bibinfo {author} {\bibfnamefont
  {A.}~\bibnamefont {{Biviano}}}, \bibinfo {author} {\bibfnamefont
  {T.}~\bibnamefont {{Broadhurst}}}, \bibinfo {author} {\bibfnamefont
  {H.}~\bibnamefont {{Ford}}}, \bibinfo {author} {\bibfnamefont
  {C.}~\bibnamefont {{Grillo}}}, \bibinfo {author} {\bibfnamefont
  {A.}~\bibnamefont {{Koekemoer}}}, \bibinfo {author} {\bibfnamefont
  {P.}~\bibnamefont {{Melchior}}}, \bibinfo {author} {\bibfnamefont
  {A.}~\bibnamefont {{Mercurio}}}, \bibinfo {author} {\bibfnamefont
  {J.}~\bibnamefont {{Moustakas}}}, \bibinfo {author} {\bibfnamefont
  {P.}~\bibnamefont {{Rosati}}},\ and\ \bibinfo {author} {\bibfnamefont
  {A.}~\bibnamefont {{Zitrin}}},\ }\bibfield  {title} {\bibinfo {title}
  {{CLASH: Weak-lensing Shear-and-magnification Analysis of 20 Galaxy
  Clusters}},\ }\href {https://doi.org/10.1088/0004-637X/795/2/163} {\bibfield
  {journal} {\bibinfo  {journal} {\apj}\ }\textbf {\bibinfo {volume} {795}},\
  \bibinfo {eid} {163} (\bibinfo {year} {2014})},\ \Eprint
  {https://arxiv.org/abs/1404.1375} {arXiv:1404.1375 [astro-ph.CO]}
  \BibitemShut {NoStop}%
\bibitem [{\citenamefont {{Aihara}}\ \emph
  {et~al.}(2018{\natexlab{b}})\citenamefont {{Aihara}}, \citenamefont
  {{Armstrong}}, \citenamefont {{Bickerton}}, \citenamefont {{Bosch}},
  \citenamefont {{Coupon}}, \citenamefont {{Furusawa}}, \citenamefont
  {{Hayashi}}, \citenamefont {{Ikeda}}, \citenamefont {{Kamata}}, \citenamefont
  {{Karoji}}, \citenamefont {{Kawanomoto}}, \citenamefont {{Koike}},
  \citenamefont {{Komiyama}}, \citenamefont {{Lang}}, \citenamefont {{Lupton}},
  \citenamefont {{Mineo}}, \citenamefont {{Miyatake}}, \citenamefont
  {{Miyazaki}}, \citenamefont {{Morokuma}}, \citenamefont {{Obuchi}},
  \citenamefont {{Oishi}}, \citenamefont {{Okura}}, \citenamefont {{Price}},
  \citenamefont {{Takata}}, \citenamefont {{Tanaka}}, \citenamefont {{Tanaka}},
  \citenamefont {{Tanaka}}, \citenamefont {{Uchida}}, \citenamefont
  {{Uraguchi}}, \citenamefont {{Utsumi}}, \citenamefont {{Wang}}, \citenamefont
  {{Yamada}}, \citenamefont {{Yamanoi}}, \citenamefont {{Yasuda}},
  \citenamefont {{Arimoto}}, \citenamefont {{Chiba}}, \citenamefont {{Finet}},
  \citenamefont {{Fujimori}}, \citenamefont {{Fujimoto}}, \citenamefont
  {{Furusawa}}, \citenamefont {{Goto}}, \citenamefont {{Goulding}},
  \citenamefont {{Gunn}}, \citenamefont {{Harikane}}, \citenamefont
  {{Hattori}}, \citenamefont {{Hayashi}}, \citenamefont {{He{\l}miniak}},
  \citenamefont {{Higuchi}}, \citenamefont {{Hikage}}, \citenamefont {{Ho}},
  \citenamefont {{Hsieh}}, \citenamefont {{Huang}}, \citenamefont {{Huang}},
  \citenamefont {{Imanishi}}, \citenamefont {{Iwata}}, \citenamefont
  {{Jaelani}}, \citenamefont {{Jian}}, \citenamefont {{Kashikawa}},
  \citenamefont {{Katayama}}, \citenamefont {{Kojima}}, \citenamefont
  {{Konno}}, \citenamefont {{Koshida}}, \citenamefont {{Kusakabe}},
  \citenamefont {{Leauthaud}}, \citenamefont {{Lee}}, \citenamefont {{Lin}},
  \citenamefont {{Lin}}, \citenamefont {{Mandelbaum}}, \citenamefont
  {{Matsuoka}}, \citenamefont {{Medezinski}}, \citenamefont {{Miyama}},
  \citenamefont {{Momose}}, \citenamefont {{More}}, \citenamefont {{More}},
  \citenamefont {{Mukae}}, \citenamefont {{Murata}}, \citenamefont
  {{Murayama}}, \citenamefont {{Nagao}}, \citenamefont {{Nakata}},
  \citenamefont {{Niida}}, \citenamefont {{Niikura}}, \citenamefont
  {{Nishizawa}}, \citenamefont {{Oguri}}, \citenamefont {{Okabe}},
  \citenamefont {{Ono}}, \citenamefont {{Onodera}}, \citenamefont {{Onoue}},
  \citenamefont {{Ouchi}}, \citenamefont {{Pyo}}, \citenamefont {{Shibuya}},
  \citenamefont {{Shimasaku}}, \citenamefont {{Simet}}, \citenamefont
  {{Speagle}}, \citenamefont {{Spergel}}, \citenamefont {{Strauss}},
  \citenamefont {{Sugahara}}, \citenamefont {{Sugiyama}}, \citenamefont
  {{Suto}}, \citenamefont {{Suzuki}}, \citenamefont {{Tait}}, \citenamefont
  {{Takada}}, \citenamefont {{Terai}}, \citenamefont {{Toba}}, \citenamefont
  {{Turner}}, \citenamefont {{Uchiyama}}, \citenamefont {{Umetsu}},
  \citenamefont {{Urata}}, \citenamefont {{Usuda}}, \citenamefont {{Yeh}},\
  and\ \citenamefont {{Yuma}}}]{HSCPDR1}%
  \BibitemOpen
  \bibfield  {author} {\bibinfo {author} {\bibfnamefont {H.}~\bibnamefont
  {{Aihara}}}, \bibinfo {author} {\bibfnamefont {R.}~\bibnamefont
  {{Armstrong}}}, \bibinfo {author} {\bibfnamefont {S.}~\bibnamefont
  {{Bickerton}}}, \bibinfo {author} {\bibfnamefont {J.}~\bibnamefont
  {{Bosch}}}, \bibinfo {author} {\bibfnamefont {J.}~\bibnamefont {{Coupon}}},
  \bibinfo {author} {\bibfnamefont {H.}~\bibnamefont {{Furusawa}}}, \bibinfo
  {author} {\bibfnamefont {Y.}~\bibnamefont {{Hayashi}}}, \bibinfo {author}
  {\bibfnamefont {H.}~\bibnamefont {{Ikeda}}}, \bibinfo {author} {\bibfnamefont
  {Y.}~\bibnamefont {{Kamata}}}, \bibinfo {author} {\bibfnamefont
  {H.}~\bibnamefont {{Karoji}}}, \bibinfo {author} {\bibfnamefont
  {S.}~\bibnamefont {{Kawanomoto}}}, \bibinfo {author} {\bibfnamefont
  {M.}~\bibnamefont {{Koike}}}, \bibinfo {author} {\bibfnamefont
  {Y.}~\bibnamefont {{Komiyama}}}, \bibinfo {author} {\bibfnamefont
  {D.}~\bibnamefont {{Lang}}}, \bibinfo {author} {\bibfnamefont {R.~H.}\
  \bibnamefont {{Lupton}}}, \bibinfo {author} {\bibfnamefont {S.}~\bibnamefont
  {{Mineo}}}, \bibinfo {author} {\bibfnamefont {H.}~\bibnamefont {{Miyatake}}},
  \bibinfo {author} {\bibfnamefont {S.}~\bibnamefont {{Miyazaki}}}, \bibinfo
  {author} {\bibfnamefont {T.}~\bibnamefont {{Morokuma}}}, \bibinfo {author}
  {\bibfnamefont {Y.}~\bibnamefont {{Obuchi}}}, \bibinfo {author}
  {\bibfnamefont {Y.}~\bibnamefont {{Oishi}}}, \bibinfo {author} {\bibfnamefont
  {Y.}~\bibnamefont {{Okura}}}, \bibinfo {author} {\bibfnamefont {P.~A.}\
  \bibnamefont {{Price}}}, \bibinfo {author} {\bibfnamefont {T.}~\bibnamefont
  {{Takata}}}, \bibinfo {author} {\bibfnamefont {M.~M.}\ \bibnamefont
  {{Tanaka}}}, \bibinfo {author} {\bibfnamefont {M.}~\bibnamefont {{Tanaka}}},
  \bibinfo {author} {\bibfnamefont {Y.}~\bibnamefont {{Tanaka}}}, \bibinfo
  {author} {\bibfnamefont {T.}~\bibnamefont {{Uchida}}}, \bibinfo {author}
  {\bibfnamefont {F.}~\bibnamefont {{Uraguchi}}}, \bibinfo {author}
  {\bibfnamefont {Y.}~\bibnamefont {{Utsumi}}}, \bibinfo {author}
  {\bibfnamefont {S.-Y.}\ \bibnamefont {{Wang}}}, \bibinfo {author}
  {\bibfnamefont {Y.}~\bibnamefont {{Yamada}}}, \bibinfo {author}
  {\bibfnamefont {H.}~\bibnamefont {{Yamanoi}}}, \bibinfo {author}
  {\bibfnamefont {N.}~\bibnamefont {{Yasuda}}}, \bibinfo {author}
  {\bibfnamefont {N.}~\bibnamefont {{Arimoto}}}, \bibinfo {author}
  {\bibfnamefont {M.}~\bibnamefont {{Chiba}}}, \bibinfo {author} {\bibfnamefont
  {F.}~\bibnamefont {{Finet}}}, \bibinfo {author} {\bibfnamefont
  {H.}~\bibnamefont {{Fujimori}}}, \bibinfo {author} {\bibfnamefont
  {S.}~\bibnamefont {{Fujimoto}}}, \bibinfo {author} {\bibfnamefont
  {J.}~\bibnamefont {{Furusawa}}}, \bibinfo {author} {\bibfnamefont
  {T.}~\bibnamefont {{Goto}}}, \bibinfo {author} {\bibfnamefont
  {A.}~\bibnamefont {{Goulding}}}, \bibinfo {author} {\bibfnamefont {J.~E.}\
  \bibnamefont {{Gunn}}}, \bibinfo {author} {\bibfnamefont {Y.}~\bibnamefont
  {{Harikane}}}, \bibinfo {author} {\bibfnamefont {T.}~\bibnamefont
  {{Hattori}}}, \bibinfo {author} {\bibfnamefont {M.}~\bibnamefont
  {{Hayashi}}}, \bibinfo {author} {\bibfnamefont {K.~G.}\ \bibnamefont
  {{He{\l}miniak}}}, \bibinfo {author} {\bibfnamefont {R.}~\bibnamefont
  {{Higuchi}}}, \bibinfo {author} {\bibfnamefont {C.}~\bibnamefont {{Hikage}}},
  \bibinfo {author} {\bibfnamefont {P.~T.~P.}\ \bibnamefont {{Ho}}}, \bibinfo
  {author} {\bibfnamefont {B.-C.}\ \bibnamefont {{Hsieh}}}, \bibinfo {author}
  {\bibfnamefont {K.}~\bibnamefont {{Huang}}}, \bibinfo {author} {\bibfnamefont
  {S.}~\bibnamefont {{Huang}}}, \bibinfo {author} {\bibfnamefont
  {M.}~\bibnamefont {{Imanishi}}}, \bibinfo {author} {\bibfnamefont
  {I.}~\bibnamefont {{Iwata}}}, \bibinfo {author} {\bibfnamefont {A.~T.}\
  \bibnamefont {{Jaelani}}}, \bibinfo {author} {\bibfnamefont {H.-Y.}\
  \bibnamefont {{Jian}}}, \bibinfo {author} {\bibfnamefont {N.}~\bibnamefont
  {{Kashikawa}}}, \bibinfo {author} {\bibfnamefont {N.}~\bibnamefont
  {{Katayama}}}, \bibinfo {author} {\bibfnamefont {T.}~\bibnamefont
  {{Kojima}}}, \bibinfo {author} {\bibfnamefont {A.}~\bibnamefont {{Konno}}},
  \bibinfo {author} {\bibfnamefont {S.}~\bibnamefont {{Koshida}}}, \bibinfo
  {author} {\bibfnamefont {H.}~\bibnamefont {{Kusakabe}}}, \bibinfo {author}
  {\bibfnamefont {A.}~\bibnamefont {{Leauthaud}}}, \bibinfo {author}
  {\bibfnamefont {C.-H.}\ \bibnamefont {{Lee}}}, \bibinfo {author}
  {\bibfnamefont {L.}~\bibnamefont {{Lin}}}, \bibinfo {author} {\bibfnamefont
  {Y.-T.}\ \bibnamefont {{Lin}}}, \bibinfo {author} {\bibfnamefont
  {R.}~\bibnamefont {{Mandelbaum}}}, \bibinfo {author} {\bibfnamefont
  {Y.}~\bibnamefont {{Matsuoka}}}, \bibinfo {author} {\bibfnamefont
  {E.}~\bibnamefont {{Medezinski}}}, \bibinfo {author} {\bibfnamefont
  {S.}~\bibnamefont {{Miyama}}}, \bibinfo {author} {\bibfnamefont
  {R.}~\bibnamefont {{Momose}}}, \bibinfo {author} {\bibfnamefont
  {A.}~\bibnamefont {{More}}}, \bibinfo {author} {\bibfnamefont
  {S.}~\bibnamefont {{More}}}, \bibinfo {author} {\bibfnamefont
  {S.}~\bibnamefont {{Mukae}}}, \bibinfo {author} {\bibfnamefont
  {R.}~\bibnamefont {{Murata}}}, \bibinfo {author} {\bibfnamefont
  {H.}~\bibnamefont {{Murayama}}}, \bibinfo {author} {\bibfnamefont
  {T.}~\bibnamefont {{Nagao}}}, \bibinfo {author} {\bibfnamefont
  {F.}~\bibnamefont {{Nakata}}}, \bibinfo {author} {\bibfnamefont
  {M.}~\bibnamefont {{Niida}}}, \bibinfo {author} {\bibfnamefont
  {H.}~\bibnamefont {{Niikura}}}, \bibinfo {author} {\bibfnamefont {A.~J.}\
  \bibnamefont {{Nishizawa}}}, \bibinfo {author} {\bibfnamefont
  {M.}~\bibnamefont {{Oguri}}}, \bibinfo {author} {\bibfnamefont
  {N.}~\bibnamefont {{Okabe}}}, \bibinfo {author} {\bibfnamefont
  {Y.}~\bibnamefont {{Ono}}}, \bibinfo {author} {\bibfnamefont
  {M.}~\bibnamefont {{Onodera}}}, \bibinfo {author} {\bibfnamefont
  {M.}~\bibnamefont {{Onoue}}}, \bibinfo {author} {\bibfnamefont
  {M.}~\bibnamefont {{Ouchi}}}, \bibinfo {author} {\bibfnamefont {T.-S.}\
  \bibnamefont {{Pyo}}}, \bibinfo {author} {\bibfnamefont {T.}~\bibnamefont
  {{Shibuya}}}, \bibinfo {author} {\bibfnamefont {K.}~\bibnamefont
  {{Shimasaku}}}, \bibinfo {author} {\bibfnamefont {M.}~\bibnamefont
  {{Simet}}}, \bibinfo {author} {\bibfnamefont {J.}~\bibnamefont {{Speagle}}},
  \bibinfo {author} {\bibfnamefont {D.~N.}\ \bibnamefont {{Spergel}}}, \bibinfo
  {author} {\bibfnamefont {M.~A.}\ \bibnamefont {{Strauss}}}, \bibinfo {author}
  {\bibfnamefont {Y.}~\bibnamefont {{Sugahara}}}, \bibinfo {author}
  {\bibfnamefont {N.}~\bibnamefont {{Sugiyama}}}, \bibinfo {author}
  {\bibfnamefont {Y.}~\bibnamefont {{Suto}}}, \bibinfo {author} {\bibfnamefont
  {N.}~\bibnamefont {{Suzuki}}}, \bibinfo {author} {\bibfnamefont {P.~J.}\
  \bibnamefont {{Tait}}}, \bibinfo {author} {\bibfnamefont {M.}~\bibnamefont
  {{Takada}}}, \bibinfo {author} {\bibfnamefont {T.}~\bibnamefont {{Terai}}},
  \bibinfo {author} {\bibfnamefont {Y.}~\bibnamefont {{Toba}}}, \bibinfo
  {author} {\bibfnamefont {E.~L.}\ \bibnamefont {{Turner}}}, \bibinfo {author}
  {\bibfnamefont {H.}~\bibnamefont {{Uchiyama}}}, \bibinfo {author}
  {\bibfnamefont {K.}~\bibnamefont {{Umetsu}}}, \bibinfo {author}
  {\bibfnamefont {Y.}~\bibnamefont {{Urata}}}, \bibinfo {author} {\bibfnamefont
  {T.}~\bibnamefont {{Usuda}}}, \bibinfo {author} {\bibfnamefont
  {S.}~\bibnamefont {{Yeh}}},\ and\ \bibinfo {author} {\bibfnamefont
  {S.}~\bibnamefont {{Yuma}}},\ }\bibfield  {title} {\bibinfo {title} {{First
  data release of the Hyper Suprime-Cam Subaru Strategic Program}},\ }\href
  {https://doi.org/10.1093/pasj/psx081} {\bibfield  {journal} {\bibinfo
  {journal} {\pasj}\ }\textbf {\bibinfo {volume} {70}},\ \bibinfo {eid} {S8}
  (\bibinfo {year} {2018}{\natexlab{b}})},\ \Eprint
  {https://arxiv.org/abs/1702.08449} {arXiv:1702.08449 [astro-ph.IM]}
  \BibitemShut {NoStop}%
\bibitem [{\citenamefont {{Aihara}}\ \emph {et~al.}(2019)\citenamefont
  {{Aihara}}, \citenamefont {{AlSayyad}}, \citenamefont {{Ando}}, \citenamefont
  {{Armstrong}}, \citenamefont {{Bosch}}, \citenamefont {{Egami}},
  \citenamefont {{Furusawa}}, \citenamefont {{Furusawa}}, \citenamefont
  {{Goulding}}, \citenamefont {{Harikane}}, \citenamefont {{Hikage}},
  \citenamefont {{Ho}}, \citenamefont {{Hsieh}}, \citenamefont {{Huang}},
  \citenamefont {{Ikeda}}, \citenamefont {{Imanishi}}, \citenamefont {{Ito}},
  \citenamefont {{Iwata}}, \citenamefont {{Jaelani}}, \citenamefont {{Kakuma}},
  \citenamefont {{Kawana}}, \citenamefont {{Kikuta}}, \citenamefont
  {{Kobayashi}}, \citenamefont {{Koike}}, \citenamefont {{Komiyama}},
  \citenamefont {{Li}}, \citenamefont {{Liang}}, \citenamefont {{Lin}},
  \citenamefont {{Luo}}, \citenamefont {{Lupton}}, \citenamefont {{Lust}},
  \citenamefont {{MacArthur}}, \citenamefont {{Matsuoka}}, \citenamefont
  {{Mineo}}, \citenamefont {{Miyatake}}, \citenamefont {{Miyazaki}},
  \citenamefont {{More}}, \citenamefont {{Murata}}, \citenamefont {{Namiki}},
  \citenamefont {{Nishizawa}}, \citenamefont {{Oguri}}, \citenamefont
  {{Okabe}}, \citenamefont {{Okamoto}}, \citenamefont {{Okura}}, \citenamefont
  {{Ono}}, \citenamefont {{Onodera}}, \citenamefont {{Onoue}}, \citenamefont
  {{Osato}}, \citenamefont {{Ouchi}}, \citenamefont {{Shibuya}}, \citenamefont
  {{Strauss}}, \citenamefont {{Sugiyama}}, \citenamefont {{Suto}},
  \citenamefont {{Takada}}, \citenamefont {{Takagi}}, \citenamefont {{Takata}},
  \citenamefont {{Takita}}, \citenamefont {{Tanaka}}, \citenamefont {{Terai}},
  \citenamefont {{Toba}}, \citenamefont {{Uchiyama}}, \citenamefont {{Utsumi}},
  \citenamefont {{Wang}}, \citenamefont {{Wang}},\ and\ \citenamefont
  {{Yamada}}}]{HSCPDR2}%
  \BibitemOpen
  \bibfield  {author} {\bibinfo {author} {\bibfnamefont {H.}~\bibnamefont
  {{Aihara}}}, \bibinfo {author} {\bibfnamefont {Y.}~\bibnamefont
  {{AlSayyad}}}, \bibinfo {author} {\bibfnamefont {M.}~\bibnamefont {{Ando}}},
  \bibinfo {author} {\bibfnamefont {R.}~\bibnamefont {{Armstrong}}}, \bibinfo
  {author} {\bibfnamefont {J.}~\bibnamefont {{Bosch}}}, \bibinfo {author}
  {\bibfnamefont {E.}~\bibnamefont {{Egami}}}, \bibinfo {author} {\bibfnamefont
  {H.}~\bibnamefont {{Furusawa}}}, \bibinfo {author} {\bibfnamefont
  {J.}~\bibnamefont {{Furusawa}}}, \bibinfo {author} {\bibfnamefont
  {A.}~\bibnamefont {{Goulding}}}, \bibinfo {author} {\bibfnamefont
  {Y.}~\bibnamefont {{Harikane}}}, \bibinfo {author} {\bibfnamefont
  {C.}~\bibnamefont {{Hikage}}}, \bibinfo {author} {\bibfnamefont {P.~T.~P.}\
  \bibnamefont {{Ho}}}, \bibinfo {author} {\bibfnamefont {B.-C.}\ \bibnamefont
  {{Hsieh}}}, \bibinfo {author} {\bibfnamefont {S.}~\bibnamefont {{Huang}}},
  \bibinfo {author} {\bibfnamefont {H.}~\bibnamefont {{Ikeda}}}, \bibinfo
  {author} {\bibfnamefont {M.}~\bibnamefont {{Imanishi}}}, \bibinfo {author}
  {\bibfnamefont {K.}~\bibnamefont {{Ito}}}, \bibinfo {author} {\bibfnamefont
  {I.}~\bibnamefont {{Iwata}}}, \bibinfo {author} {\bibfnamefont {A.~T.}\
  \bibnamefont {{Jaelani}}}, \bibinfo {author} {\bibfnamefont {R.}~\bibnamefont
  {{Kakuma}}}, \bibinfo {author} {\bibfnamefont {K.}~\bibnamefont {{Kawana}}},
  \bibinfo {author} {\bibfnamefont {S.}~\bibnamefont {{Kikuta}}}, \bibinfo
  {author} {\bibfnamefont {U.}~\bibnamefont {{Kobayashi}}}, \bibinfo {author}
  {\bibfnamefont {M.}~\bibnamefont {{Koike}}}, \bibinfo {author} {\bibfnamefont
  {Y.}~\bibnamefont {{Komiyama}}}, \bibinfo {author} {\bibfnamefont
  {X.}~\bibnamefont {{Li}}}, \bibinfo {author} {\bibfnamefont {Y.}~\bibnamefont
  {{Liang}}}, \bibinfo {author} {\bibfnamefont {Y.-T.}\ \bibnamefont {{Lin}}},
  \bibinfo {author} {\bibfnamefont {W.}~\bibnamefont {{Luo}}}, \bibinfo
  {author} {\bibfnamefont {R.}~\bibnamefont {{Lupton}}}, \bibinfo {author}
  {\bibfnamefont {N.~B.}\ \bibnamefont {{Lust}}}, \bibinfo {author}
  {\bibfnamefont {L.~A.}\ \bibnamefont {{MacArthur}}}, \bibinfo {author}
  {\bibfnamefont {Y.}~\bibnamefont {{Matsuoka}}}, \bibinfo {author}
  {\bibfnamefont {S.}~\bibnamefont {{Mineo}}}, \bibinfo {author} {\bibfnamefont
  {H.}~\bibnamefont {{Miyatake}}}, \bibinfo {author} {\bibfnamefont
  {S.}~\bibnamefont {{Miyazaki}}}, \bibinfo {author} {\bibfnamefont
  {S.}~\bibnamefont {{More}}}, \bibinfo {author} {\bibfnamefont
  {R.}~\bibnamefont {{Murata}}}, \bibinfo {author} {\bibfnamefont {S.~V.}\
  \bibnamefont {{Namiki}}}, \bibinfo {author} {\bibfnamefont {A.~J.}\
  \bibnamefont {{Nishizawa}}}, \bibinfo {author} {\bibfnamefont
  {M.}~\bibnamefont {{Oguri}}}, \bibinfo {author} {\bibfnamefont
  {N.}~\bibnamefont {{Okabe}}}, \bibinfo {author} {\bibfnamefont
  {S.}~\bibnamefont {{Okamoto}}}, \bibinfo {author} {\bibfnamefont
  {Y.}~\bibnamefont {{Okura}}}, \bibinfo {author} {\bibfnamefont
  {Y.}~\bibnamefont {{Ono}}}, \bibinfo {author} {\bibfnamefont
  {M.}~\bibnamefont {{Onodera}}}, \bibinfo {author} {\bibfnamefont
  {M.}~\bibnamefont {{Onoue}}}, \bibinfo {author} {\bibfnamefont
  {K.}~\bibnamefont {{Osato}}}, \bibinfo {author} {\bibfnamefont
  {M.}~\bibnamefont {{Ouchi}}}, \bibinfo {author} {\bibfnamefont
  {T.}~\bibnamefont {{Shibuya}}}, \bibinfo {author} {\bibfnamefont {M.~A.}\
  \bibnamefont {{Strauss}}}, \bibinfo {author} {\bibfnamefont {N.}~\bibnamefont
  {{Sugiyama}}}, \bibinfo {author} {\bibfnamefont {Y.}~\bibnamefont {{Suto}}},
  \bibinfo {author} {\bibfnamefont {M.}~\bibnamefont {{Takada}}}, \bibinfo
  {author} {\bibfnamefont {Y.}~\bibnamefont {{Takagi}}}, \bibinfo {author}
  {\bibfnamefont {T.}~\bibnamefont {{Takata}}}, \bibinfo {author}
  {\bibfnamefont {S.}~\bibnamefont {{Takita}}}, \bibinfo {author}
  {\bibfnamefont {M.}~\bibnamefont {{Tanaka}}}, \bibinfo {author}
  {\bibfnamefont {T.}~\bibnamefont {{Terai}}}, \bibinfo {author} {\bibfnamefont
  {Y.}~\bibnamefont {{Toba}}}, \bibinfo {author} {\bibfnamefont
  {H.}~\bibnamefont {{Uchiyama}}}, \bibinfo {author} {\bibfnamefont
  {Y.}~\bibnamefont {{Utsumi}}}, \bibinfo {author} {\bibfnamefont {S.-Y.}\
  \bibnamefont {{Wang}}}, \bibinfo {author} {\bibfnamefont {W.}~\bibnamefont
  {{Wang}}},\ and\ \bibinfo {author} {\bibfnamefont {Y.}~\bibnamefont
  {{Yamada}}},\ }\bibfield  {title} {\bibinfo {title} {{Second data release of
  the Hyper Suprime-Cam Subaru Strategic Program}},\ }\href
  {https://doi.org/10.1093/pasj/psz103} {\bibfield  {journal} {\bibinfo
  {journal} {\pasj}\ }\textbf {\bibinfo {volume} {71}},\ \bibinfo {eid} {114}
  (\bibinfo {year} {2019})},\ \Eprint {https://arxiv.org/abs/1905.12221}
  {arXiv:1905.12221 [astro-ph.IM]} \BibitemShut {NoStop}%
\bibitem [{\citenamefont {{Mandelbaum}}\ \emph {et~al.}(2018)\citenamefont
  {{Mandelbaum}}, \citenamefont {{Miyatake}}, \citenamefont {{Hamana}},
  \citenamefont {{Oguri}}, \citenamefont {{Simet}}, \citenamefont
  {{Armstrong}}, \citenamefont {{Bosch}}, \citenamefont {{Murata}},
  \citenamefont {{Lanusse}}, \citenamefont {{Leauthaud}}, \citenamefont
  {{Coupon}}, \citenamefont {{More}}, \citenamefont {{Takada}}, \citenamefont
  {{Miyazaki}}, \citenamefont {{Speagle}}, \citenamefont {{Shirasaki}},
  \citenamefont {{Sif{\'o}n}}, \citenamefont {{Huang}}, \citenamefont
  {{Nishizawa}}, \citenamefont {{Medezinski}}, \citenamefont {{Okura}},
  \citenamefont {{Okabe}}, \citenamefont {{Czakon}}, \citenamefont
  {{Takahashi}}, \citenamefont {{Coulton}}, \citenamefont {{Hikage}},
  \citenamefont {{Komiyama}}, \citenamefont {{Lupton}}, \citenamefont
  {{Strauss}}, \citenamefont {{Tanaka}},\ and\ \citenamefont
  {{Utsumi}}}]{HSCshear}%
  \BibitemOpen
  \bibfield  {author} {\bibinfo {author} {\bibfnamefont {R.}~\bibnamefont
  {{Mandelbaum}}}, \bibinfo {author} {\bibfnamefont {H.}~\bibnamefont
  {{Miyatake}}}, \bibinfo {author} {\bibfnamefont {T.}~\bibnamefont
  {{Hamana}}}, \bibinfo {author} {\bibfnamefont {M.}~\bibnamefont {{Oguri}}},
  \bibinfo {author} {\bibfnamefont {M.}~\bibnamefont {{Simet}}}, \bibinfo
  {author} {\bibfnamefont {R.}~\bibnamefont {{Armstrong}}}, \bibinfo {author}
  {\bibfnamefont {J.}~\bibnamefont {{Bosch}}}, \bibinfo {author} {\bibfnamefont
  {R.}~\bibnamefont {{Murata}}}, \bibinfo {author} {\bibfnamefont
  {F.}~\bibnamefont {{Lanusse}}}, \bibinfo {author} {\bibfnamefont
  {A.}~\bibnamefont {{Leauthaud}}}, \bibinfo {author} {\bibfnamefont
  {J.}~\bibnamefont {{Coupon}}}, \bibinfo {author} {\bibfnamefont
  {S.}~\bibnamefont {{More}}}, \bibinfo {author} {\bibfnamefont
  {M.}~\bibnamefont {{Takada}}}, \bibinfo {author} {\bibfnamefont
  {S.}~\bibnamefont {{Miyazaki}}}, \bibinfo {author} {\bibfnamefont {J.~S.}\
  \bibnamefont {{Speagle}}}, \bibinfo {author} {\bibfnamefont {M.}~\bibnamefont
  {{Shirasaki}}}, \bibinfo {author} {\bibfnamefont {C.}~\bibnamefont
  {{Sif{\'o}n}}}, \bibinfo {author} {\bibfnamefont {S.}~\bibnamefont
  {{Huang}}}, \bibinfo {author} {\bibfnamefont {A.~J.}\ \bibnamefont
  {{Nishizawa}}}, \bibinfo {author} {\bibfnamefont {E.}~\bibnamefont
  {{Medezinski}}}, \bibinfo {author} {\bibfnamefont {Y.}~\bibnamefont
  {{Okura}}}, \bibinfo {author} {\bibfnamefont {N.}~\bibnamefont {{Okabe}}},
  \bibinfo {author} {\bibfnamefont {N.}~\bibnamefont {{Czakon}}}, \bibinfo
  {author} {\bibfnamefont {R.}~\bibnamefont {{Takahashi}}}, \bibinfo {author}
  {\bibfnamefont {W.~R.}\ \bibnamefont {{Coulton}}}, \bibinfo {author}
  {\bibfnamefont {C.}~\bibnamefont {{Hikage}}}, \bibinfo {author}
  {\bibfnamefont {Y.}~\bibnamefont {{Komiyama}}}, \bibinfo {author}
  {\bibfnamefont {R.~H.}\ \bibnamefont {{Lupton}}}, \bibinfo {author}
  {\bibfnamefont {M.~A.}\ \bibnamefont {{Strauss}}}, \bibinfo {author}
  {\bibfnamefont {M.}~\bibnamefont {{Tanaka}}},\ and\ \bibinfo {author}
  {\bibfnamefont {Y.}~\bibnamefont {{Utsumi}}},\ }\bibfield  {title} {\bibinfo
  {title} {{The first-year shear catalog of the Subaru Hyper Suprime-Cam Subaru
  Strategic Program Survey}},\ }\href {https://doi.org/10.1093/pasj/psx130}
  {\bibfield  {journal} {\bibinfo  {journal} {\pasj}\ }\textbf {\bibinfo
  {volume} {70}},\ \bibinfo {eid} {S25} (\bibinfo {year} {2018})},\ \Eprint
  {https://arxiv.org/abs/1705.06745} {arXiv:1705.06745 [astro-ph.CO]}
  \BibitemShut {NoStop}%
\bibitem [{\citenamefont {{Tanaka}}\ \emph {et~al.}(2018)\citenamefont
  {{Tanaka}}, \citenamefont {{Coupon}}, \citenamefont {{Hsieh}}, \citenamefont
  {{Mineo}}, \citenamefont {{Nishizawa}}, \citenamefont {{Speagle}},
  \citenamefont {{Furusawa}}, \citenamefont {{Miyazaki}},\ and\ \citenamefont
  {{Murayama}}}]{HSCpz1}%
  \BibitemOpen
  \bibfield  {author} {\bibinfo {author} {\bibfnamefont {M.}~\bibnamefont
  {{Tanaka}}}, \bibinfo {author} {\bibfnamefont {J.}~\bibnamefont {{Coupon}}},
  \bibinfo {author} {\bibfnamefont {B.-C.}\ \bibnamefont {{Hsieh}}}, \bibinfo
  {author} {\bibfnamefont {S.}~\bibnamefont {{Mineo}}}, \bibinfo {author}
  {\bibfnamefont {A.~J.}\ \bibnamefont {{Nishizawa}}}, \bibinfo {author}
  {\bibfnamefont {J.}~\bibnamefont {{Speagle}}}, \bibinfo {author}
  {\bibfnamefont {H.}~\bibnamefont {{Furusawa}}}, \bibinfo {author}
  {\bibfnamefont {S.}~\bibnamefont {{Miyazaki}}},\ and\ \bibinfo {author}
  {\bibfnamefont {H.}~\bibnamefont {{Murayama}}},\ }\bibfield  {title}
  {\bibinfo {title} {{Photometric redshifts for Hyper Suprime-Cam Subaru
  Strategic Program Data Release 1}},\ }\href
  {https://doi.org/10.1093/pasj/psx077} {\bibfield  {journal} {\bibinfo
  {journal} {\pasj}\ }\textbf {\bibinfo {volume} {70}},\ \bibinfo {eid} {S9}
  (\bibinfo {year} {2018})},\ \Eprint {https://arxiv.org/abs/1704.05988}
  {arXiv:1704.05988 [astro-ph.GA]} \BibitemShut {NoStop}%
\bibitem [{\citenamefont {{Nishizawa}}\ \emph {et~al.}(2020)\citenamefont
  {{Nishizawa}}, \citenamefont {{Hsieh}}, \citenamefont {{Tanaka}},\ and\
  \citenamefont {{Takata}}}]{HSCpz2}%
  \BibitemOpen
  \bibfield  {author} {\bibinfo {author} {\bibfnamefont {A.~J.}\ \bibnamefont
  {{Nishizawa}}}, \bibinfo {author} {\bibfnamefont {B.-C.}\ \bibnamefont
  {{Hsieh}}}, \bibinfo {author} {\bibfnamefont {M.}~\bibnamefont {{Tanaka}}},\
  and\ \bibinfo {author} {\bibfnamefont {T.}~\bibnamefont {{Takata}}},\
  }\bibfield  {title} {\bibinfo {title} {{Photometric Redshifts for the Hyper
  Suprime-Cam Subaru Strategic Program Data Release 2}},\ }\href@noop {}
  {\bibfield  {journal} {\bibinfo  {journal} {arXiv e-prints}\ ,\ \bibinfo
  {eid} {arXiv:2003.01511}} (\bibinfo {year} {2020})},\ \Eprint
  {https://arxiv.org/abs/2003.01511} {arXiv:2003.01511 [astro-ph.GA]}
  \BibitemShut {NoStop}%
\bibitem [{\citenamefont {{Liu}}\ \emph {et~al.}(2015)\citenamefont {{Liu}},
  \citenamefont {{Pan}}, \citenamefont {{Li}}, \citenamefont {{Shan}},
  \citenamefont {{Wang}}, \citenamefont {{Fu}}, \citenamefont {{Fan}},
  \citenamefont {{Kneib}}, \citenamefont {{Leauthaud}}, \citenamefont {{Van
  Waerbeke}}, \citenamefont {{Makler}}, \citenamefont {{Moraes}}, \citenamefont
  {{Erben}},\ and\ \citenamefont {{Charbonnier}}}]{Liu2015}%
  \BibitemOpen
  \bibfield  {author} {\bibinfo {author} {\bibfnamefont {X.}~\bibnamefont
  {{Liu}}}, \bibinfo {author} {\bibfnamefont {C.}~\bibnamefont {{Pan}}},
  \bibinfo {author} {\bibfnamefont {R.}~\bibnamefont {{Li}}}, \bibinfo {author}
  {\bibfnamefont {H.}~\bibnamefont {{Shan}}}, \bibinfo {author} {\bibfnamefont
  {Q.}~\bibnamefont {{Wang}}}, \bibinfo {author} {\bibfnamefont
  {L.}~\bibnamefont {{Fu}}}, \bibinfo {author} {\bibfnamefont {Z.}~\bibnamefont
  {{Fan}}}, \bibinfo {author} {\bibfnamefont {J.-P.}\ \bibnamefont {{Kneib}}},
  \bibinfo {author} {\bibfnamefont {A.}~\bibnamefont {{Leauthaud}}}, \bibinfo
  {author} {\bibfnamefont {L.}~\bibnamefont {{Van Waerbeke}}}, \bibinfo
  {author} {\bibfnamefont {M.}~\bibnamefont {{Makler}}}, \bibinfo {author}
  {\bibfnamefont {B.}~\bibnamefont {{Moraes}}}, \bibinfo {author}
  {\bibfnamefont {T.}~\bibnamefont {{Erben}}},\ and\ \bibinfo {author}
  {\bibfnamefont {A.}~\bibnamefont {{Charbonnier}}},\ }\bibfield  {title}
  {\bibinfo {title} {{Cosmological constraints from weak lensing peak
  statistics with Canada-France-Hawaii Telescope Stripe 82 Survey}},\ }\href
  {https://doi.org/10.1093/mnras/stv784} {\bibfield  {journal} {\bibinfo
  {journal} {\mnras}\ }\textbf {\bibinfo {volume} {450}},\ \bibinfo {pages}
  {2888} (\bibinfo {year} {2015})},\ \Eprint {https://arxiv.org/abs/1412.3683}
  {arXiv:1412.3683 [astro-ph.CO]} \BibitemShut {NoStop}%
\bibitem [{\citenamefont {{Oguri}}\ \emph {et~al.}(2018)\citenamefont
  {{Oguri}}, \citenamefont {{Miyazaki}}, \citenamefont {{Hikage}},
  \citenamefont {{Mandelbaum}}, \citenamefont {{Utsumi}}, \citenamefont
  {{Miyatake}}, \citenamefont {{Takada}}, \citenamefont {{Armstrong}},
  \citenamefont {{Bosch}}, \citenamefont {{Komiyama}}, \citenamefont
  {{Leauthaud}}, \citenamefont {{More}}, \citenamefont {{Nishizawa}},
  \citenamefont {{Okabe}},\ and\ \citenamefont {{Tanaka}}}]{Oguri2018}%
  \BibitemOpen
  \bibfield  {author} {\bibinfo {author} {\bibfnamefont {M.}~\bibnamefont
  {{Oguri}}}, \bibinfo {author} {\bibfnamefont {S.}~\bibnamefont {{Miyazaki}}},
  \bibinfo {author} {\bibfnamefont {C.}~\bibnamefont {{Hikage}}}, \bibinfo
  {author} {\bibfnamefont {R.}~\bibnamefont {{Mandelbaum}}}, \bibinfo {author}
  {\bibfnamefont {Y.}~\bibnamefont {{Utsumi}}}, \bibinfo {author}
  {\bibfnamefont {H.}~\bibnamefont {{Miyatake}}}, \bibinfo {author}
  {\bibfnamefont {M.}~\bibnamefont {{Takada}}}, \bibinfo {author}
  {\bibfnamefont {R.}~\bibnamefont {{Armstrong}}}, \bibinfo {author}
  {\bibfnamefont {J.}~\bibnamefont {{Bosch}}}, \bibinfo {author} {\bibfnamefont
  {Y.}~\bibnamefont {{Komiyama}}}, \bibinfo {author} {\bibfnamefont
  {A.}~\bibnamefont {{Leauthaud}}}, \bibinfo {author} {\bibfnamefont
  {S.}~\bibnamefont {{More}}}, \bibinfo {author} {\bibfnamefont {A.~J.}\
  \bibnamefont {{Nishizawa}}}, \bibinfo {author} {\bibfnamefont
  {N.}~\bibnamefont {{Okabe}}},\ and\ \bibinfo {author} {\bibfnamefont
  {M.}~\bibnamefont {{Tanaka}}},\ }\bibfield  {title} {\bibinfo {title} {{Two-
  and three-dimensional wide-field weak lensing mass maps from the Hyper
  Suprime-Cam Subaru Strategic Program S16A data}},\ }\href
  {https://doi.org/10.1093/pasj/psx070} {\bibfield  {journal} {\bibinfo
  {journal} {\pasj}\ }\textbf {\bibinfo {volume} {70}},\ \bibinfo {eid} {S26}
  (\bibinfo {year} {2018})},\ \Eprint {https://arxiv.org/abs/1705.06792}
  {arXiv:1705.06792 [astro-ph.CO]} \BibitemShut {NoStop}%
\bibitem [{\citenamefont {{Schneider}}\ \emph {et~al.}(2002)\citenamefont
  {{Schneider}}, \citenamefont {{van Waerbeke}}, \citenamefont {{Kilbinger}},\
  and\ \citenamefont {{Mellier}}}]{Sch2002}%
  \BibitemOpen
  \bibfield  {author} {\bibinfo {author} {\bibfnamefont {P.}~\bibnamefont
  {{Schneider}}}, \bibinfo {author} {\bibfnamefont {L.}~\bibnamefont {{van
  Waerbeke}}}, \bibinfo {author} {\bibfnamefont {M.}~\bibnamefont
  {{Kilbinger}}},\ and\ \bibinfo {author} {\bibfnamefont {Y.}~\bibnamefont
  {{Mellier}}},\ }\bibfield  {title} {\bibinfo {title} {{Analysis of two-point
  statistics of cosmic shear. I. Estimators and covariances}},\ }\href
  {https://doi.org/10.1051/0004-6361:20021341} {\bibfield  {journal} {\bibinfo
  {journal} {\aap}\ }\textbf {\bibinfo {volume} {396}},\ \bibinfo {pages} {1}
  (\bibinfo {year} {2002})},\ \Eprint {https://arxiv.org/abs/astro-ph/0206182}
  {arXiv:astro-ph/0206182 [astro-ph]} \BibitemShut {NoStop}%
\bibitem [{\citenamefont {{Liu}}\ \emph {et~al.}(2014)\citenamefont {{Liu}},
  \citenamefont {{Wang}}, \citenamefont {{Pan}},\ and\ \citenamefont
  {{Fan}}}]{Liu2014}%
  \BibitemOpen
  \bibfield  {author} {\bibinfo {author} {\bibfnamefont {X.}~\bibnamefont
  {{Liu}}}, \bibinfo {author} {\bibfnamefont {Q.}~\bibnamefont {{Wang}}},
  \bibinfo {author} {\bibfnamefont {C.}~\bibnamefont {{Pan}}},\ and\ \bibinfo
  {author} {\bibfnamefont {Z.}~\bibnamefont {{Fan}}},\ }\bibfield  {title}
  {\bibinfo {title} {{Mask Effects on Cosmological Studies with Weak-lensing
  Peak Statistics}},\ }\href {https://doi.org/10.1088/0004-637X/784/1/31}
  {\bibfield  {journal} {\bibinfo  {journal} {\apj}\ }\textbf {\bibinfo
  {volume} {784}},\ \bibinfo {eid} {31} (\bibinfo {year} {2014})},\ \Eprint
  {https://arxiv.org/abs/1304.2873} {arXiv:1304.2873 [astro-ph.CO]}
  \BibitemShut {NoStop}%
\bibitem [{\citenamefont {{Hildebrandt}}(2016)}]{Hilde2016}%
  \BibitemOpen
  \bibfield  {author} {\bibinfo {author} {\bibfnamefont {H.}~\bibnamefont
  {{Hildebrandt}}},\ }\bibfield  {title} {\bibinfo {title} {{Observational
  biases in flux magnification measurements}},\ }\href
  {https://doi.org/10.1093/mnras/stv2575} {\bibfield  {journal} {\bibinfo
  {journal} {\mnras}\ }\textbf {\bibinfo {volume} {455}},\ \bibinfo {pages}
  {3943} (\bibinfo {year} {2016})},\ \Eprint {https://arxiv.org/abs/1511.01352}
  {arXiv:1511.01352 [astro-ph.GA]} \BibitemShut {NoStop}%
\bibitem [{\citenamefont {Raftery}(1995)}]{Raftery1995}%
  \BibitemOpen
  \bibfield  {author} {\bibinfo {author} {\bibfnamefont {A.~E.}\ \bibnamefont
  {Raftery}},\ }\bibfield  {title} {\bibinfo {title} {Bayesian model selection
  in social research (with discussion)},\ }\href
  {https://doi.org/10.2307/271063} {\bibfield  {journal} {\bibinfo  {journal}
  {Sociological Methodology}\ }\textbf {\bibinfo {volume} {25}},\ \bibinfo
  {pages} {111} (\bibinfo {year} {1995})}\BibitemShut {NoStop}%
\end{thebibliography}%

\end{document}